\documentclass[]{aa} % for the abstract without structuration 
\usepackage{graphicx}
\usepackage[usenames, dvipsnames]{xcolor}
\usepackage{txfonts}
\usepackage{natbib}
\usepackage{psfrag}
\usepackage[hidelinks]{hyperref}
\hypersetup{colorlinks=false}

\bibpunct{(}{)}{;}{a}{}{,} % to follow the A&A style
\def\be{\begin{equation}}
\def\ee{\end{equation}}
\def\kms{{\rm\,km\,s^{-1}}}
\def\kmskpc{{\rm\,km\,s^{-1}\,{kpc}^{-1}}}
\def\pc{{\rm\,pc}}
\def\Msun{{\rm\,M_\odot}}
\def\Rsun{{\rm\,R_\odot}}
\def\yr{{\rm\,yr}}

\def\Gyr{{\rm\,Gyr}}
\def\deg{{^\circ}}

\def\kpc{{\rm\,kpc}}
\def\masyr{{\rm\,mas/yr}}
\def \sun{{_\odot}}
\def\1s{{1$\sigma$}}
\def\2s{{2$\sigma$}}
\def\3s{{3$\sigma$}}

\newcommand{\bea}       {\begin{array}}
\newcommand{\eea}       {\end{array}}
\newcommand{\ben}       {\begin{eqnarray}}
\newcommand{\een}       {\end{eqnarray}}
\newcommand{\bsq}       {\begin{mathletters}}
\newcommand{\esq}       {\end{mathletters}}

\newcommand{\Op}        {\Omega_{\rm p}}
\newcommand{\Asp}       {A_{\rm sp}}

\newcommand{\vphi}      {V_{\phi}}
\newcommand{\vr}        {V_{R}}
\newcommand{\vt}        {V_{\rm T}}
\newcommand{\vlos}      {V_{\rm los}}

\newcommand{\Ws}        {W_\odot}
\newcommand{\Vs}        {V_\odot}
\newcommand{\Us}        {U_\odot}

\newcommand{\Sec}[1]    {Section~\ref{#1}}
\newcommand{\Fig}[1]    {Figure~\ref{#1}}
\newcommand{\fig}[1]    {Fig.~\ref{#1}}
\newcommand{\figs}[2]   {Figs.~\ref{#1} and \ref{#2}}

\newcommand{\aunits}    {{\rm\,(km\,s^{-1})^2\,{kpc}^{-1}}}
\newcommand{\D} {\Delta}
\newcommand{\De}        {\Delta_{\rm exp}}
\newcommand{\los}{{\textit{los }}}

\def\aap{A\&A}

\def\2s{2-$\sigma$}
\def\3s{3-$\sigma$}

\def\gaia{\emph{Gaia }}

\begin{document}

 \title{Kinematics of symmetric Galactic longitudes to probe\\ the spiral arms of the Milky Way with {\em Gaia}}

  % \subtitle{I. Overviewing the $\kappa$-mechanism}

   \author{T. Antoja
          \inst{1}\fnmsep\thanks{ESA Research Fellow.}
          \and
           S. Roca-F\`abrega\inst{2, 3}
          \and
          J. de Bruijne\inst{1}
 \and
          T. Prusti\inst{1}
          %    C. Ptolemy\inst{2}\fnmsep\thanks{Just to show the usage
      %    of the elements in the author field}
          }

%   \institute{Scientific Support Office, European Space Research and Technology Center, European Space Agency (ESA-ESTEC), PO Box 299, 2200 AG Noordwijk, The Netherlands\\
   \institute{Directorate of Science, European Space Agency (ESA-ESTEC), PO Box 299, 2200 AG Noordwijk, The Netherlands\\ %European Space Research and Technology Center,
              \email{tantoja@cosmos.esa.int}
         \and
             Departament d'Astronomia i Meteorologia and IEEC-UB, Institut de Ci\`encies del Cosmos de la Universitat de Barcelona,
Mart\'i i Franqu\`es, 1, E-08028 Barcelona, Spain
 \and
Racah Institute of Physics, The Hebrew University of Jerusalem, Edmond J. Safra Campus, Givat Ram, Kaplun building, office 110, Jerusalem 91904, Israel\\
% \and
%             xxxx, ...\\     %        \email{c.ptolemy@hipparch.uheaven.space}
    %         \thanks{The university of heaven temporarily does not
    %                 accept e-mails}
             }

%              \title{}% 
%           %Tracing the Hercules stream across the Galaxy \\to constrain the Galactic bar}%change
% %: \\from the Solar neighbourhood to the district
% %   \subtitle{I. Solar neighbourhood}
% 
%    \author{T. Antoja\inst{1}
% \and
% A. Helmi\inst{1}
% \and
% W. Dehnen\inst{2}
% 
% 
% \fnmsep%\thanks{Just to show the usage           of the elements in the author field}
%           }
% 
%    \institute{Kapteyn Astronomical Institute, University of Groningen, PO Box 800, 9700 AV Groningen, the Netherlands\\
%               \email{antoja@astro.rug.nl}
% %  \and
% %University of Leicester, University Road, Leicester LE1 7RH, UK
% %\and
% %Universit\'e de Strasbourg, CNRS, Observatoire, 11 rue de l'Universit\'e F-67000 Strasbourg, France
%        }
\titlerunning{Symmetric Galactic longitudes and spiral arms with {\em Gaia}}
%\authorrunning

   \date{Received XX; accepted XX}

 \abstract{}
{We model the effects of the spiral arms of the Milky Way on the disk stellar kinematics in the \gaia observable space. We also estimate the \gaia capabilities of detecting the predicted signatures. }
{We use both controlled orbital integrations in analytic potentials and self-consistent simulations. We introduce a new strategy to investigate the  effects of spiral arms, which consists of comparing the stellar kinematics of symmetric Galactic longitudes ($+l$ and $-l$), in particular the median transverse velocity as determined from parallaxes and proper motions. This approach does not require the assumption of an axisymmetric model because it involves an internal comparison of the data.}{The typical differences between the transverse velocity in symmetric longitudes in the models are of the order of $\sim2\kms$, but can be larger than $10\kms$ for certain longitudes and distances. The longitudes close to the Galactic centre and to the anti-centre are those with larger and smaller differences, respectively.  The  differences between the kinematics for $+l$ and $-l$ show clear trends that depend strongly on the properties of  spiral arms.  Thus, this method can be used to quantify the importance of the effects of  spiral arms on the orbits of stars in the different regions of the disk, and to constrain the location of the arms, main resonances and, thus, pattern speed. Moreover, the method allows us to test different origin scenarios of spiral arms and the dynamical nature of the spiral structure (e.g. grand design versus transient multiple arms). We estimate the number of stars of each spectral type that \gaia will observe in certain representative Galactic longitudes, their characteristic errors in distance and transverse velocity, and the error in computing the median velocity as a function of distance. We will be able to measure the median transverse velocity exclusively with \gaia data,  with precision smaller than $\sim1\kms$ up to distances of $\sim4$-$6\kpc $ for certain giant stars, and  up to $\sim2$-$4\kpc$ and better kinematic precision ($\lesssim0.5\kms$) for  certain sub-giants and dwarfs. These are enough to measure the typical signatures seen in the models.}
{The \gaia catalogue will allow us to use the presented approach successfully and improve significantly upon current studies of the dynamics of the spiral arms of our Galaxy. We also show that a similar strategy can be used with line-of-sight velocities, which could be applied to \gaia data and to upcoming spectroscopic surveys.}

   \maketitle

%________________________________________________________________

\section{Introduction}\label{intro}

 Spiral arms in the Milky Way (MW) and in other galaxies are conspicuous features that impact major aspects of the dynamics and evolution of the disks. For instance, spiral arms gather gas-forming massive clouds, which affect the global gas dynamics and star formation \citep{Dobbs2011}.
 The stellar orbits are also perturbed by spiral arms, with resonant trapping and orbital stochasticity primarily at corotation (CR) and in the Lindblad resonances (LR; e.g. \citealt{Contopoulos1986}). This can lead to optical imprints in the global galaxy morphology, such as gaps and bifurcations \citep{Elmegreen1990,Junqueira2011}, as well as to streaming motion in the stellar and gaseous components \citep{Font2011,ErrozFerrer2016}. The spiral structure perturbs the kinematics of star-forming regions \citep{Reid2014} and has an effect on the disruption of clusters \citep{Gieles2007,Fujii2012,MartinezBarbosa2016}. In addition, spiral arms can also cause radial migration \citep{Sellwood2002,Roskar2008,Minchev2010b}, which has an influence on the disk age-metallicity relation and the
 chemical radial and azimuthal gradients that we measure today. Even thick disks of galaxies may be affected by spiral arms \citep{Solway2012}.
%the bar \citet{Monari2013,Antoja2015} and by

 Yet, the exact impact of spiral arms on all these aspects in the MW is not well constrained. Moreover, we still long for a comprehensive theory of the spiral structure. Unknown aspects of  spiral arms are their long- or short-lived condition, their excitation mechanism and relation with the bar, the difference between gaseous and stellar arms, how they rotate with respect to the stellar disk, and their exact nature and dynamics. For a review of theories and implications, see e.g. \citet{Grand2012} and \citet{Dobbs2014}.

For the MW, despite not having as a complete a vision of the spiral structure as in external galaxies, we have the possibility of directly measuring its (nearby) kinematic influence on stellar orbits. From the theoretical side, the use of simulations (both with analytic potentials and self-consistent simulations) is now a generalised tool for the study of the nature and dynamics of the spiral structure. From modelling, we know that the dynamical moving   groups are a predicted kinematic signature of the arms. These moving groups are a set of stars trapped in peculiar orbits of the spiral gravitational potential that form over-densities in the kinematic space \citep[e.g.][]{DeSimone2004,Quillen2005,Antoja2009,Antoja2011}. These over-densities have been observed in the local vicinity \citep{Dehnen1998,Famaey2005,Antoja2008} and in the solar suburbs \citep{Antoja2012}.
%that present dispersion in the chemical and age distributions, contrary to what was expected if those groups were remnants of disrupted clusters. 
Some studies have attempted to match these observed over-densities to the modelled effects of spiral arms and, thus, constrain properties of the spiral structure such as the pattern speed. However, it has been shown that several models, for instance involving the Galactic bar \citep{Antoja2014}, can explain the same kinematic substructure \citep{Antoja2011} and, therefore, this task is not so straightforward.
% These relatively local  findings have probed useful to even derive properties of the global spiral structure,a \citet{} and similarly of the other main non=-axymetri, namely the galactic bar \citet{Antoja2014}

Another testable signature of spiral arms is the perturbation of the moments of the disk velocity distribution compared to an axisymmetric case. \citet{Minchev2007b} showed with test particle simulations how the Oort's constants are modified by the presence of spiral arms following the density wave theory. \citet{Vorobyov2008} showed with simulations how moments, such as velocity dispersions or vertex deviation, followed peculiarities, mainly in the outer edges of the arms. \citet{Minchev2008} simulated the perturbations in the line-of-sight (\textit los) velocities, finding indicators of resonance locations, such as an increase of velocity dispersion in the 2:1 Outer LR (OLR) compared to an axisymmetric disk. \citet{RocaFabrega2014} found with N-body and test particle simulations that the vertex deviation changed its sign at the CR resonance and characterised these changes for different spiral arm types. \citet{Faure2014} showed that spiral arms can create not only radial velocity gradients across the disk but also  vertical velocity gradients and gradients in the vertical direction (see also \citealt{Debattista2014}).

Recent spectroscopic surveys show evidence that such velocity gradients, especially in the \los velocities, exist in our Galaxy (\citealt{Siebert2011a,Widrow2012,Williams2013}, \citealt{Carlin2013}).%,,Bovy2015}. 
%\citet{Liu2012b} \citet{McMillan2013}
These gradients are a clear quantitative observable that can be compared to models to constraint properties of the arms such as \citet{Siebert2012} accomplished with RAVE data. However, \citet{Bovy2015} showed that the power spectrum of the observed gradients is more compatible with the bar perturbation (see also \citealt{Monari2014}), while in other studies the gradients were related to the oscillations excited by orbiting Galaxy satellites (\citealt{Purcell2011,Gomez2013}, \citealt{Carlin2013}, \citealt{Widrow2014},\citealt{Widrow2015}). 

\gaia \citep{Perryman2001} is a cornerstone mission of the European Space Agency successfully launched in December 2013 to produce the largest and most precise census of positions, distances, velocities and other stellar properties for a billion stars. \gaia will extend without precedent the horizon of our kinematic studies. 
%The mission will allow us to measure with great precision the velocity gradients in the MW. 
In the \gaia era, it is crucial to model in detail all mechanisms that can create kinematic perturbations to i) disentangle the causes of the observed gradients, ii) establish the relative importance of all the phenomena affecting the disk dynamics, and iii) use the measured signatures of all these processes to put constraints on them.  
 
 In this work, we extend the modelling of the effects of spiral arms on the disk kinematics that is necessary for the above task. We study the typical trends and magnitudes of the effects of the arms with both controlled orbital integrations in analytic potentials and self-consistent simulations (N-body). Thus we use different types of spiral arms, namely transient arms, strong arms linked to a central bar, and arms in the classical density wave theory. While previous studies focused on \los velocities, here we explore the transverse velocity for the first time. In the \gaia catalogue, these will be available for a larger number of stars and for farther distances in the disk compared to the \los velocities. We use the velocity moments, mostly the median transverse velocity, to quantify the effects of spiral arms  since these have well-defined errors and can be statistically compared to observations straightforwardly. We present a novel approach to measure the effects of  spiral arms that compares symmetric Galactic longitudes and does not require the assumption of an axisymmetric model as in previous studies mentioned above. We also use the simulated catalogue \gaia Universe Model Snapshot  (GUMS; \citealt{Robin2012}) to estimate the number of stars and kinematic precision of the \gaia sources as a function of distance to determine the detectability of the spiral arm effects that we find in the models.
 
 The paper is organised as follows: Section \ref{sim} presents the models; Section \ref{effects} shows a global view of the effects of the arms on the stellar kinematics; Section \ref{sym} presents the method of exploring symmetric longitudes; in Section \ref{gaia} we explore whether \gaia will be able to detect the modelled signatures; 
 %in Section \ref{others} we comment on the possibility of using our approach to other kinematic moments and to \los velocities; 
 and we conclude in Section \ref{conclusions}.

%\gaia 
%\citet{Antoja2011b}
%\citet{RomeroGomez2015}
%\citet{Hunt2015}

 %%%%%%%%%%%%%%%%%%%%%%%%%%%%%%%%%%%%%%%%%%%%%%%%%%%%
 \section{The simulations}\label{sim}
\subsection{Simulations with analytic potentials}\label{simTP}

We use a set of eight models that consist of orbital integrations of massless particles under an analytic gravitational potential (Table~\ref{t_sim}). These are controlled simulations that we tune within the estimated limits of the MW spiral structure. These simulations have $2\times10^7$ particles and are two-dimensional. It has been shown in \citet{Faure2014} that the kinematic effects of spiral arms do not depend on the height above the plane up to $500\pc$. We assume here that our results are, therefore, valid up to this height. We focus on the in-plane components of the velocity as the vertical components are studied elsewhere \citep[e.g.][]{Faure2014}.

The potential includes an axisymmetric part and spiral arms. We use the model of \citet{Allen1991}, which has a flattened disk, a bulge, and a spherical halo, for the axisymmetric
part. 
The first two are modelled as Miyamoto--Nagai potentials \citep{Miyamoto1975}. 
In this model, a value of $\Rsun=8.5\kpc$ for the Sun's galactocentric distance and a circular speed of $V_c(R\sun)=220\kms$ are adopted. 
The potential of the spiral arms follows the tight-winding approximation (TWA) model \citep{Bible2008,Lin1969} as described in \citet{Antoja2011}. The only difference is that here the potential of the spiral arms  is introduced gradually in time (see below) instead of abruptly. 

\begin{figure}
   \centering
\includegraphics[width=0.29\textwidth]{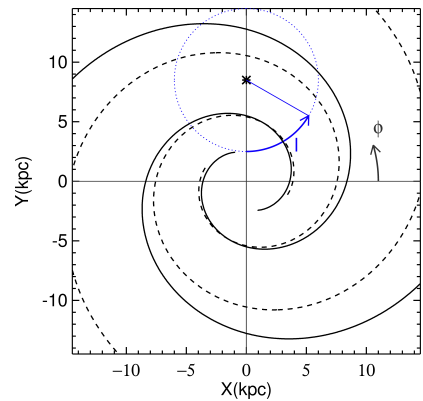}
  \caption{Scheme of the MW plane with the locus 1 (solid) and locus 2 (dashed) employed in the simulations with analytic potentials. The Galaxy rotates clockwise in this picture.  }
         \label{coord}
   \end{figure}

In the models TWA0 to TWA3, we only vary the pattern speed $\Op$ between 12 and $30\kmskpc$, which covers the MW range of different determinations \citep{Gerhard2011}. The other models have the same pattern speed as TWA1 but different initial conditions (TWA10), spiral arms amplitude (TWA11 and TWA12) and locus (TWA13). %, using as standard reference the model TWA0 with $\Op=18\kmskpc$. 
We use two different loci with pitch angles $i$ of 15.5 (locus 1) and $12.8\deg$ (locus 2), following \citet{Drimmel2001}  and \citet{Vallee2008}, respectively. \Fig{coord} shows locus 1 (solid) and locus 2 (dashed). Both loci have two arms representing Perseus and Scutum, with tangencies consistent with observations within the errors: $l=-21\pm2\deg$ for Perseus, and at $l=-51\pm3\deg$ and $l=32\pm3\deg$ for Scutum \citep{Vallee2008}. 
%,  (sometimes called Scutum--Centaurus or Scutum--Crux) arms 
These are the major arms seen in the infrared Spitzer/GLIMPSE survey \citep{Churchwell2009}. 
  The main difference between the two loci is in the outer Galaxy, with the Perseus arm closer to the Sun in locus 2. The top left panel in \Fig{XY} shows the density output of model TWA1.
We vary the amplitude of the spiral arm potential $\Asp$ between the lower and upper MW limits estimated in \citet{Antoja2011}, which are $\Asp=[850,1300]\aunits$ for locus 1 and $\Asp=[650,1100]\aunits$ for locus 2. In most cases, we use the lower limit, except for model TWA12 and TWA13, for which we use  higher limits, and model TWA11, for which we adopt a limit that is smaller than the lower limit.%The arm-interarm density contrast $K\equiv(\sigma_0+\delta\sigma)/(\sigma_0-\delta\sigma)$, where $\sigma_0$ is the axisymmetric surface density and $\delta\sigma$ is the density enhancement on the spiral arms, is also varied between the lower ($K=1.32$) and upper ($K=1.6$) MW limits estimated in \citet{Antoja2011}. Most of our models use the lower limit. 

The initial conditions are generated as detailed in Appendix A from \citet{RomeroGomez2015}, except that here they are two-dimensional. They consist of the Miyamoto-Nagai disk as in the axisymmetric part of the potential. The velocities are approximated as Gaussian distributions, with the dispersion in the radial velocity component decaying as a function of radius with a scale length that is twice that of the density. The azimuthal component is related to the radial component through the epicyclic relation. We also include the asymmetric drift. We use two sets of initial conditions with radial velocity dispersion in the solar neighbourhood $\sigma_{\vr}(\Rsun)$ of $20\kms$ (IC20) and $40\kms$ (IC40).

As the initial conditions are just approximated, stationarity is not guaranteed. 
%not fully consistent with the potential, 
To ensure that the system is in steady state and has achieved complete phase mixing, we first integrate the orbits in the axisymmetric potential for $6.1 \Gyr$. After this, the spiral arms forces are gradually introduced following equation 4 of \citet{Dehnen2000} used for a similar purpose but for the Galactic bar. We do this in four revolutions of the arms, which corresponds to 0.9-$1.2\Gyr$ depending on the pattern speed of the model. From the moment that the spiral arms begin to grow, the orbits are integrated for another 14.9 $\Gyr$. Thus, in the integration the spiral arms are fully grown during $\sim14\Gyr$. We note that the MW may not be in a stationary state. Our integration scheme does not try to mimic the evolution of the MW, but aims to obtain a set of final conditions in which the particles are fully phase mixed. 

%\red{density destroyed beyond 2:1 OLR and inner to 1:2 ILR}. 

 \begin{table}
  \setlength{\tabcolsep}{1.5pt}
\caption{Models used in this study and main parameters: type of locus, pattern speed, spiral amplitude, initial conditions, and number of particles in the simulation.}\label{t_sim}      
 \centering          
% \begin{tabular}{llllll}     % 11 columns 
 \begin{tabular}{lrcccc}     % 11 columns 
  %   \tabcolsep 3.pt
 \hline\hline    
 Model& locus&$\Op$ & $\Asp$  &IC    &$N$\\
      &      &($\kmskpc$)  &($\aunits$)&      &($10^6$)\\\hline 
TWA0  & 1    &12           &850        & IC20 &$20$\\
TWA1  & 1    &18           &850        & IC20 &$20$\\
TWA2  & 1    &24           &850        & IC20 &$20$\\
TWA3  & 1    &30           &850        & IC20 &$20$\\
TWA10 & 1    &18           &850        & IC40 &$20$\\
TWA11 & 1    &18           &400        & IC20 &$20$\\
TWA12 & 1    &18           &1300       & IC20 &$20$\\
TWA13 & 2    &18           &1100       & IC20 &$20$\\
%TWA11 & 1    &18           &1300       & IC20 &$20$\\
B1    & $\sim2$   &30-40        &-          & - &$1$\\
B5    & -  &20-25        &-          & - &$5$\\
U5, U5b    & -   &-            &-          & - &$5$\\
 \hline                  
 \end{tabular}
 \end{table}
% \multicolumn{2}{l}{Input}&\multicolumn{1}{c}{$\pb$(MAX)}&\multicolumn{1}{c}{E($\pb$)}&\multicolumn{1}{c}{$\sigma_{\pb}$}&\multicolumn{1}{c}{$\Ob/\Oo$(MAX)}&\multicolumn{1}{c}{E($\Ob/\Oo$)}&\multicolumn{1}{c}{$\sigma_{\Ob/\Oo}$}&\multicolumn{1}{c}{$\rho_{\pb\Ob}$}&\multicolumn{1}{c}{E($\Ob/\Oo|\pb=input$)}&\multicolumn{1}{c}{E($\Ob|\pb=input$)}\\ 

\subsection{N-body simulations}\label{simNB}

 \begin{figure}
   \centering
 \includegraphics[width=0.24\textwidth]{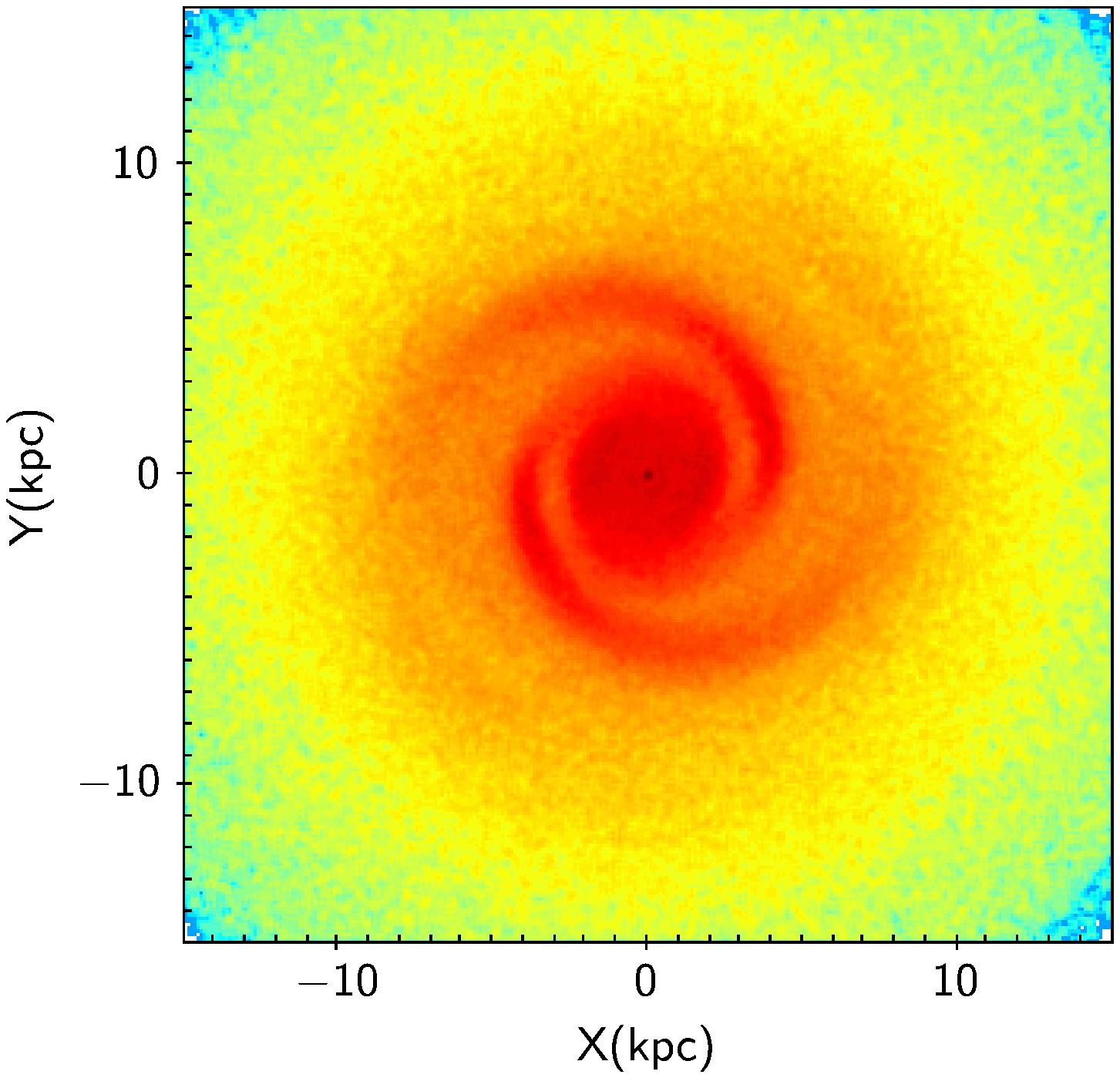}
 \includegraphics[width=0.24\textwidth]{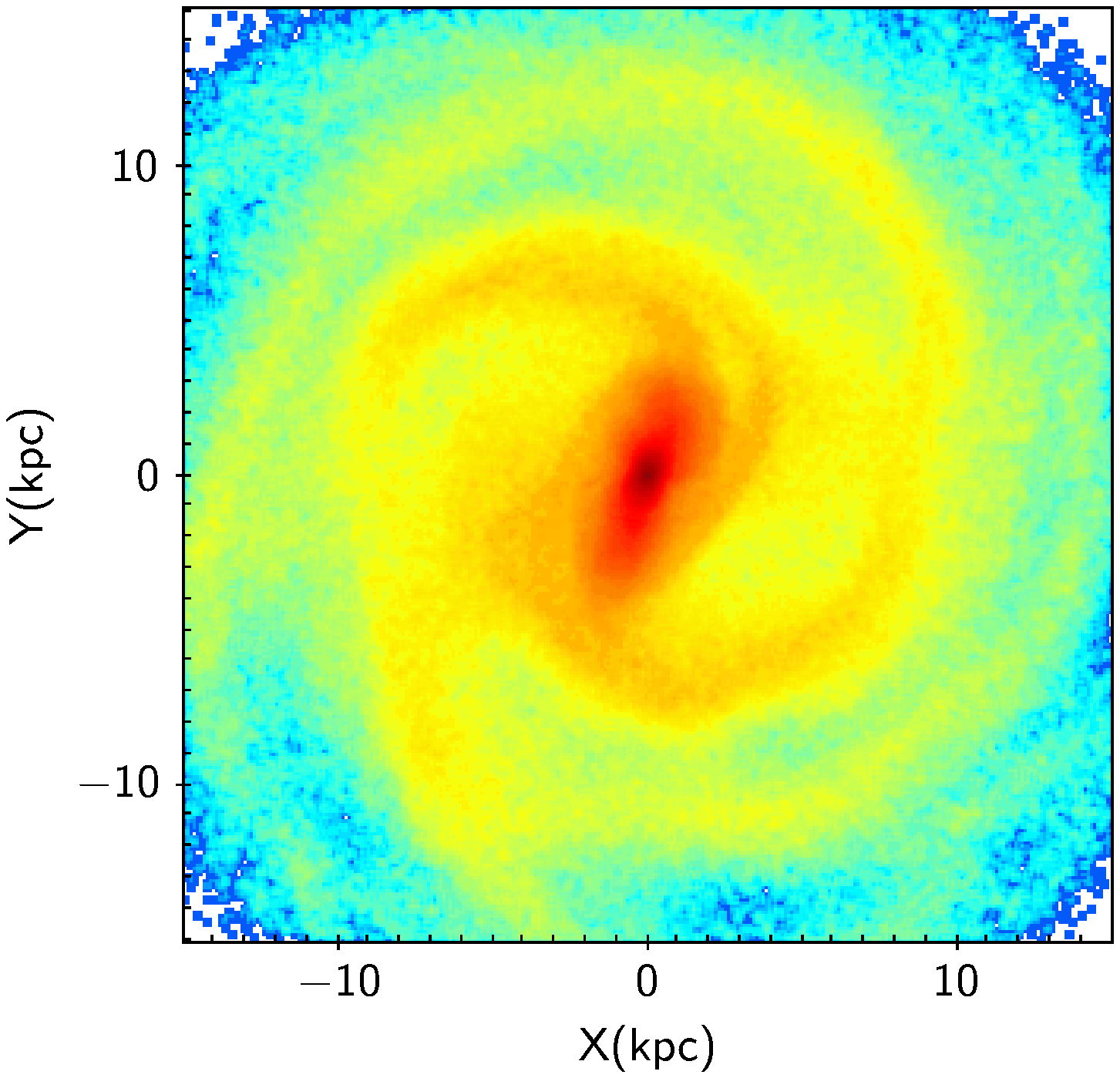}
 
 \includegraphics[width=0.24\textwidth]{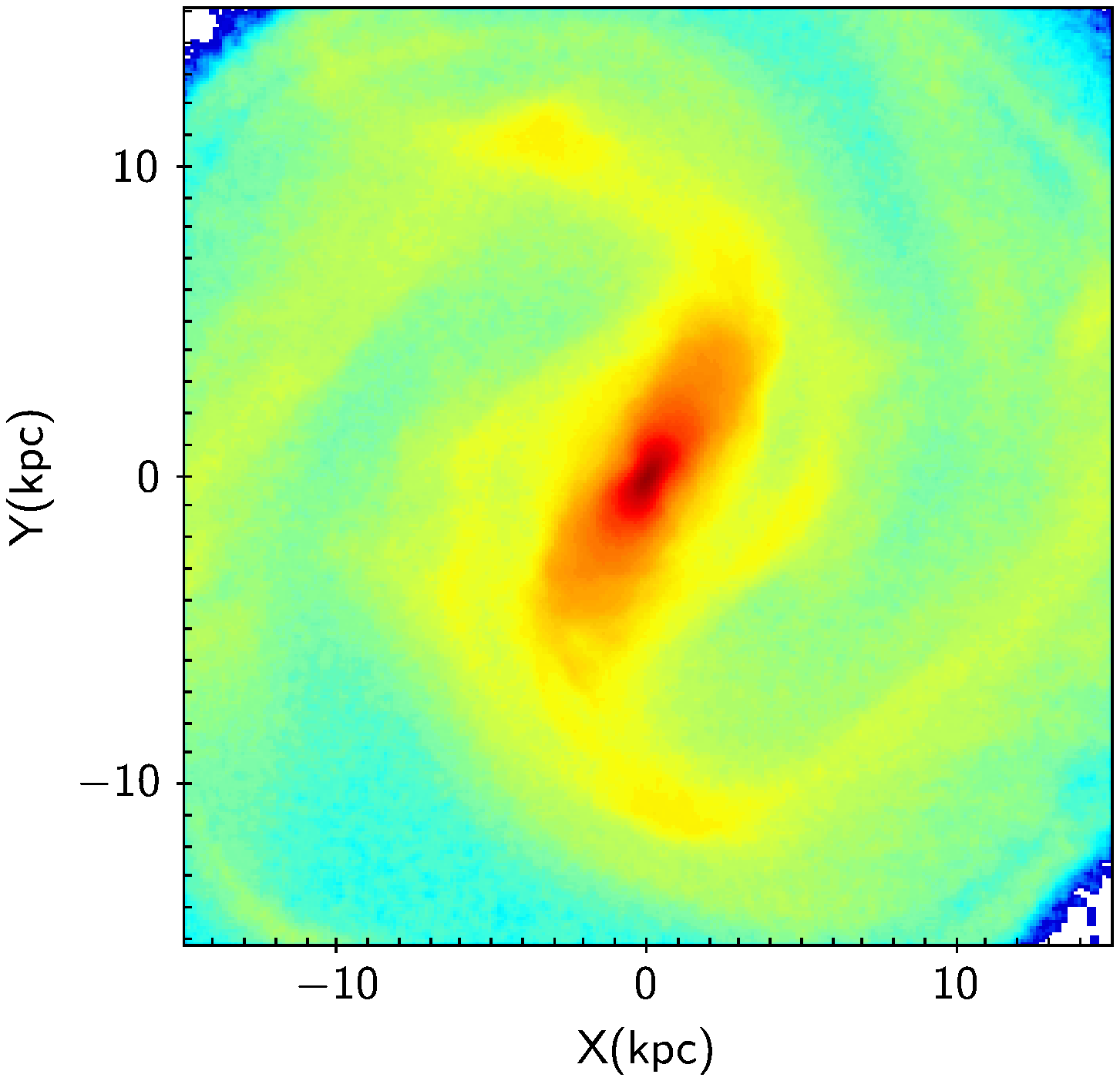}
 \includegraphics[width=0.24\textwidth]{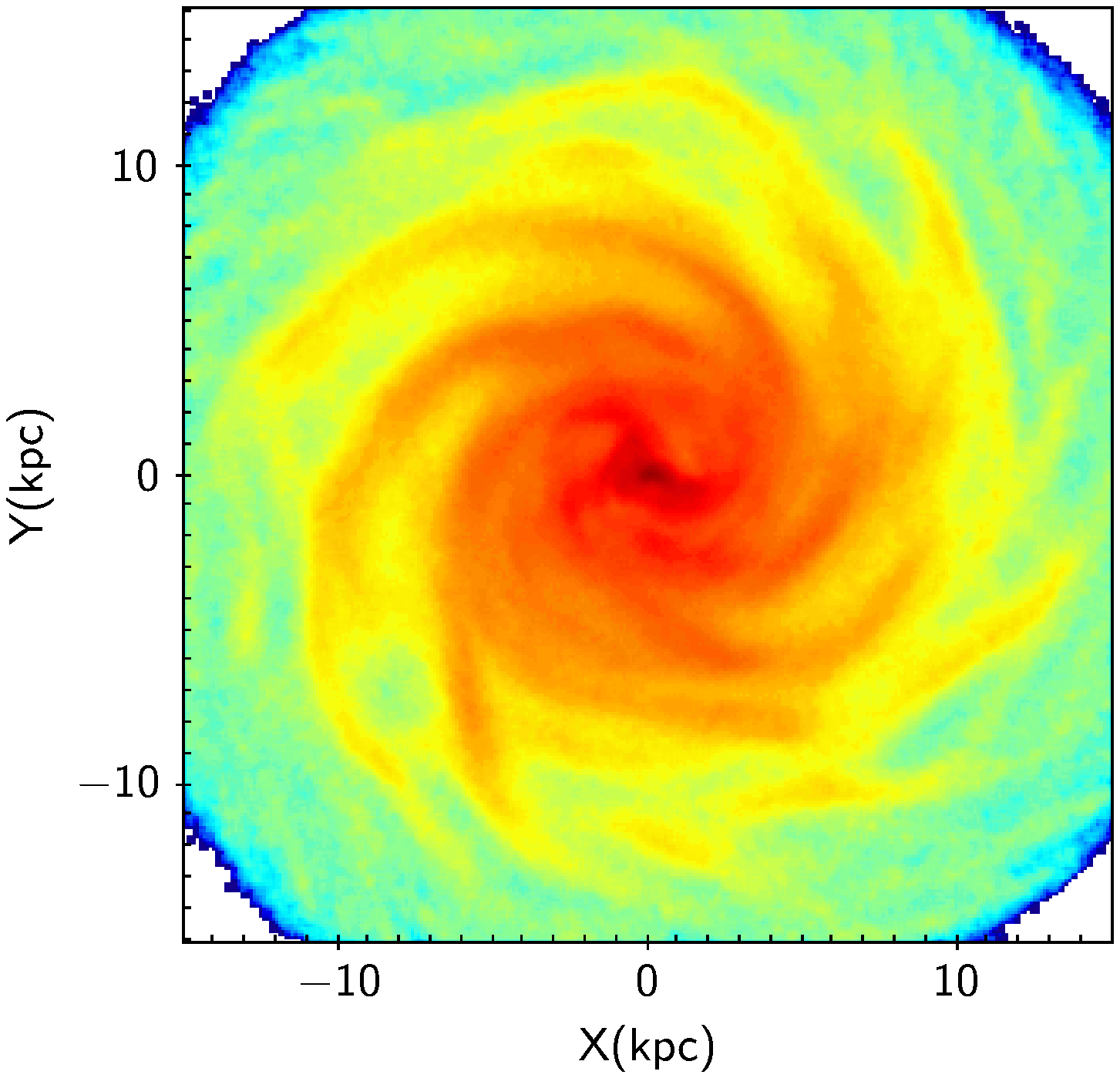}
  \caption{Disk density distribution of models TWA1 (top left), B1 (top right), B5 (bottom left), and U5 (bottom right).  }
         \label{XY}
   \end{figure}

We use the models B1, B5, and U5 from \citet{RocaFabrega2013,RocaFabrega2014}. These are fully self-consistent models with a live exponential disk and live dark matter Navarro–Frenk–White (NFW) halo, run with the pure N-body adaptive refinement tree (ART) code \citep{Kravtsov1997}. The spatial resolution is $44\pc$ for B1 and $11\pc$ for B5 and U5 models. The number of disk particles is 1 million for B1, and 5 million for B5 and U5 models. The differences between B and U models are their initial disk mass ($5\times10^{10}$ and $3.75\times10^{10}$ $\Msun$,  for B and U models, respectively) and rotation curve. B models have heavier disks and, thus, a similar contribution from the halo and the disk components to the total circular velocity curve. B models develop strong bars in their central regions, which drives the formation of dominant bi-symmetric spiral arms. In the U5 model, the halo contribution is much higher, which prevents the formation of a central barred structure in the first $\Gyr$. The U5 model develops weak multiple-armed structures. The scale length of the disk is 4.2 and 4.1 $\kpc$ for B and U models, respectively, and the scale height is $0.2\kpc$ for both. The stability Toomre parameter Q at solar radius is 3.3 and 1.4, for models B and U, respectively,  and $>1$, thus locally stable, for most of the disk. Details on the initial conditions, initial parameters, and code convergence tests can be found in \citet{RocaFabrega2013}.

The top right and bottom panels of \fig{XY} show the disk density of these three models. These models do not necessarily resemble the MW in terms of their rotations curve, total mass, or disk mass, but we use them here as interesting case studies.
Model B1 has a  bar with a length similar to that of the MW and also a consistent orientation of the spiral arms with respect to the bar that resembles locus 1.  The spiral arms of this model rotate at an approximately constant pattern speed at all radii of  $\sim30$-$40\kmskpc$. Model B5 does not resemble the MW because it has a bar that is too strong and long and the tangencies of the arms are not consistent with observations, but this model serves as a comparison model. The pattern speed in this case is $\sim20$-$25\kmskpc$. The CR resonance for B1 and B5 is inside the solar radius, contrary to most of our analytic models. U5 corresponds to a galaxy with floculent spiral structure with five major arms that do not rotate with a fixed pattern speed but corotate at the velocity of the disk particles. This is in total opposition to all previous models (both analytic and N-body) and it is an interesting case to be tested for the MW. 

For the models B1 and B5, we have oriented the bar at $\sim20\deg$ with respect to the line Sun-Galactic centre (GC), similar to the COBE/DIRBE bar \citep{Gerhard2002}. We orient model U5 to have the Sun between two arms. We also use the case U5b, which is model U5 but for a different orientation of the arms in which the Sun is on top of an arm. In all these simulations we have selected disk particles with height $|Z|<0.6\kpc$, which corresponds to 3 scale heights.

\subsection{Definitions}\label{def}

We take the location of the Sun at $X=0$ and $Y=8.5\kpc$, which is indicated with a star in \Fig{coord}. The Galaxy and the arms rotate clockwise in this plot. Throughout the paper we use Galactocentric cylindrical coordinates $R$ and $\phi$  (see convention in \Fig{coord}) and velocity components $\vr$ (positive with increasing $R$) and $\vphi$ (positive opposite to rotation). 

Here we focus on the \gaia observable space and as explained before, on the in-plane components of the velocity. With the usual transformations (e.g. \citealt{BinneyMerrifield1998}), we turn the simulations into the observables: sky positions  $l$ and $b$, parallax $\varpi$, \los velocity $\vlos$, and proper motions $\mu_l*\equiv\mu_l\cos b$ and $\mu_b$. For this transformation, we assume that the Sun moves at a peculiar velocity of $(\Us,\Vs,\Ws)=(9,12,7)\kms$ (\citealt{Dehnen1998}; see discussion in Section~\ref{sym3}) with $U$ positive towards the Galactic centre. We will use the transverse velocity in Galactic longitude $\vt\equiv k\, D\, \mu_l*$, where $k=4.7404705 \kms\yr {\rm s}^{-1}$, the distance $D$ is in $\pc$, and $\mu_l*$ is in $\masyr$. %Here $\mu_b$ is not relevant as most of our models are two-dimensional.

 %%%%%%%%%%%%%%%%%%%%%%%%%%%%%%%%%%%%%%%%%%%%%%%%%%%%

\section{Spiral arm effects on the stellar velocities}\label{effects}

 \begin{figure*}
   \centering
 \includegraphics[width=0.32\textwidth]{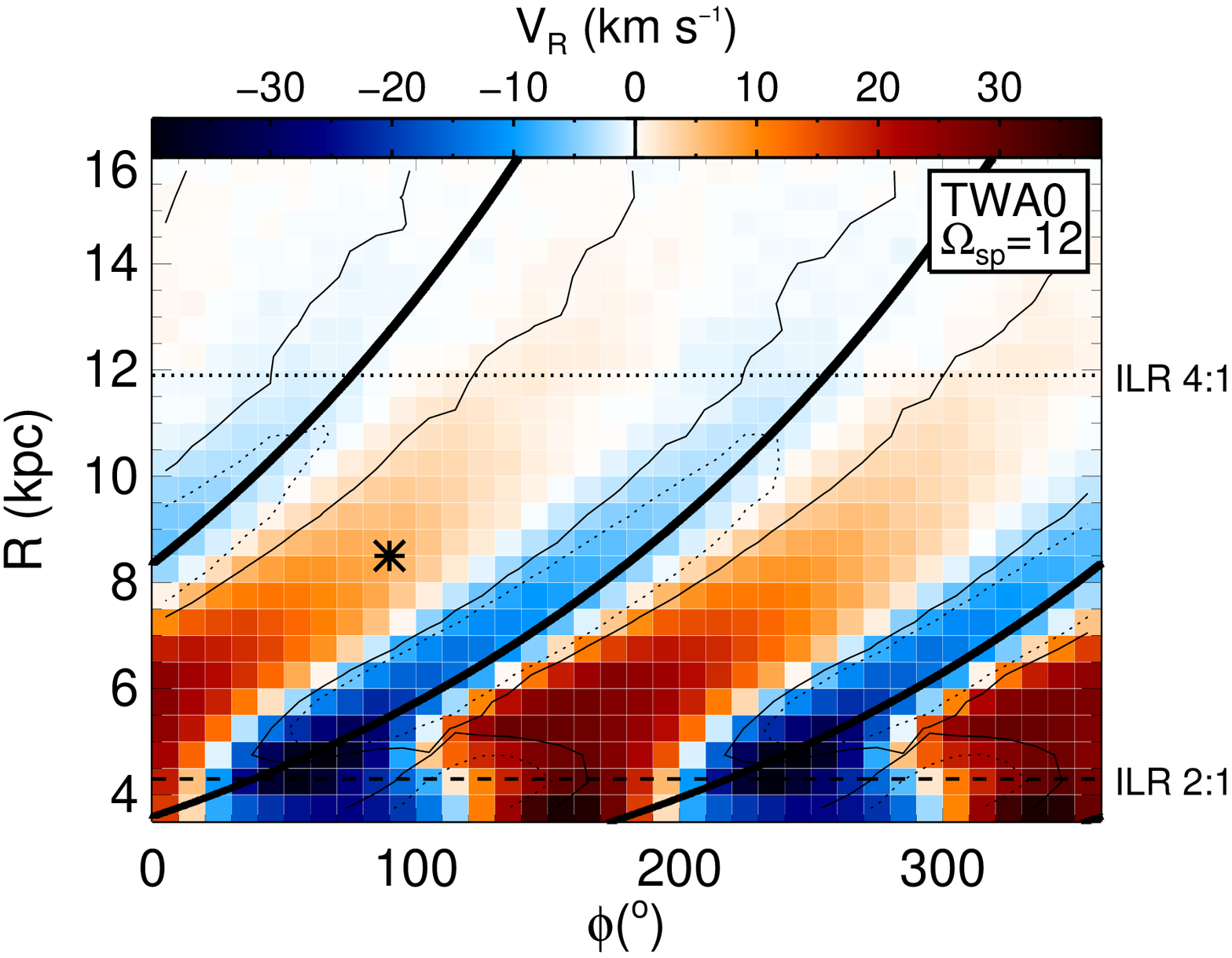}
 \includegraphics[width=0.32\textwidth]{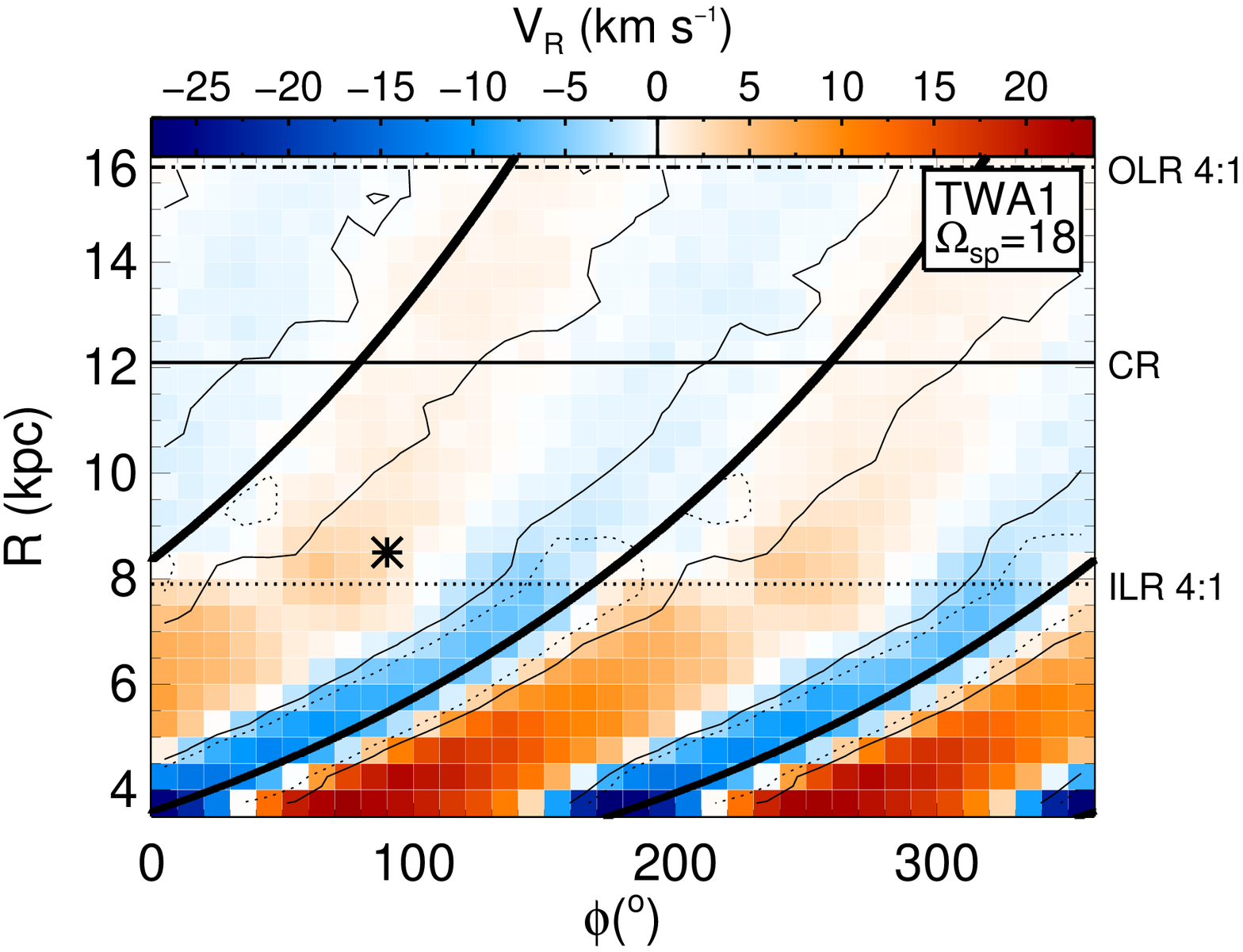}
 \includegraphics[width=0.32\textwidth]{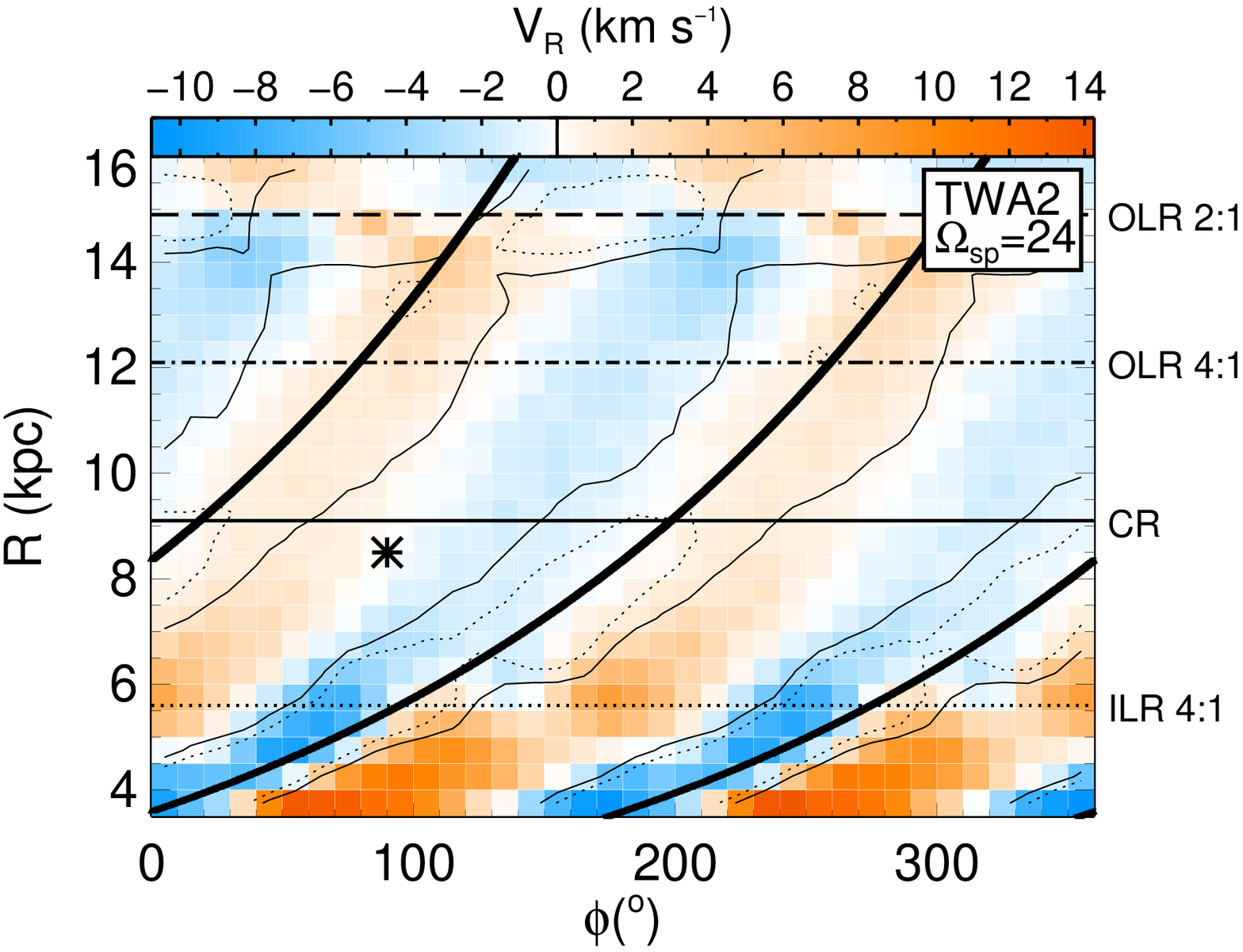}
 
 \includegraphics[width=0.32\textwidth]{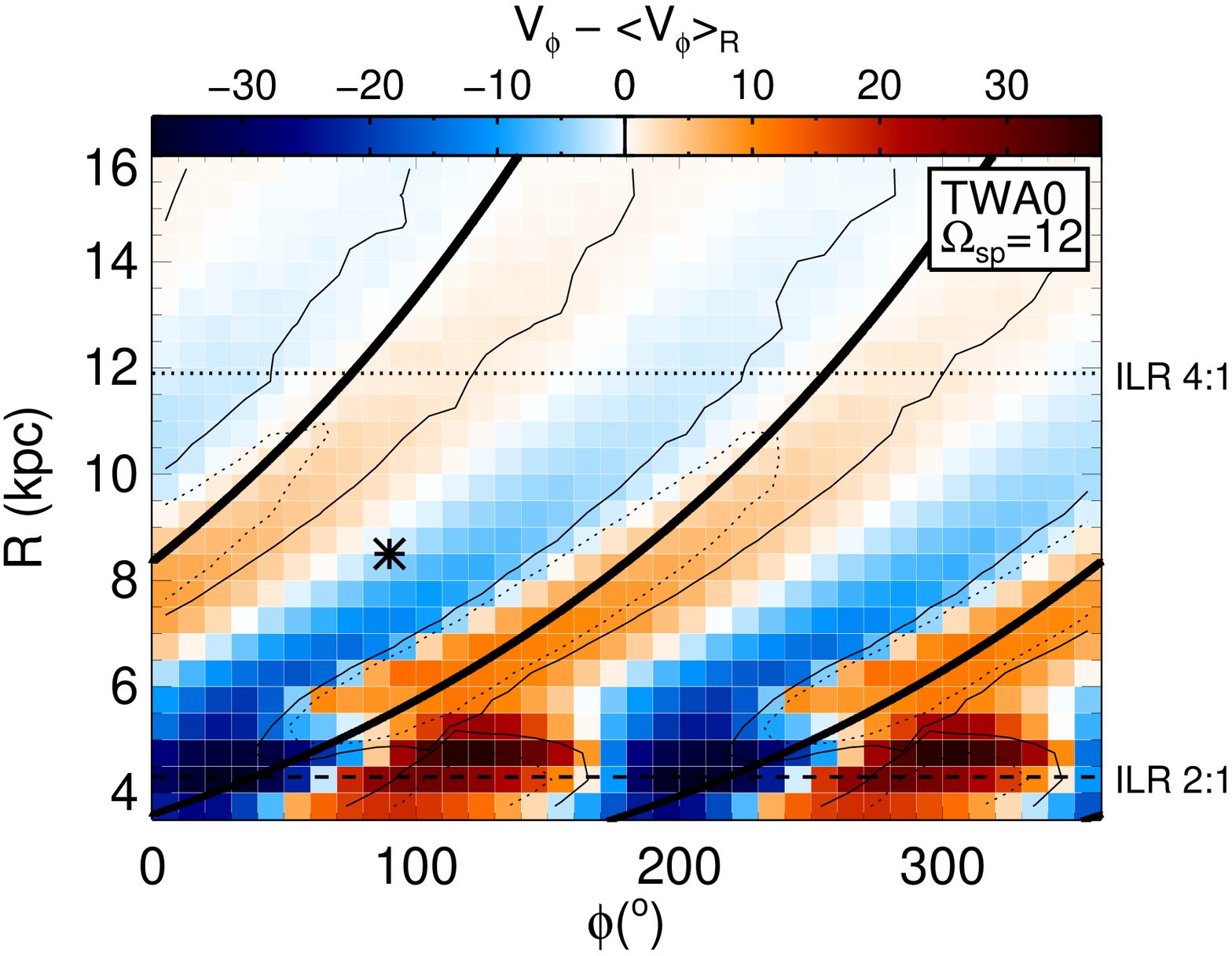}
 \includegraphics[width=0.32\textwidth]{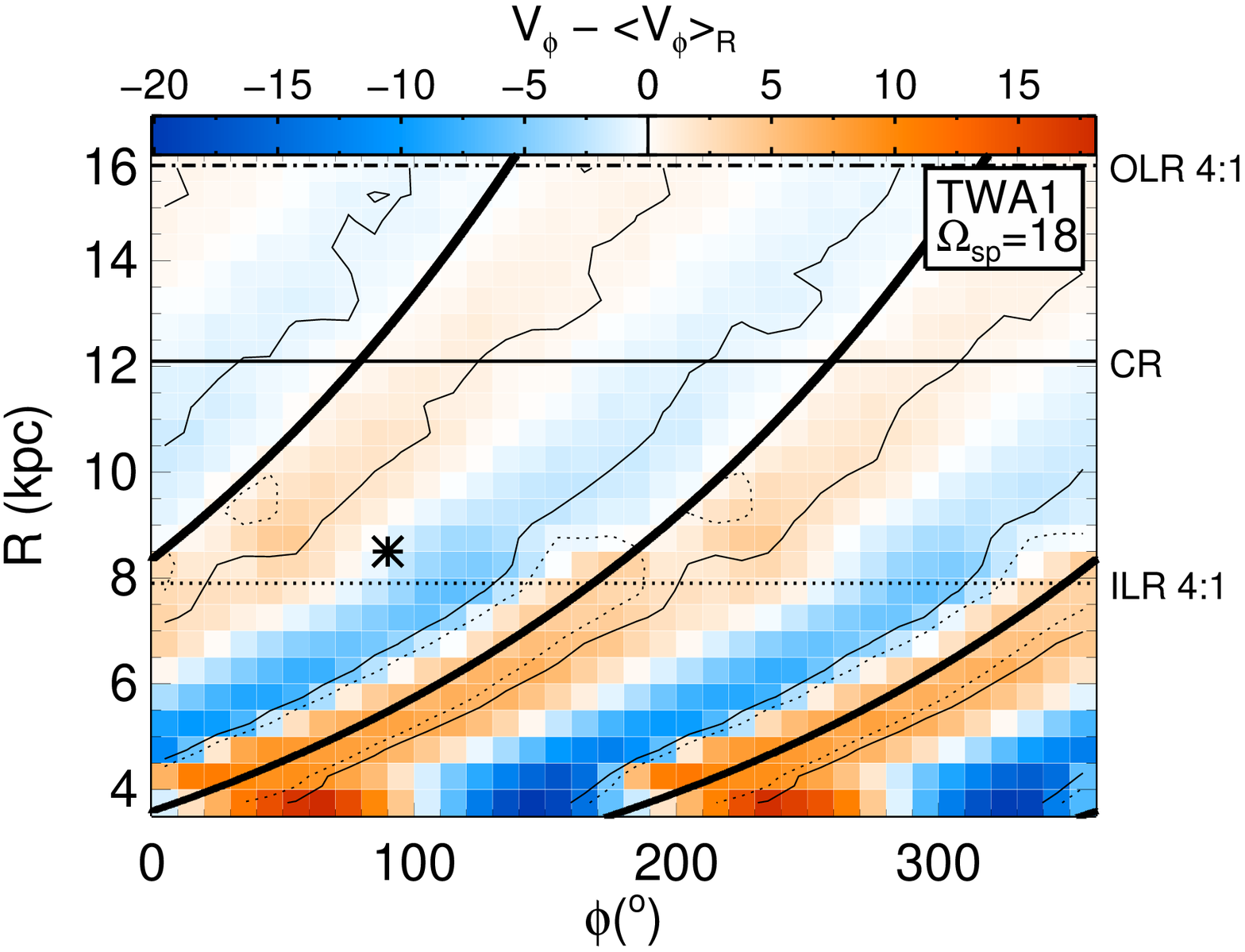}
 \includegraphics[width=0.32\textwidth]{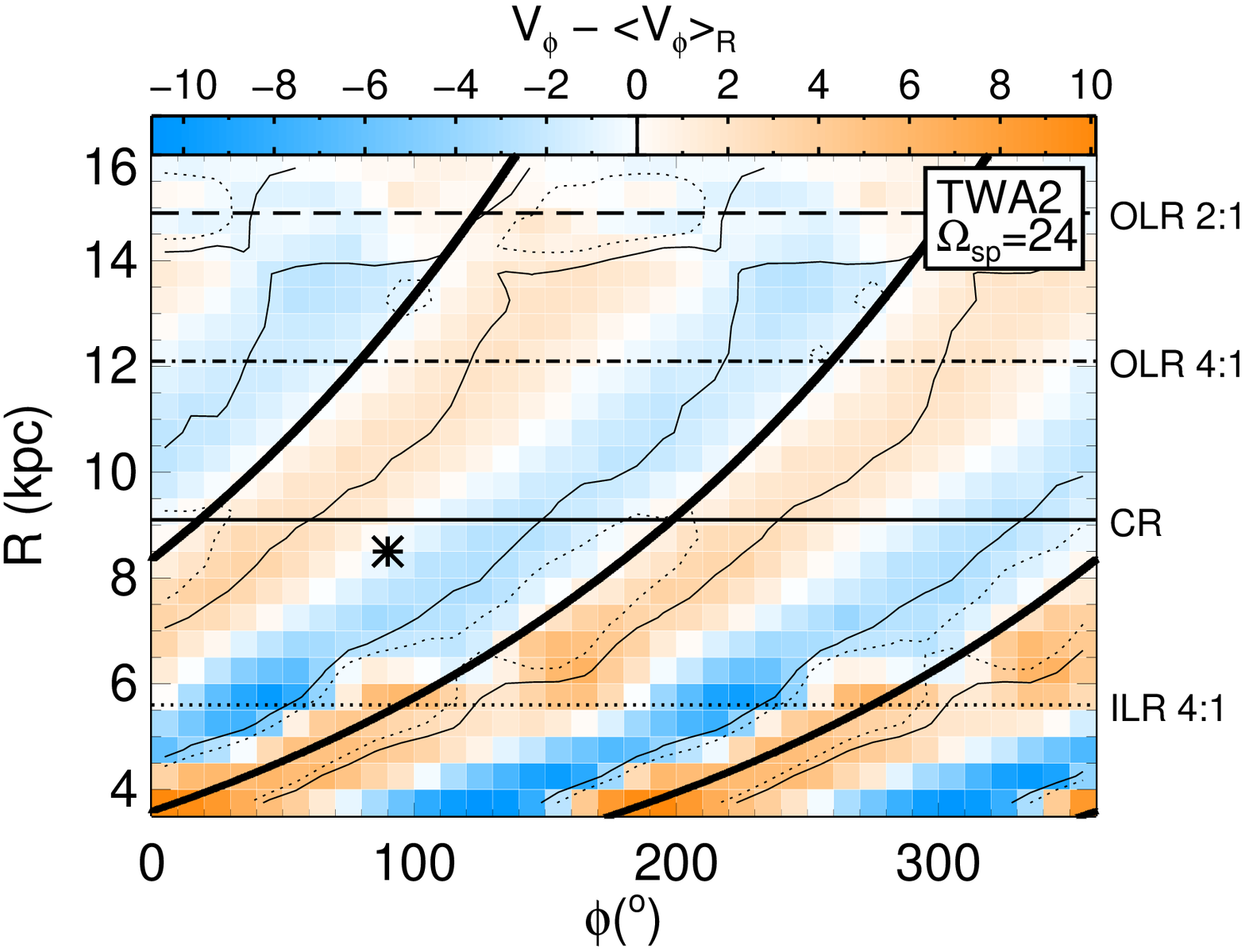}

  \caption{Galactocentric radial $\vr$ (top) and azimuthal $\vphi$ (bottom) velocities as a function of cylindrical coordinates for models TWA0, TWA1, and TWA2 (left, middle, and right). Colours show the median velocity in bins in cylindrical coordinates of size $\Delta R=0.5\kpc$ and $\Delta\phi=10\deg$. For $\vphi$ we plot $\vphi-<\vphi>_R$ where $<\vphi>_R$ is the average median over all bins at the same radius $R$, that is we subtract the average rotation velocity at each radius. To make the comparison easier, the colour scale is the same for all panels but the scale indicated above each panel shows only the range for that particular model. The theoretical locus of the arms is shown as a thick black line. The solid and dotted curves indicate the over-density of the spiral arms where the density contrast $(N-<N>_R)/<N>_R$ is 0 and 0.2 of the maximum value, respectively, where $N$ is the number of particles in each pixel of the grid and $<N>_R$ is the azimuthal average. The locations of the main resonances CR, ILR 2:1, ILR 4:1, OLR 4:1, OLR 2:1 are shown with black horizontal lines (solid, dashed, dotted, dashed-dotted, and long-dashed, respectively). The rotation of the Galaxy is towards the left. The black asterisk shows the location of the Sun.}
         \label{vgalTP}
   \end{figure*}
   
  \begin{figure*}
   \centering

 \includegraphics[width=0.32\textwidth]{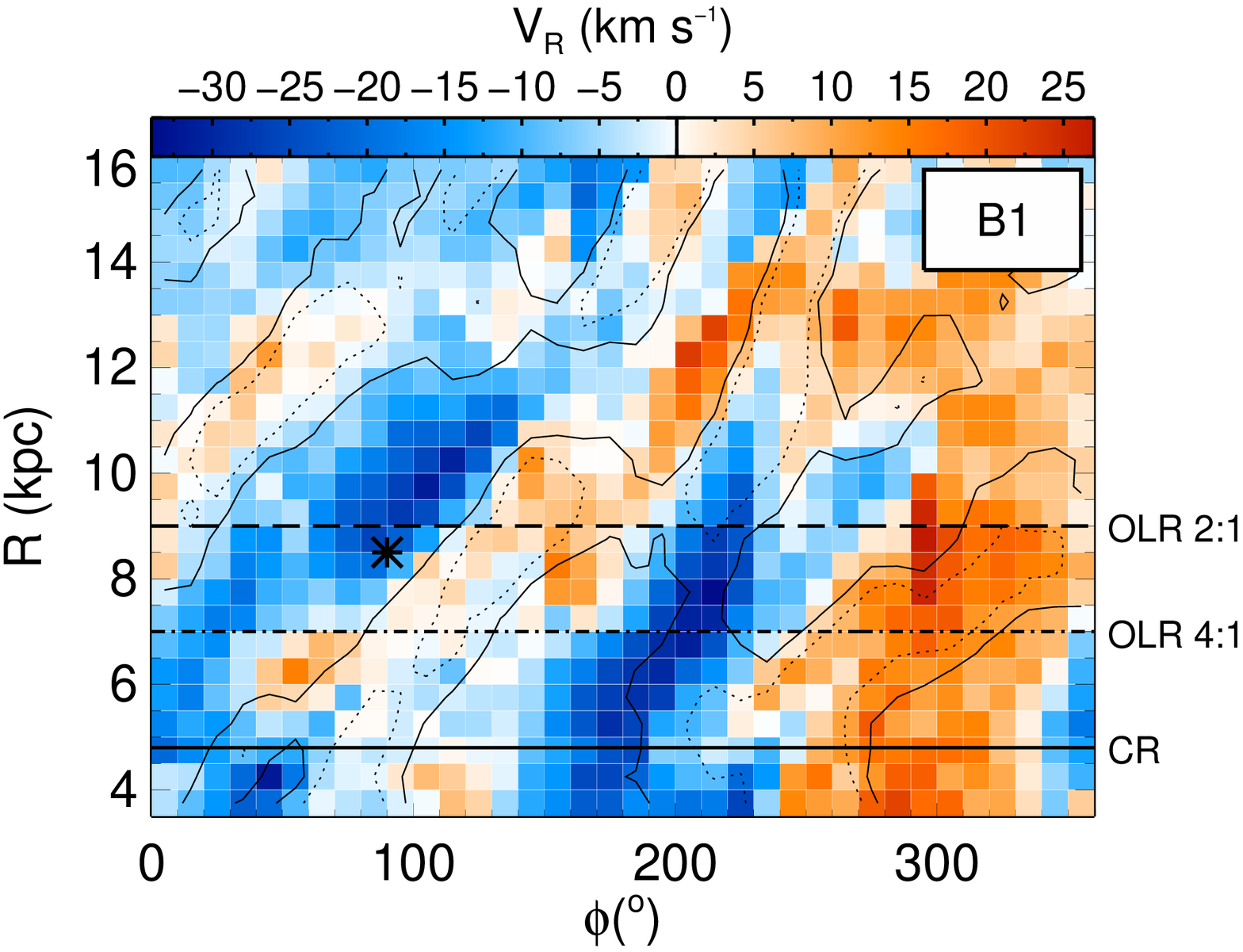}
 \includegraphics[width=0.32\textwidth]{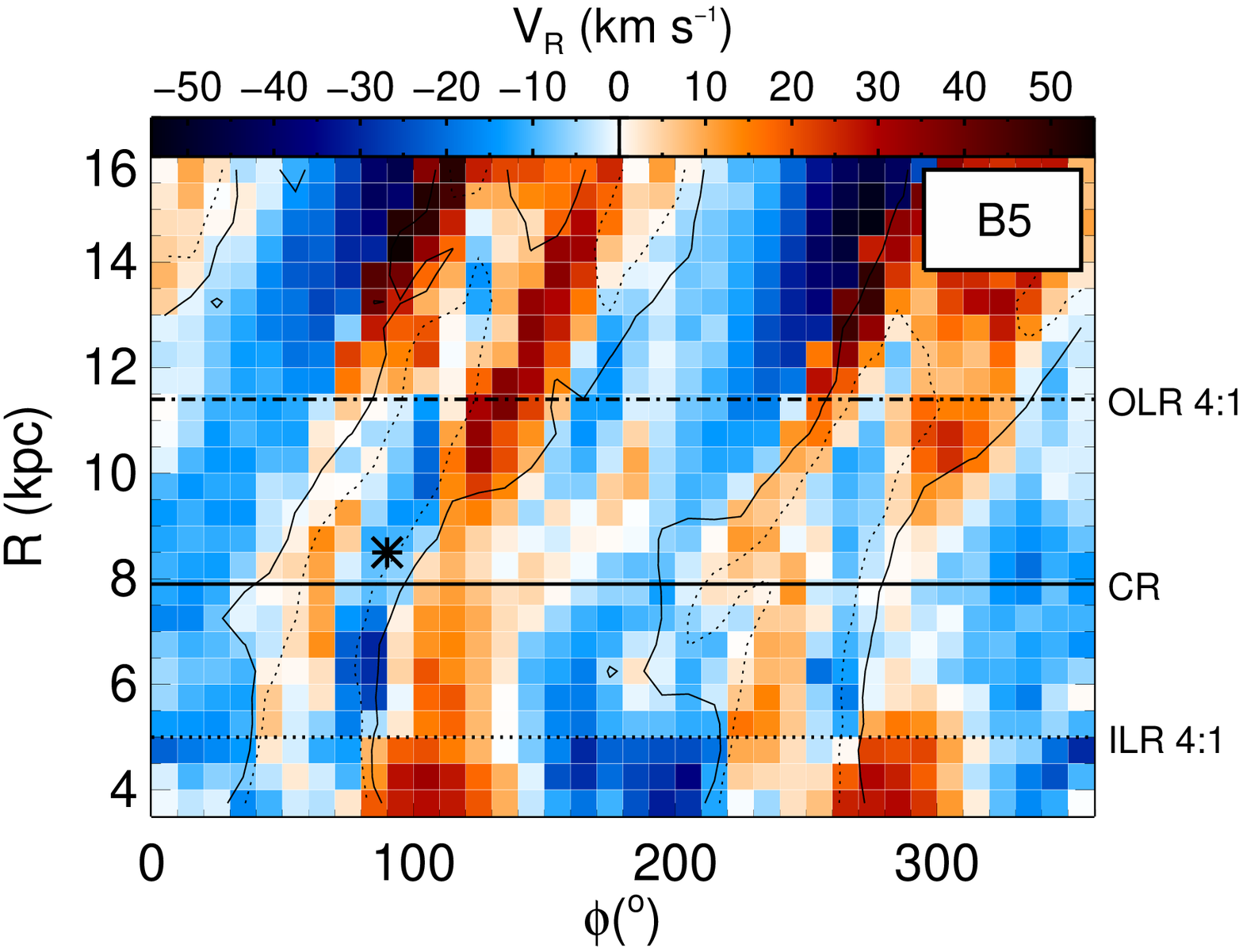}
 \includegraphics[width=0.32\textwidth]{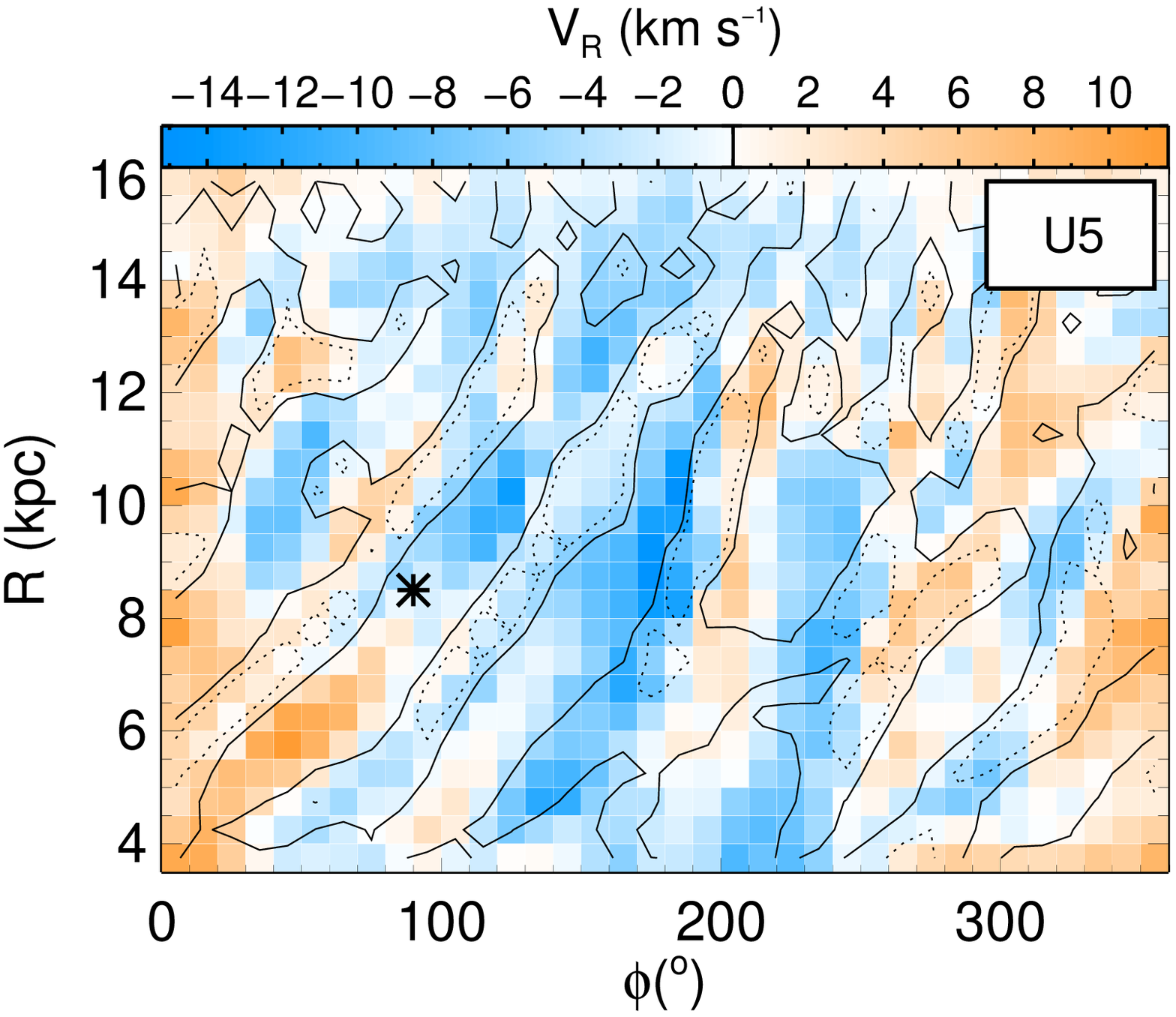}
 
 \includegraphics[width=0.32\textwidth]{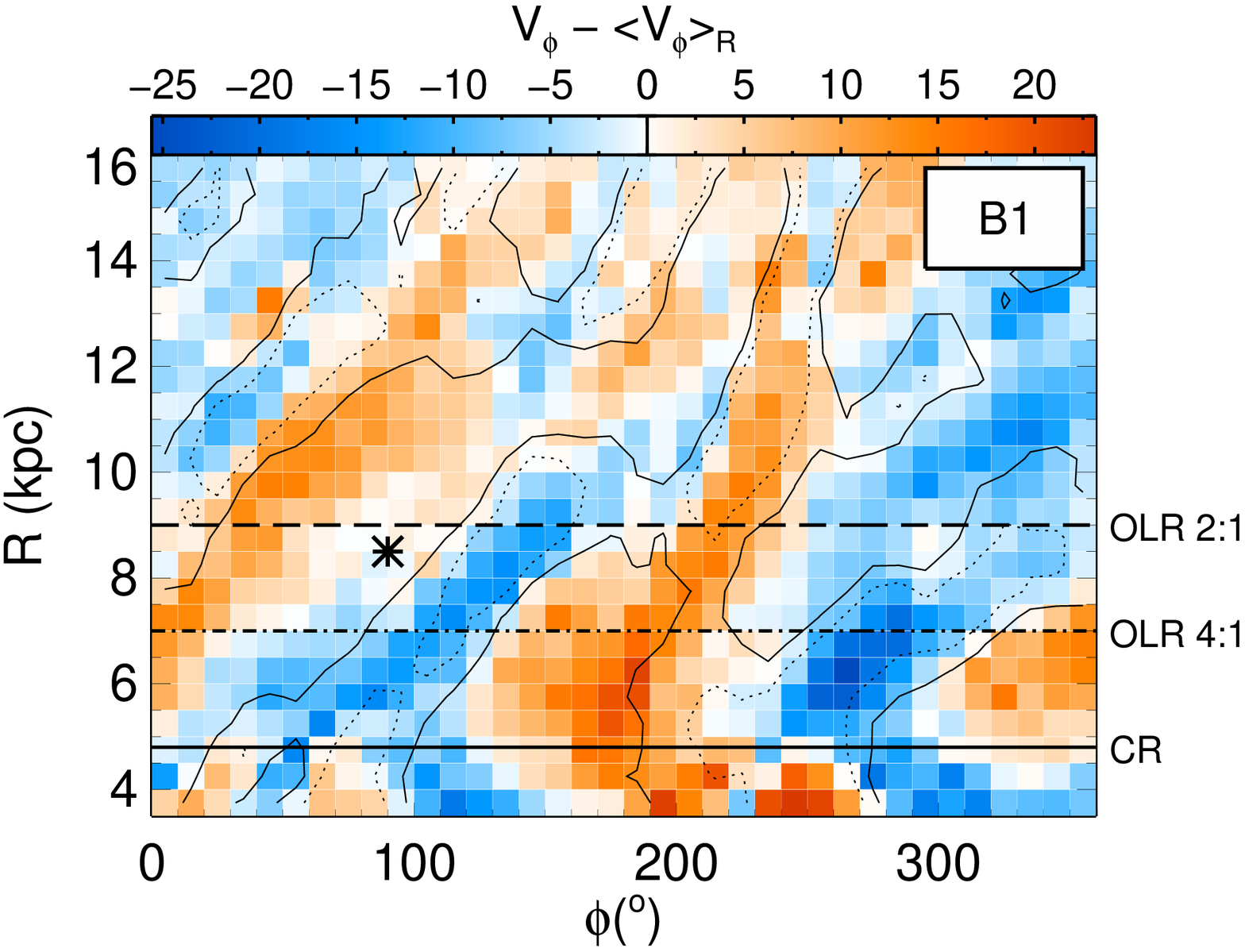} 
 \includegraphics[width=0.32\textwidth]{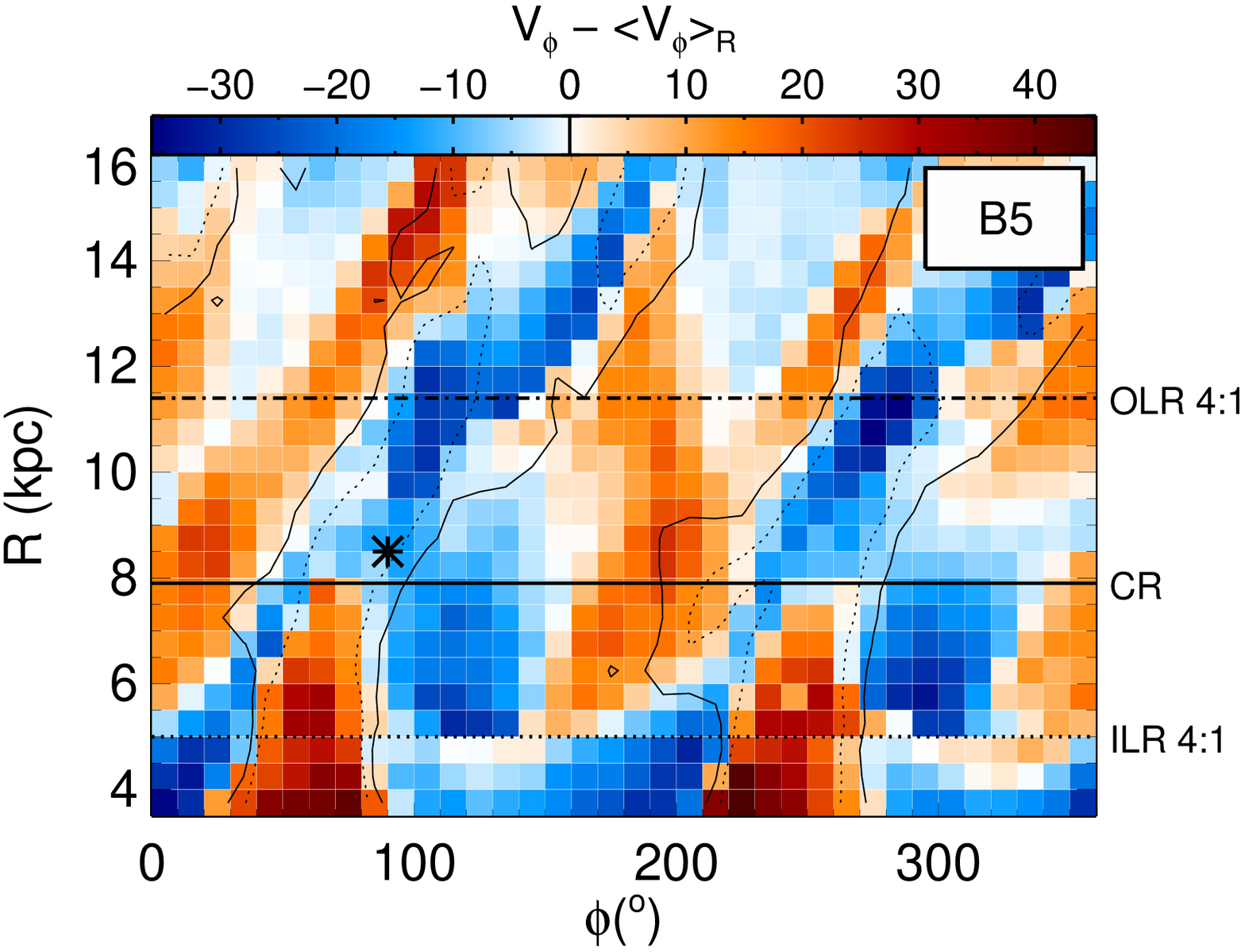}
 \includegraphics[width=0.32\textwidth]{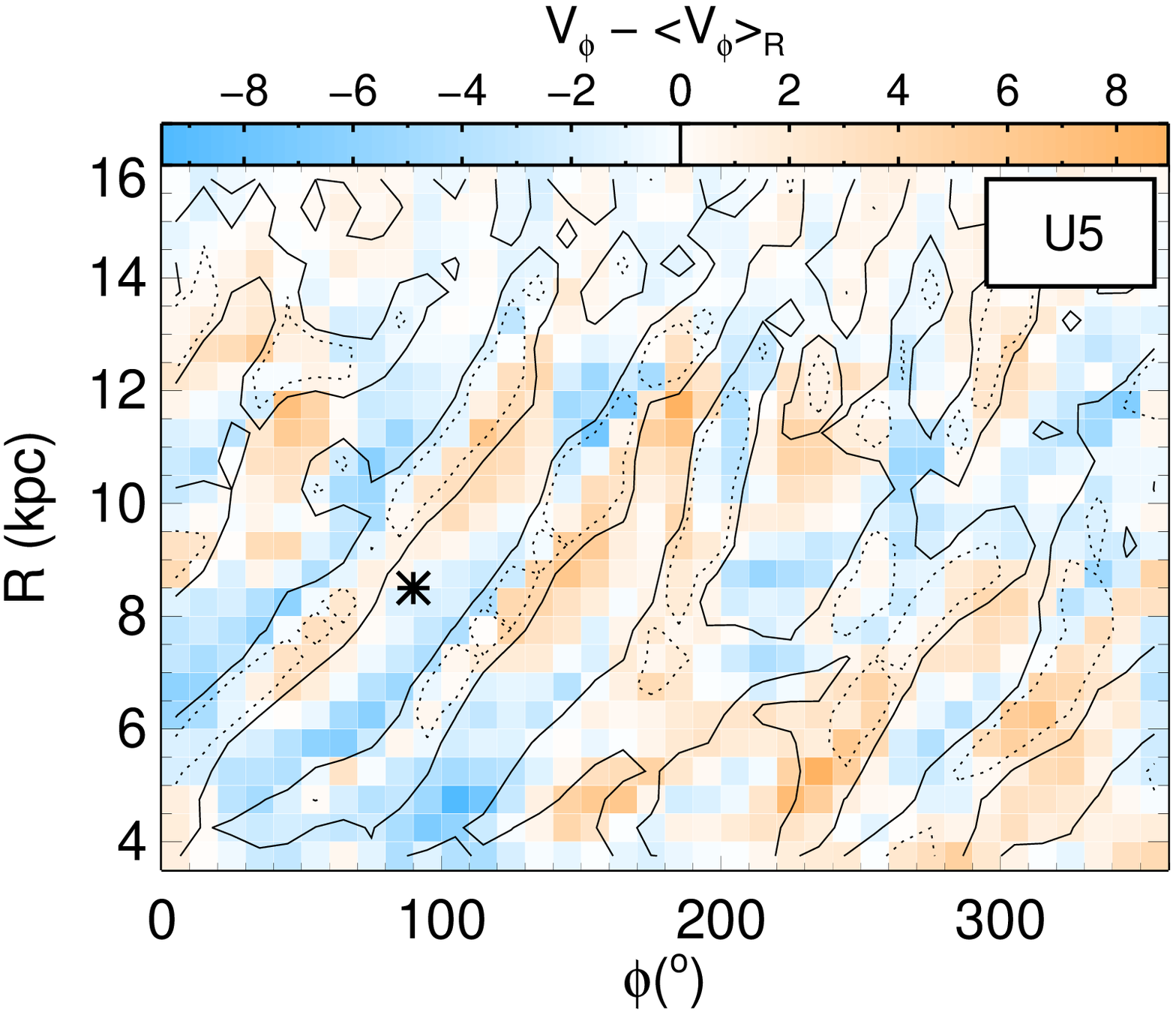}
  \caption{Same as \fig{vgalTP} but for models B1, B5, and U5. The locus of these models is not predefined and thus we only plot the over-density of the spiral arms. The colour scale is the same for all panels but is slightly different from the scale of Fig.~\ref{vgalTP}.}
         \label{vgalNB}
   \end{figure*}

   \Fig{vgalTP} shows  the median galactocentric radial $\vr$ (top) and azimuthal $\vphi$ (bottom) velocities as a function of disk position in cylindrical coordinates for models TWA0, TWA1, and TWA2. The rotation of the Galaxy is towards the left part of the plots. In the case of $\vphi$ we subtract the average rotation velocity at each radius ($\vphi-<{\vphi}>_R$). The theoretical locus of the arms is shown as a thick black line. Surrounding the locus, there are solid and dotted curves which indicate the over-density of the spiral arms (see caption). The locations of the main resonances CR, the Inner LR (ILR 2:1, ILR 4:1) and Outer LR (OLR 4:1, OLR 2:1) are shown as black horizontal lines and the position of the Sun is indicated by a black asterisk.

  The median $\vr$ and $\vphi-<{\vphi}>_R$ are expected to be $\sim0$ in an axisymmetric disk. For our models in  \fig{vgalTP} these are clearly not null. Moreover, these velocities follow a pattern related to the location of the spiral arms and their main resonances and are different for each of these models. For instance, $\vr$ is negative (blue colours) on top of the arms before the CR resonance, it is 0 around CR, and positive (red) beyond. On the other hand, $\vphi-<{\vphi}>_R$ is positive in the trailing part of the arm (right of the locus) for most of the radius and negative in the leading part (left). This quantity is null along the locus except inside the ILR 4:1 where it is positive.
   
    For the TWA model, the mean velocities can be  estimated analytically (Appendix A3.2 of \citealt{RocaFabrega2014}). These predictions agree well with our findings above; the main  difference is the more abrupt changes of sign in the analytic predictions compared to more gradual changes of sign in the simulations.
   
   %For the TWA model, the mean velocities can be analytically estimated (Appendix A3.2 of \citealt{RocaFabrega2014}). According to these approximations, $\vr$ should be negative on the arm, positive in the interarm regions inside CR and outside the OLR, and the other way around between CR and the OLR. For $\vphi-<{\vphi}>_R$ we expect positive values in the trailing part of the arm and negative values in front of the arm (leading part), with null values on the locus itself and in the interarm region between the 2:1 ILR and the 2:1 OLR, with a reverse of sign close to the 2:1 ILR. \red{SANTI: it's difficult to see in the analytic predictions what exactly happens.} These predictions agree well with our findings in the simulations, with the main important change being the more abrupt changes in the analytic predictions compared to more gradual changes in the simulations. 

  \fig{vgalNB} is the same as \fig{vgalTP} but for the N-body models. In these cases the velocity patterns also have a spiral shape but they are related to the arms in a more complex way. For model B1, the inter-arm region seems dominated by negative $\vr$ , similar to what happens after CR for the analytic models. This is also the case for model B5 in the regions where the spiral arms, and not the bar, dominate (beyond $6\kpc$). For these two models there seems to be a mix of negative and positive values of $\vr$ on the spiral arms. For model B1, there is positive  $\vphi-<{\vphi}>_R$ in the trailing parts of the arms, except for the segment of arm at $\phi\sim250\deg$. Model U5 presents tangled velocity patterns, although they also follow some spiral shape. The goal of our study is not to determine the cause of the patterns in these models, but to explore the kinematic features of more complex models.

 \begin{figure*}
   \centering
 \includegraphics[width=0.32\textwidth]{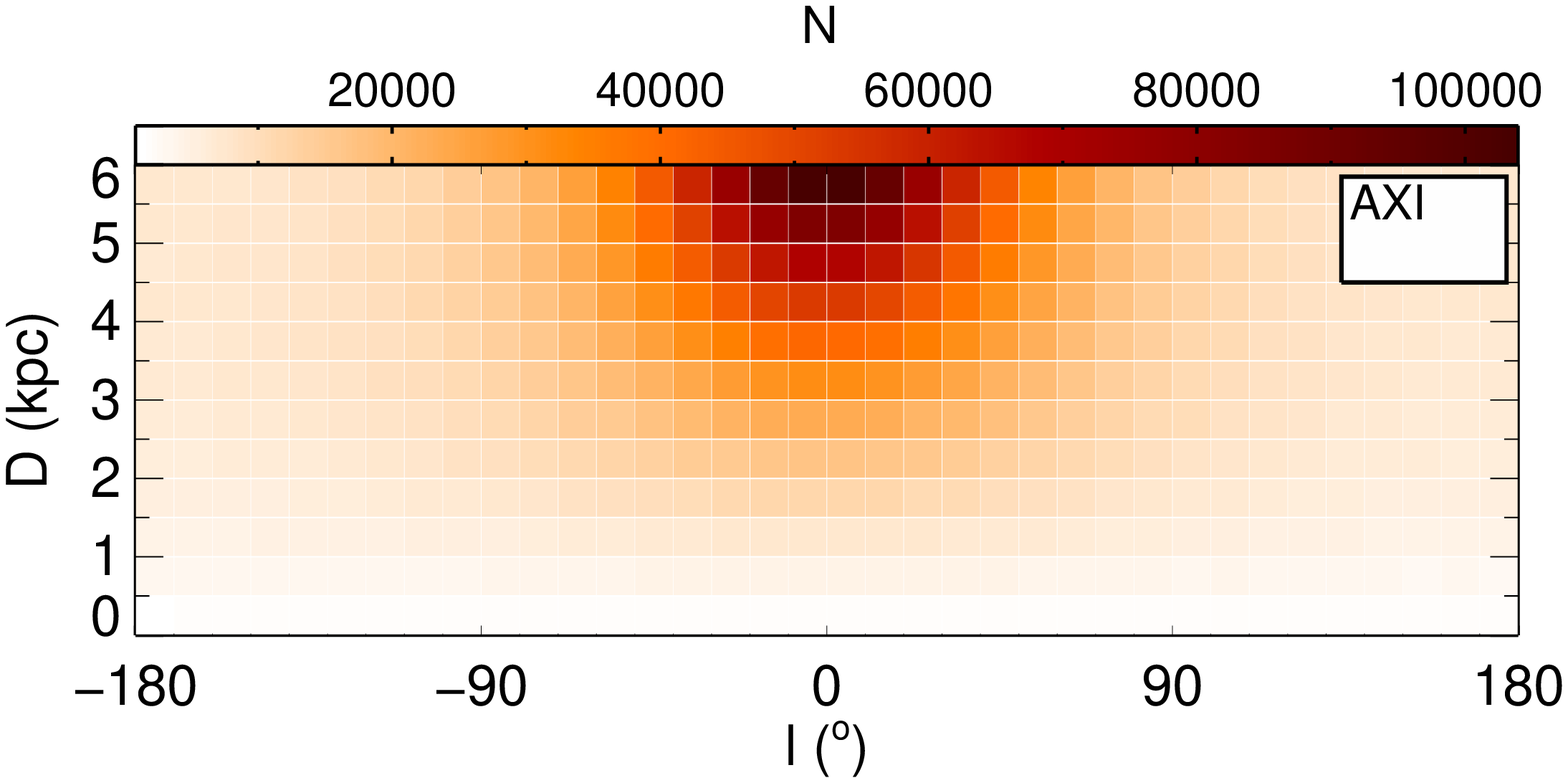}
 \includegraphics[width=0.32\textwidth]{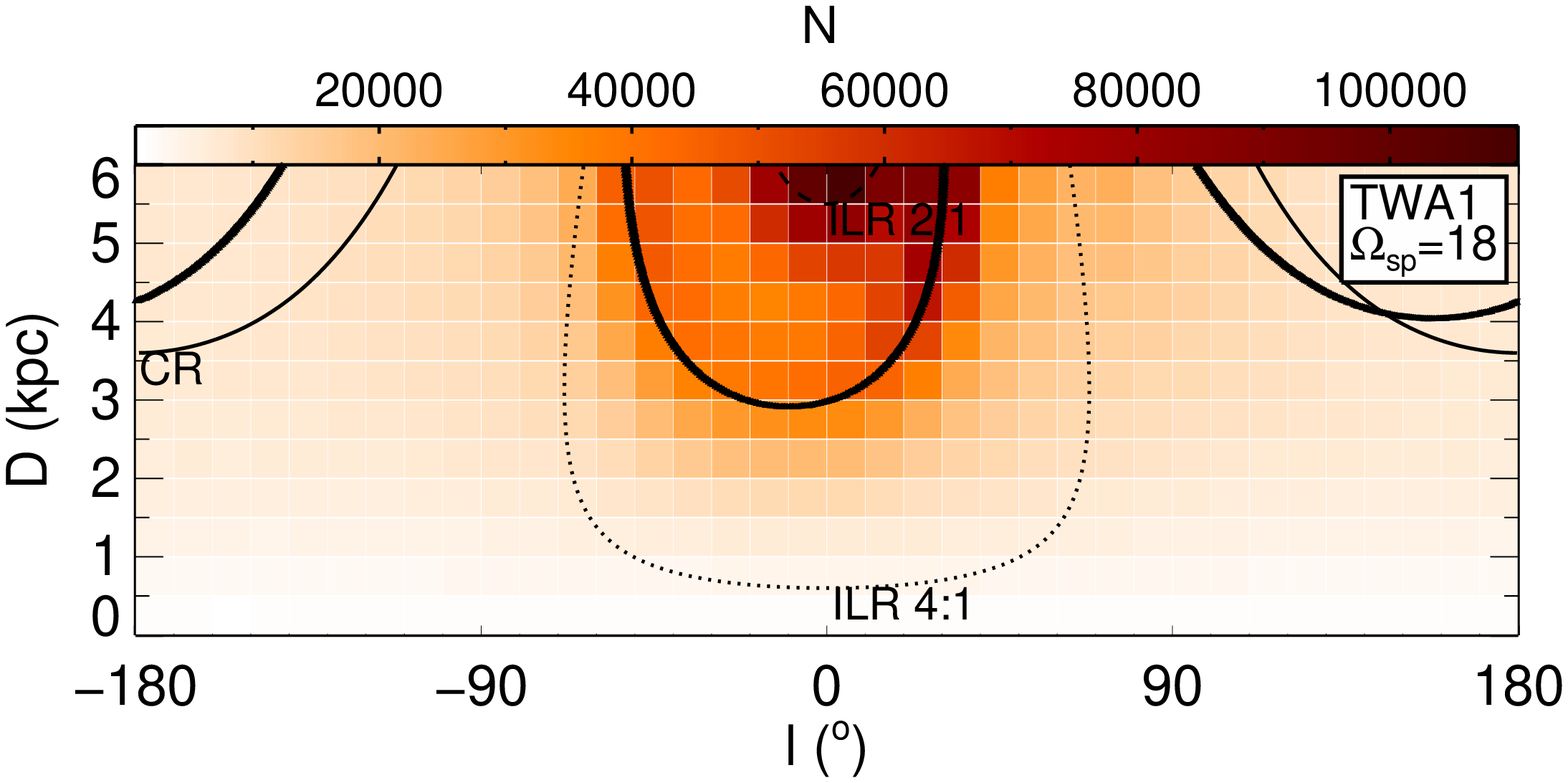}
 \includegraphics[width=0.32\textwidth]{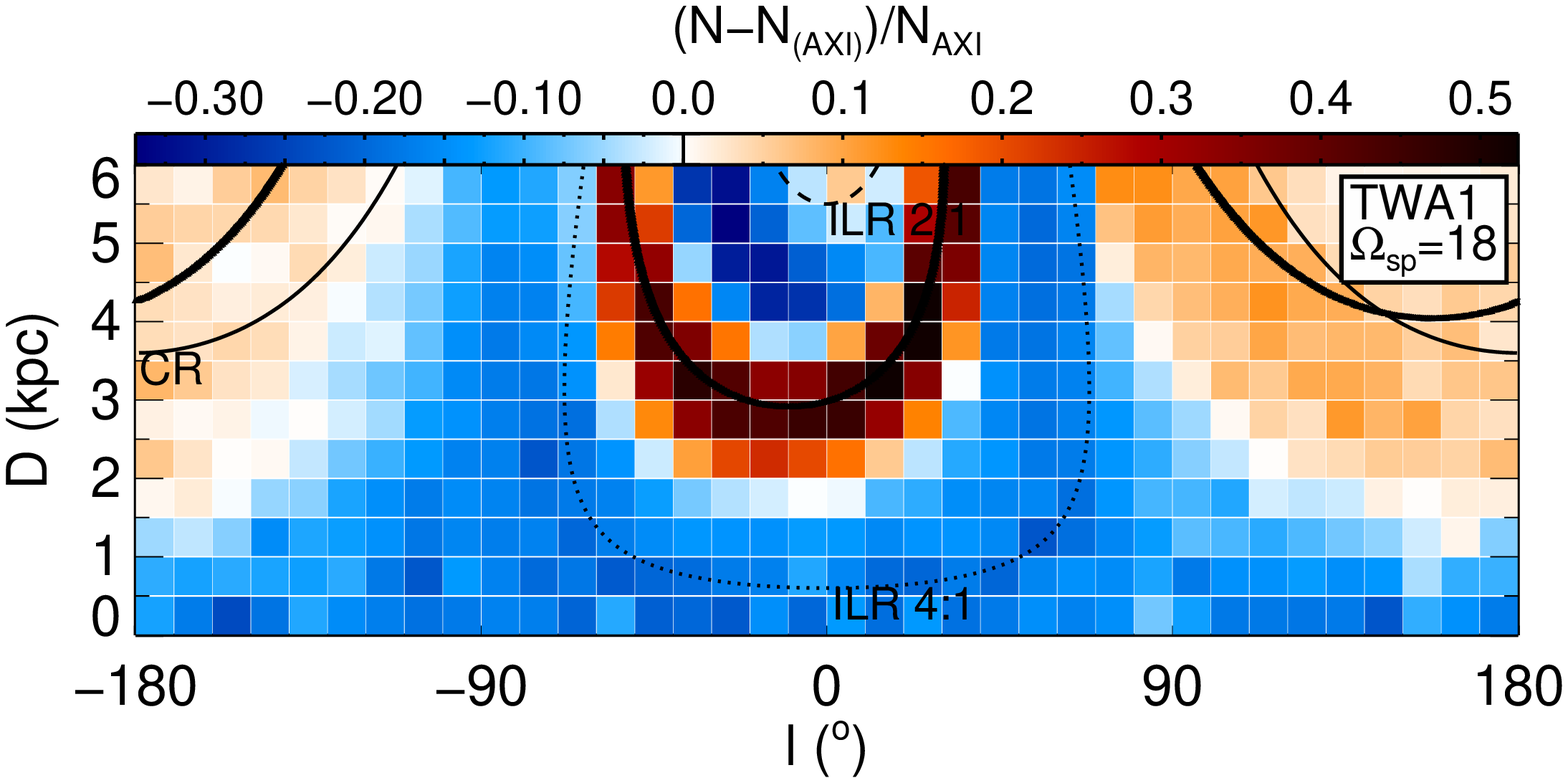}
 
 \includegraphics[width=0.32\textwidth]{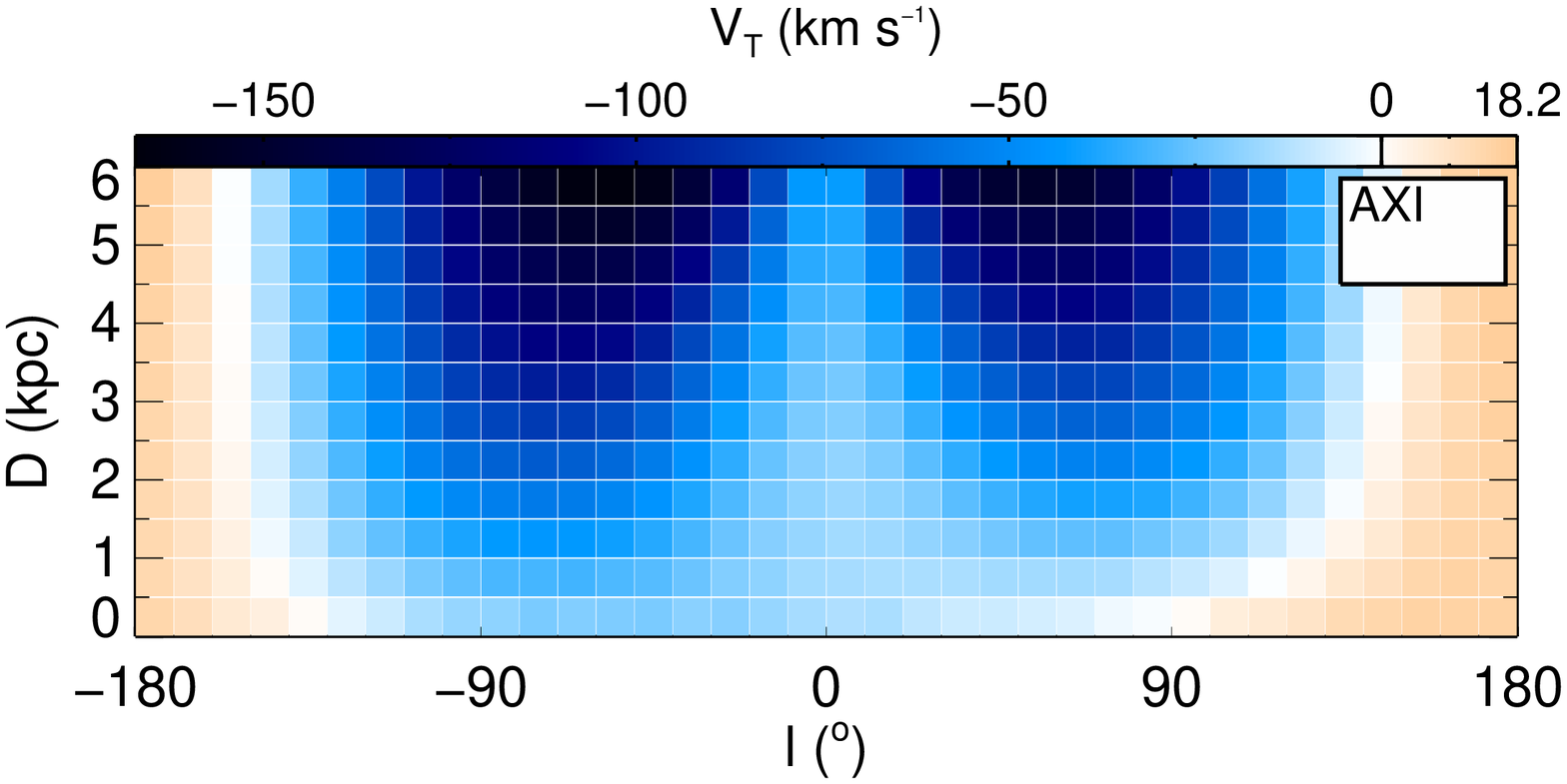}
 \includegraphics[width=0.32\textwidth]{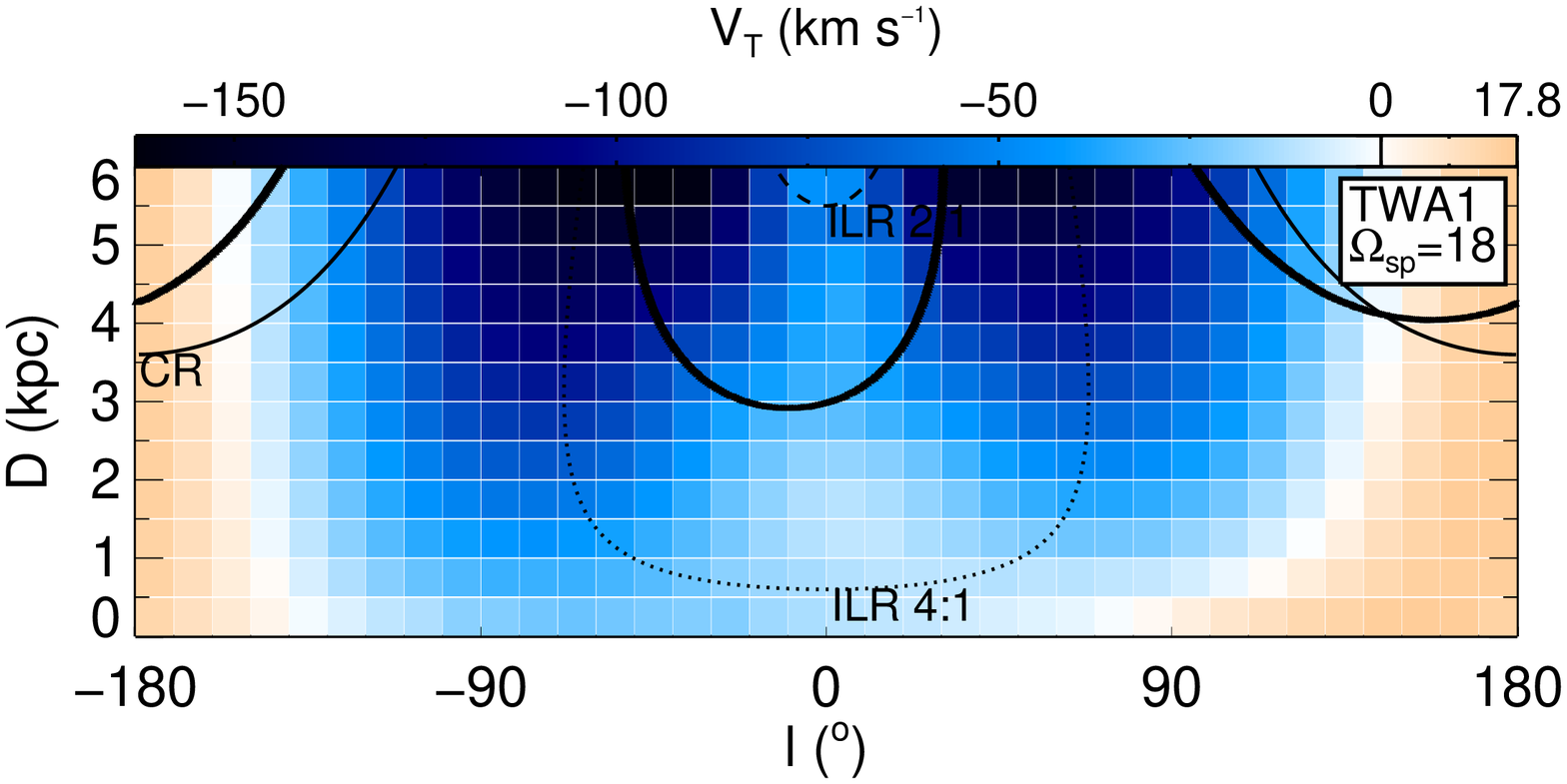}
 \includegraphics[width=0.32\textwidth]{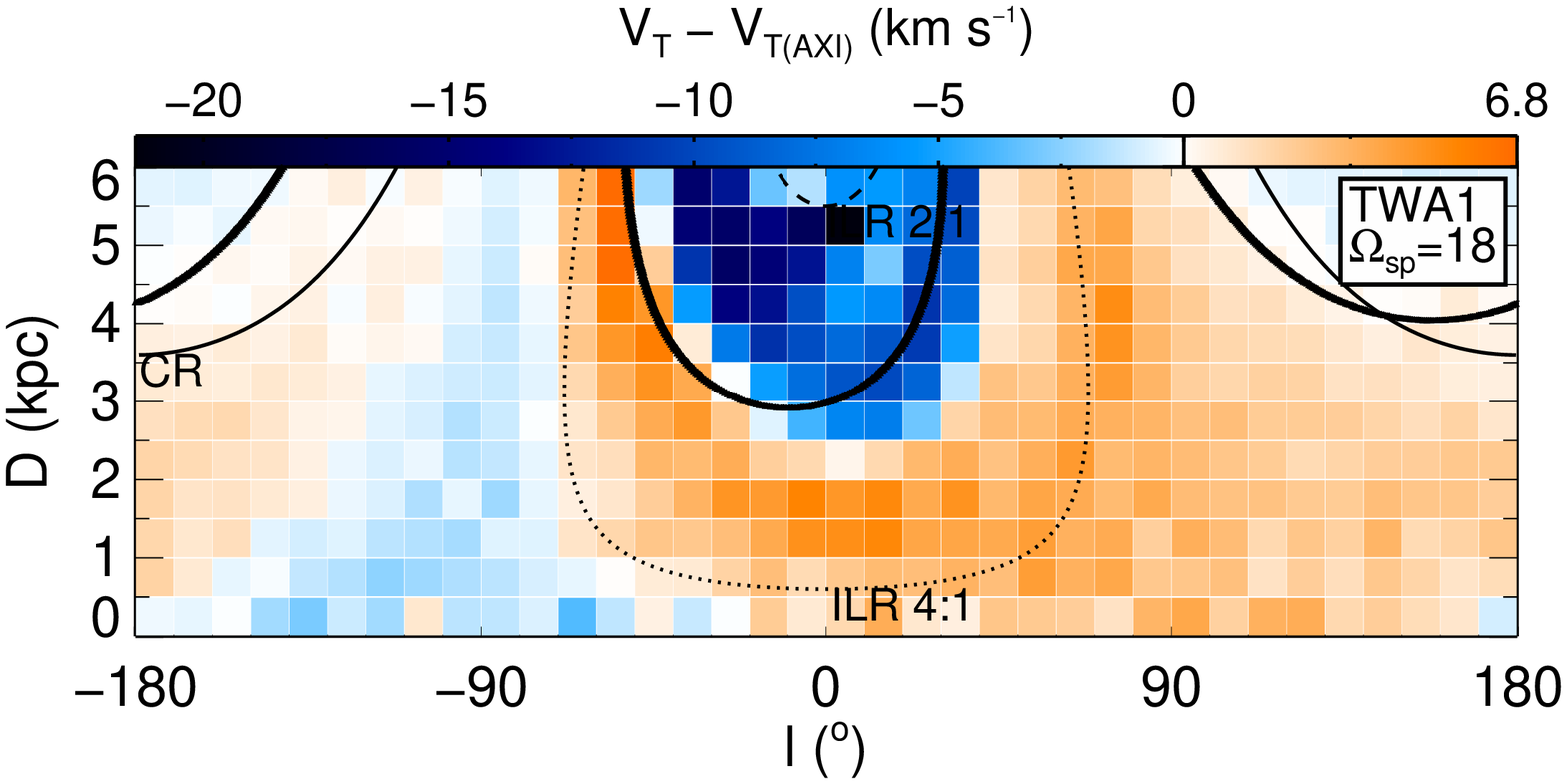}

  \caption{Number of particles per bin (top) and transverse velocities $\vt$ (bottom) as a function of longitude and distance from the Sun. We have binned this space in bins of $\Delta l=10\deg$ and $\Delta D=0.5\kpc$. We show the axisymmetric model (left), the TWA1 model (middle), and the difference between these two (right). We superpose the spiral arms locus (black thick line) in these coordinates and the main resonances with curves as in \fig{vgalTP}.}
         \label{obs}
   \end{figure*}
   
   The patterns in velocity that we observe in \figs{vgalTP}{vgalNB} translate into patterns in the \gaia observable space. We present in \fig{obs} several quantities in the longitude-distance plane. The area covered in these plots corresponds to the blue circle in \fig{coord}. For the analytic models, we also use the symmetric circle (at $Y=-8.5\kpc$) to obtain better particle statistics since these models are by definition symmetric. We do not do this with the N-body models since they are not strictly symmetric (\fig{XY}).  The top row shows the number of particles per bin. We plot here the axisymmetric model (left), that is before the spiral arms are introduced in the simulation, the model TWA1 (middle), and the difference between these two (right). We also plot the spiral arms locus (black thick line), which coincides with the main arm over-densities in the middle and right top panels (red colours).
   
   In the bottom of \fig{obs} we plot the transverse velocity $\vt$. The axisymmetric case (left) and the TWA1 model (middle) do not seem to differ significantly.   The effect of the spiral arms only becomes evident in the difference between the two (right). At closer distances ($1\kpc$) the arms increase the transverse velocity with respect to the axisymmetric case at positive longitudes, and decrease the transverse velocity for negative longitudes. Beyond this distance, the arms and  resonances delimit regions of enhanced and reduced $\vt$. For instance, towards the Galactic centre ($l=0\deg$) $\vt$ is enhanced at close distances, but it decreases beyond the spiral arm ($d\sim3\kpc$), and similarly towards the anti-centre ($l=\pm180\deg$). The pattern that we observe here is related to that seen in $\vr$ and $\vphi$ since the transverse velocity is a combination of these two. For example, around $l=0\deg$ and $l=180\deg$, the $\vt$ velocity is mainly $-\vphi$ and $\vphi$, respectively.
   
   Table \ref{t:results0} quantifies how much the kinematics of the spiral arm models deviate from the axisymmetric case. We only consider here the analytic models for which we have the corresponding axisymmetric model. In columns 2 and 3 we show the maximum  values of $|\vt-{\vt}_{\rm (AXI)}|$ (maximum absolute value in the colour scale of the right bottom panel of \fig{obs}) and the median value. In all cases the maximum deviation is at least $10\kms$ and up to $38\kms$. In all but one model, $50\%$ of the bins differ from the axisymmetric models by at least $1\kms$, and in most of the cases by $\sim2\kms$. In the final columns we repeat this computation for $\vlos$ velocities, and find very similar values.
   
 \begin{table}
 \caption{Kinematic deviations of the spiral models compared to the axisymmetric models. Columns show: 1) model; 2) maximum absolute difference between the spiral arms model and the axisymmetric case for the transverse velocity $|\vt-{\vt}_{\rm (AXI)}|$; 3) median difference; and 4) and 5) the same as 2) 3) but for \los velocities.}
  \label{t:results0}      
 \centering          
     \tabcolsep 3.pt
 \begin{tabular}{lcccc}     % 11 columns 
 \hline\hline       
Model& \multicolumn{2}{c}{$|\vt-{\vt}_{\rm (AXI)}|$}&\multicolumn{2}{c}{$|\vlos-{\vlos}_{\rm (AXI)}|$}\\     
& \multicolumn{2}{c}{($\kms$)}&\multicolumn{2}{c}{($\kms$)}\\     
& max&med&max&med\\      \hline     
%     &($\kms$) &($\kms$)&($\kms$)&($\kms$)\\ \hline  
             TWA0&38.& 4.0&39.& 2.8\\
             TWA1&21.& 1.7&30.& 1.5\\
             TWA2&16.& 1.4&19.& 1.5\\
             TWA3&10.& 2.5& 8.& 2.1\\
            TWA10&13.& 1.2&14.& 1.1\\
            TWA11&14.& 0.9&19.& 0.7\\
            TWA12&24.& 2.8&33.& 2.3\\
            TWA13&34.& 2.2&31.& 2.5 \\
              \hline                  
 \end{tabular}
 \end{table}

   Given that the patterns in $\vr$ and $\vphi$,  in \figs{vgalTP}{vgalNB}, depend on the properties of the spiral arms, the patterns seen in $\vt$ are also different for the different models. However, the patterns in $\vt$ are not very conspicuous unless we dispose of and subtract the exact axisymmetric model, which is not the case for the real MW. In \Sec{sym} we overcome this limitation.

 %\begin{figure}
 %  \centering
%\includegraphics[width=0.32\textwidth]{/user_data/tantoja/ana_sim_t/moments_plots/medVT_l_ldist_r10_v12_a85_i15_nrev4_ic2my2d_a15000_r_t0tf}
% \includegraphics[width=0.32\textwidth]{/user_data/tantoja/ana_sim_t/moments_plots/medVT_l_ldist_r10_v24_a85_i15_nrev4_ic2my2d_a15000_r_t0tf}
%  \caption{}
%         \label{obs}
%   \end{figure}

   %%%%%%%%%%%%%%%%%%%%%%%%%%%%%%%%%%%%%%%%%%%%%%%%%%%%%%%%%%%%%%%%%%%%%%%%%%
   %%%%%%%%%%%%%%%%%%%%%%%%%%%%%%%%%%%%%%%%%%%%%%%%%%%%%%%%%%%%%%%%%%%%%%%%%%
   %%%%%%%%%%%%%%%%%%%%%%%%%%%%%%%%%%%%%%%%%%%%%%%%%%%%%%%%%%%%%%%%%%%%%%%%%%
\section{Symmetric Galactic longitudes}\label{sym}

\subsection{The method}\label{sym0}

   Here we compare our simulated data of symmetric Galactic longitudes, that is $l$ and $-l$, instead of comparing with 
 an axisymmetric model. In particular, we look at the difference between the median transverse velocity $\vt$ at symmetric longitudes   \be\label{e_d}
   \D\equiv\vt\,(l>0)-\vt\,(l<0)
   .\ee
   For any axisymmetric model, we expect this quantity to be 
   \be\label{e_d2}
    \De\equiv\vt{_{\rm (AXI)}}\,(l>0)-\vt{_{(\rm AXI)}}\,(l<0)= 2\Us\sin l
   .\ee
   This is because the transverse velocities are the same in symmetric longitudes except for the Sun's motion, which in one direction adds up to the stellar velocity and in the other subtracts from the stellar velocity. However, as seen in the bottom right panel of \fig{obs}, the spiral arm models present kinematic features that are not  longitude symmetric.
   
\begin{figure*}
   \centering
\includegraphics[width=0.24\textwidth]{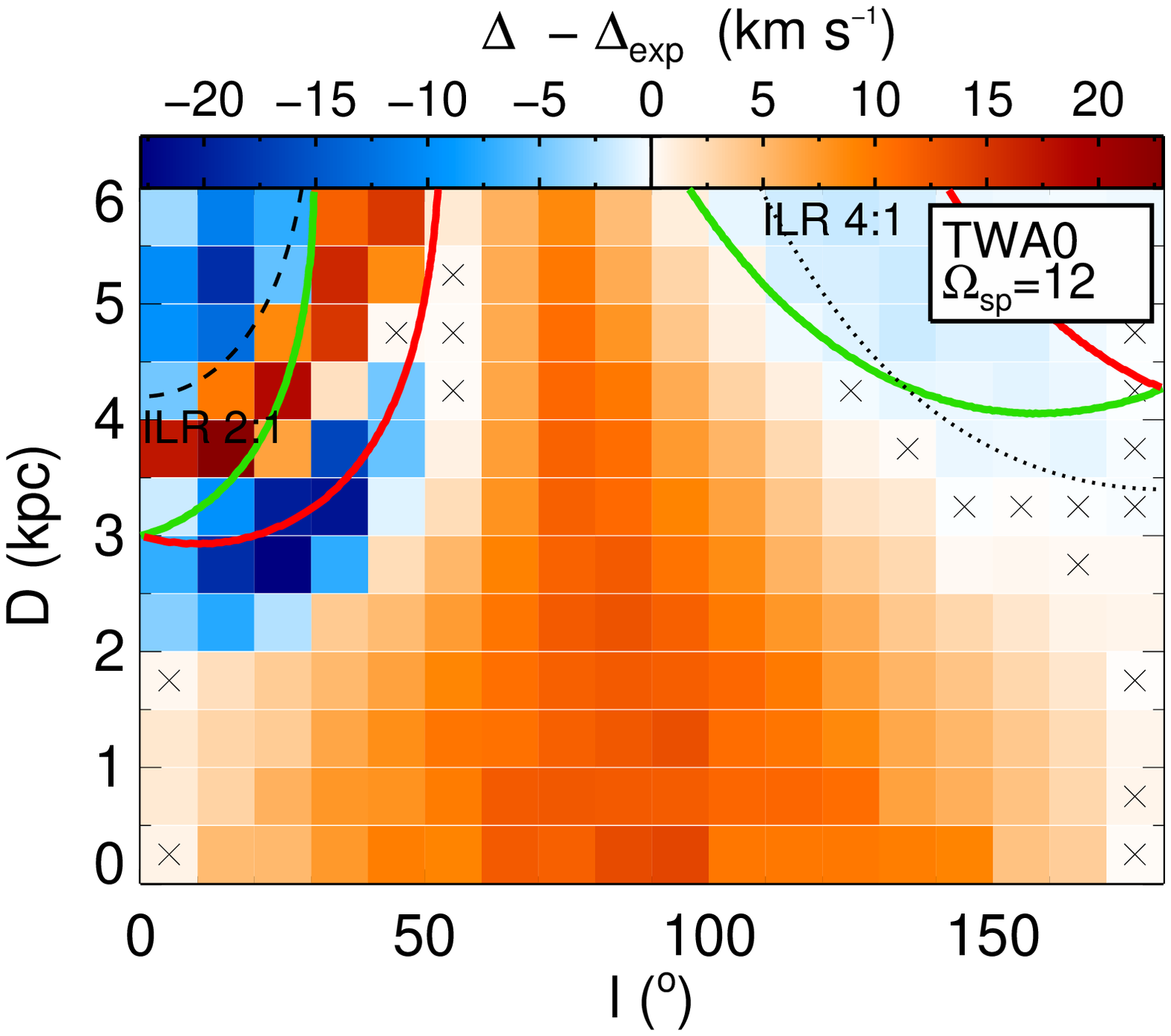}
\includegraphics[width=0.24\textwidth]{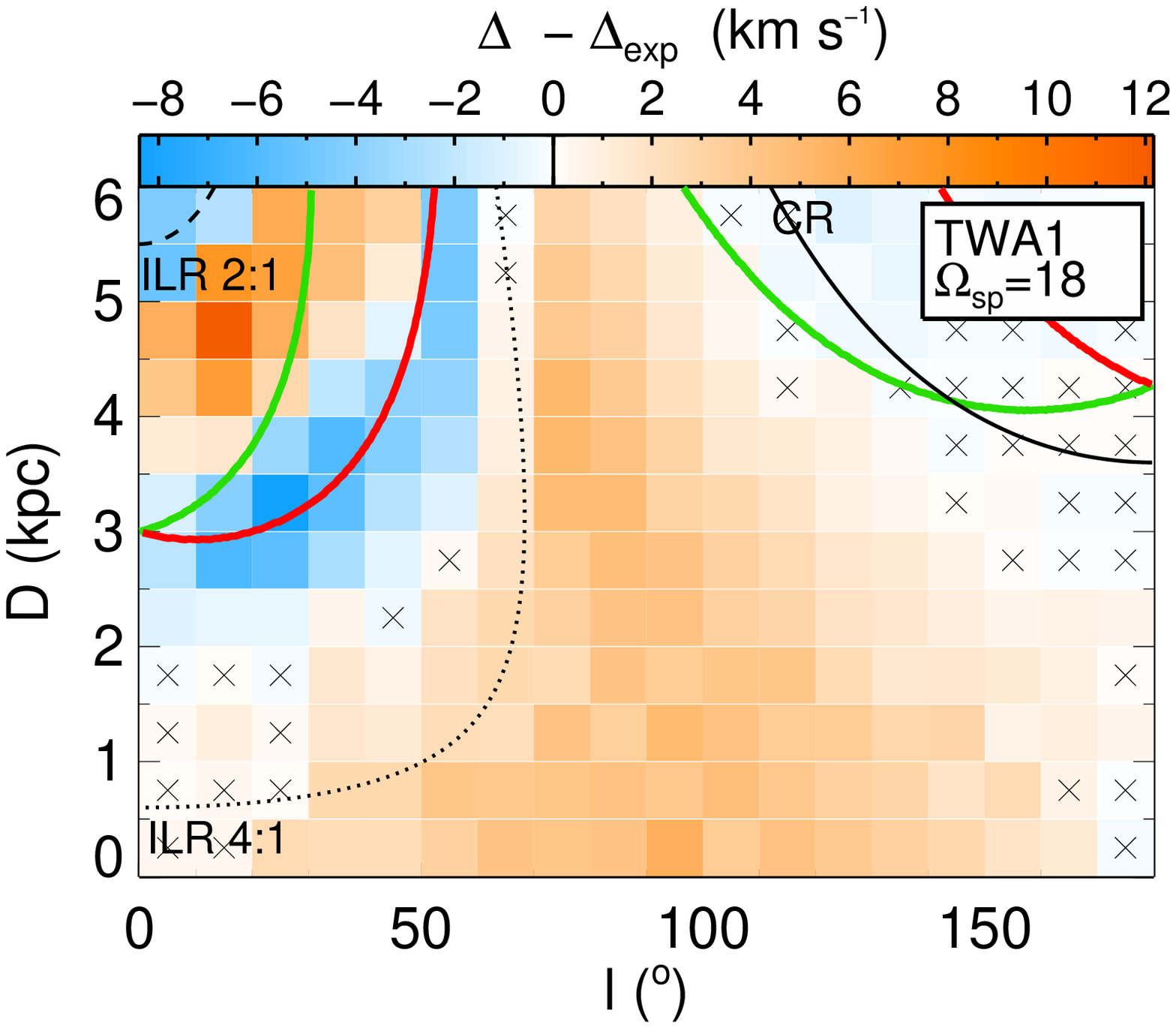}
\includegraphics[width=0.24\textwidth]{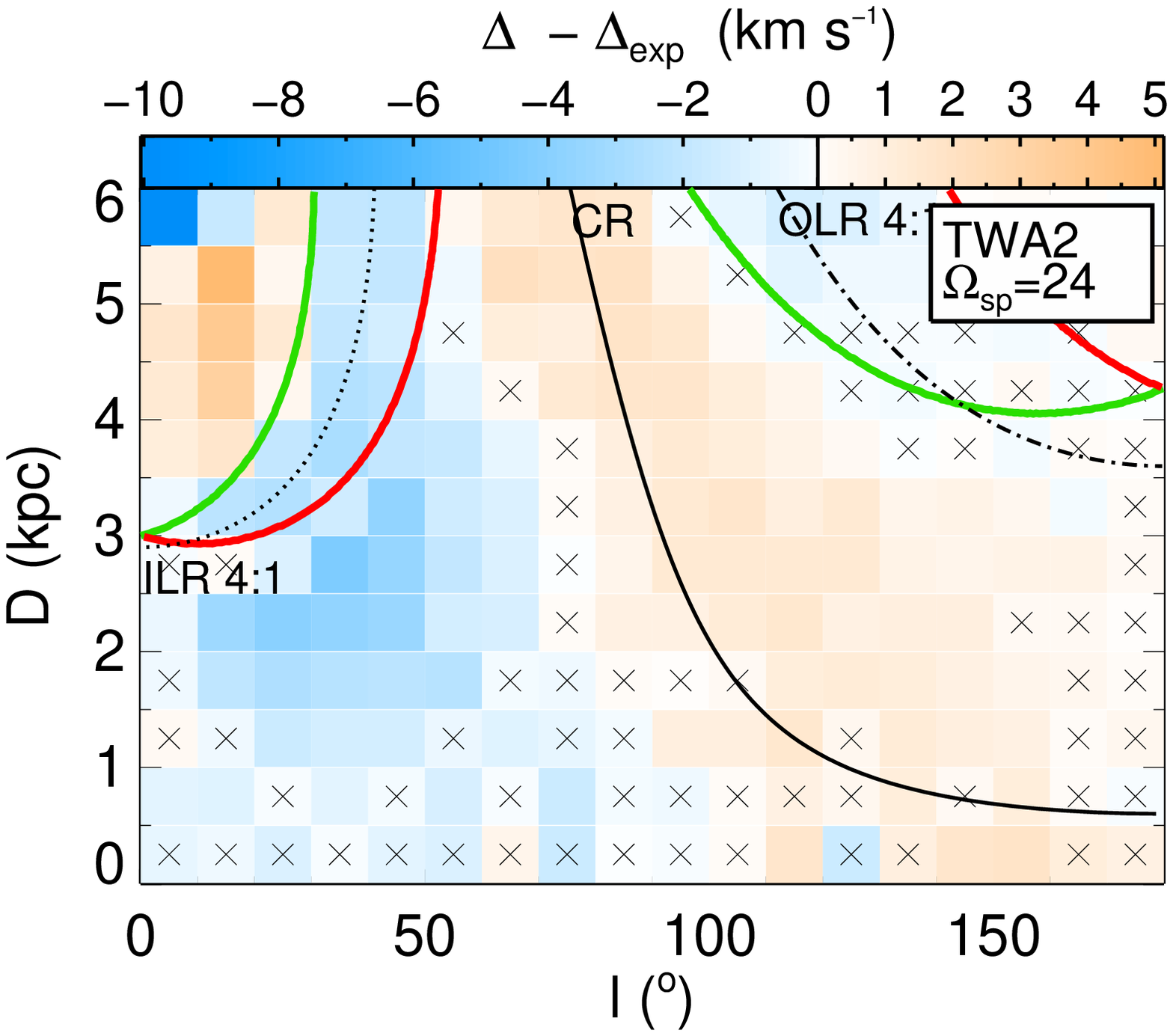}
\includegraphics[width=0.24\textwidth]{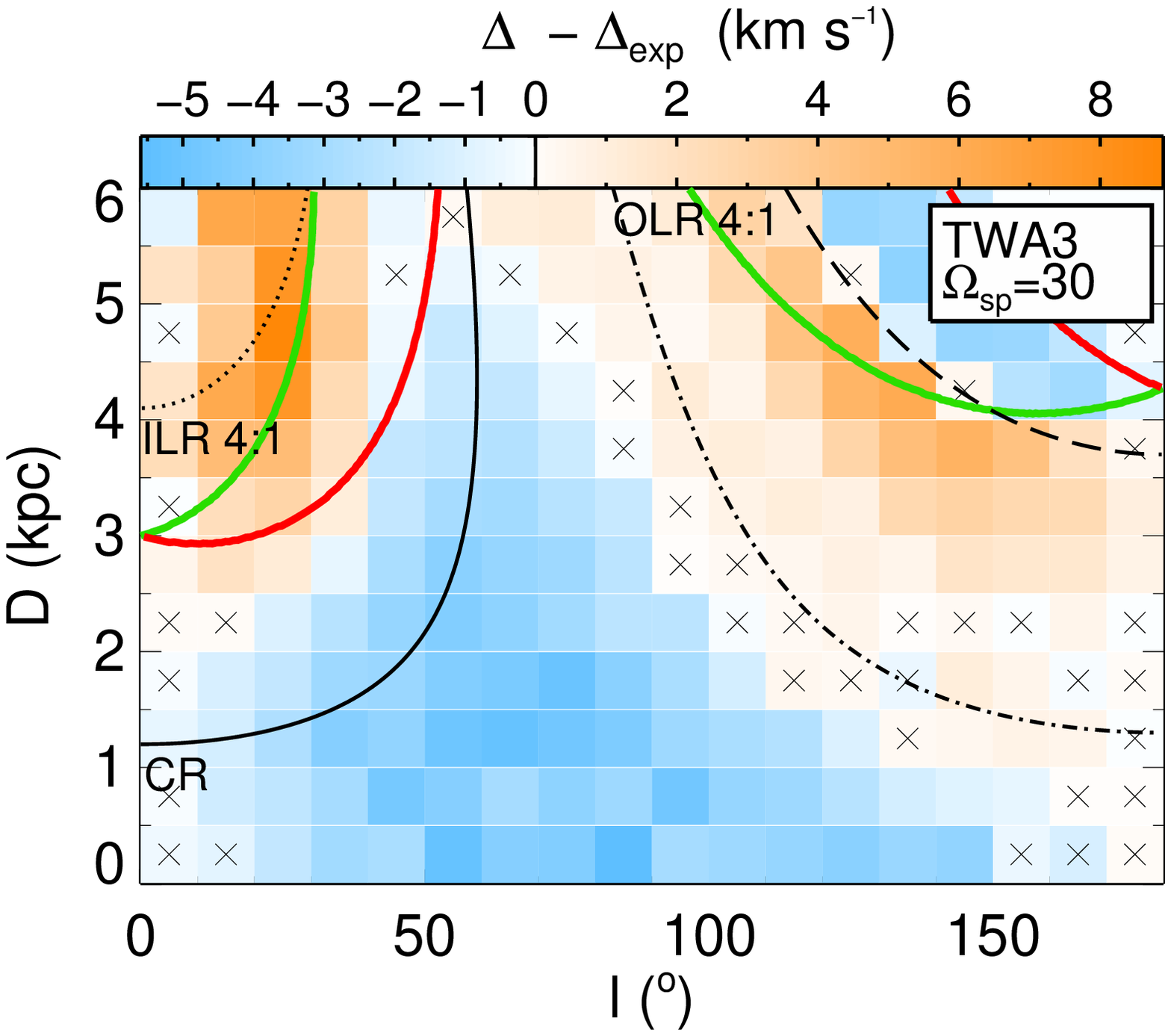}

\includegraphics[width=0.24\textwidth]{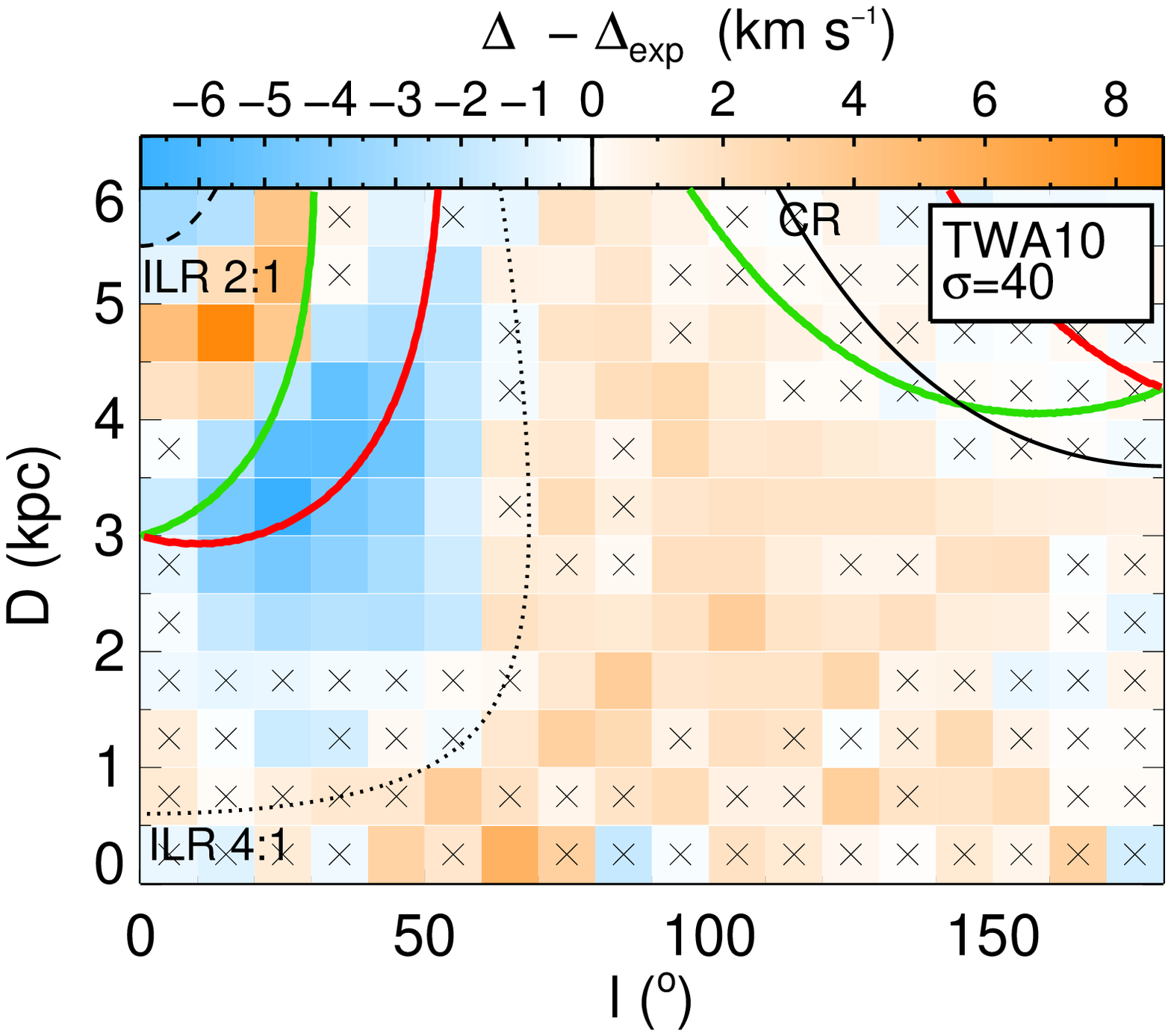}
\includegraphics[width=0.24\textwidth]{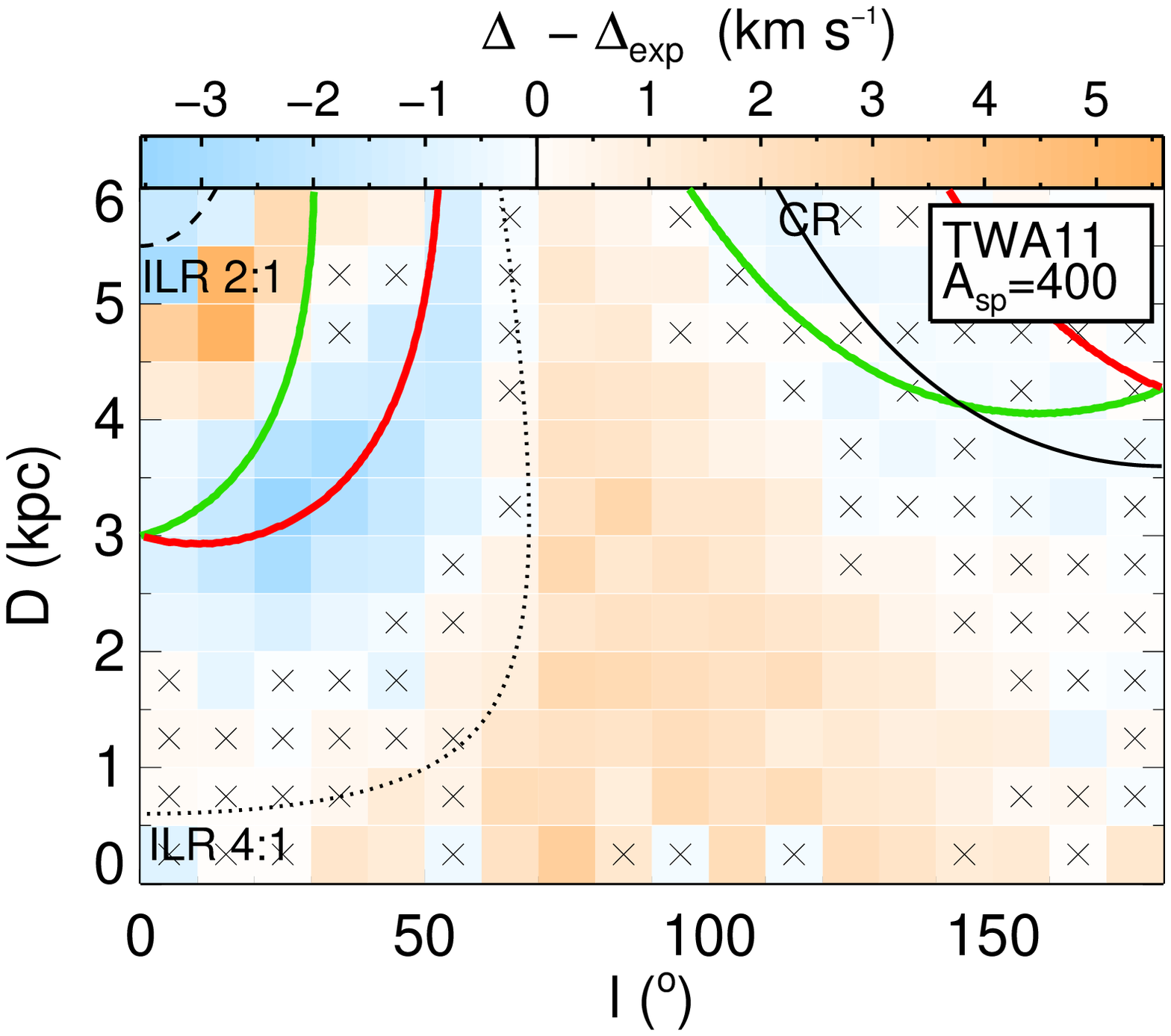}
\includegraphics[width=0.24\textwidth]{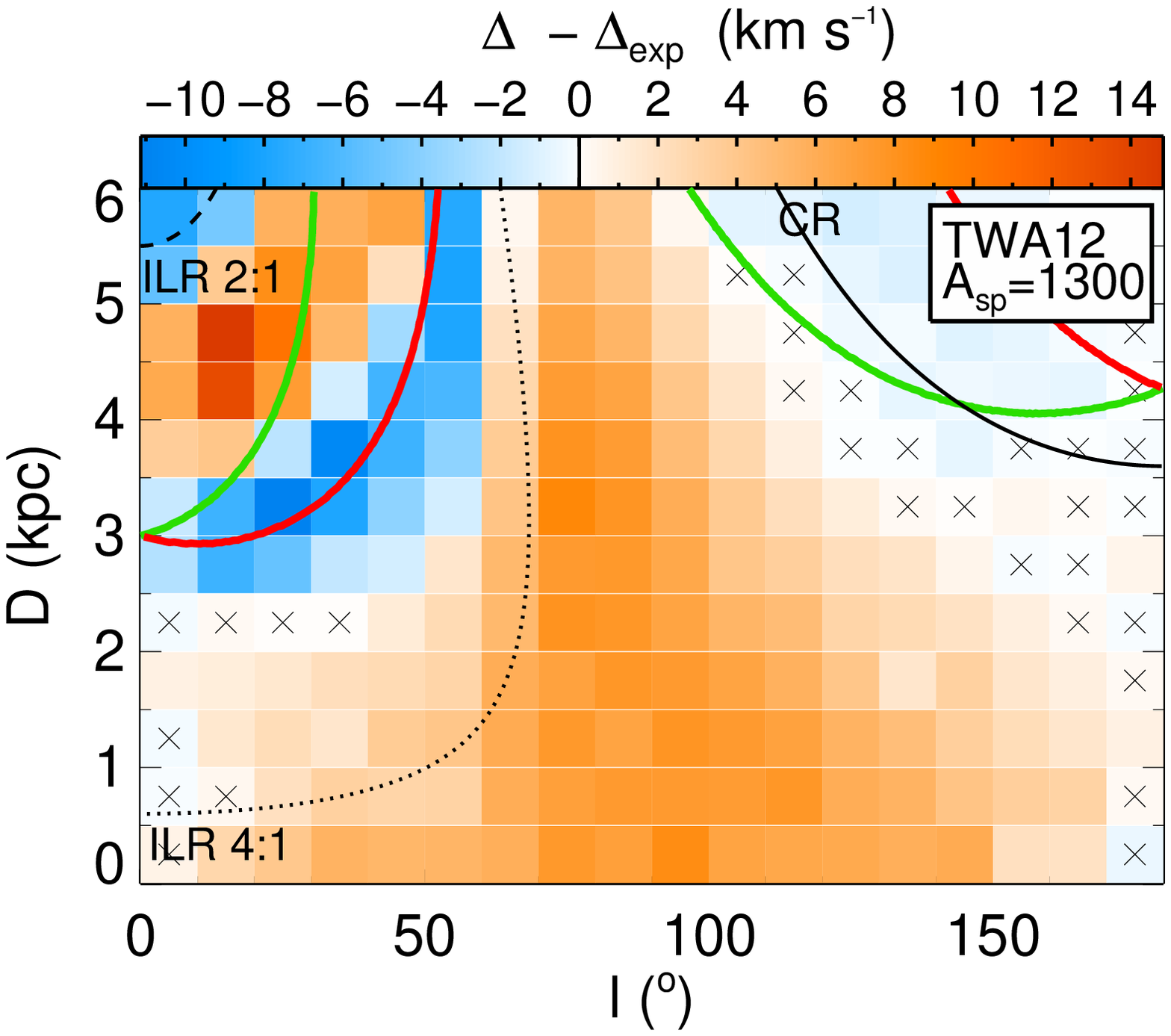}
\includegraphics[width=0.24\textwidth]{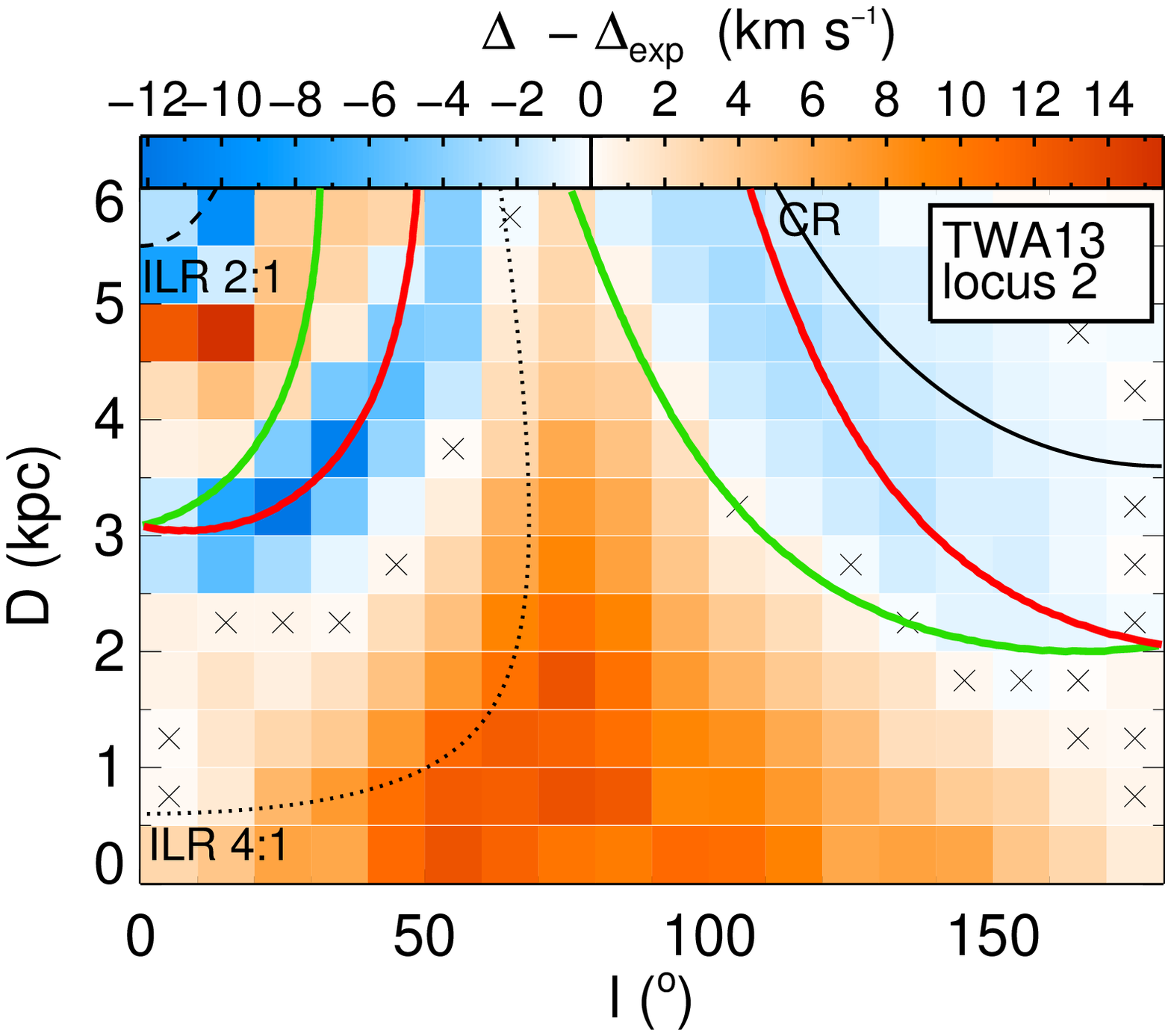}

\includegraphics[width=0.24\textwidth]{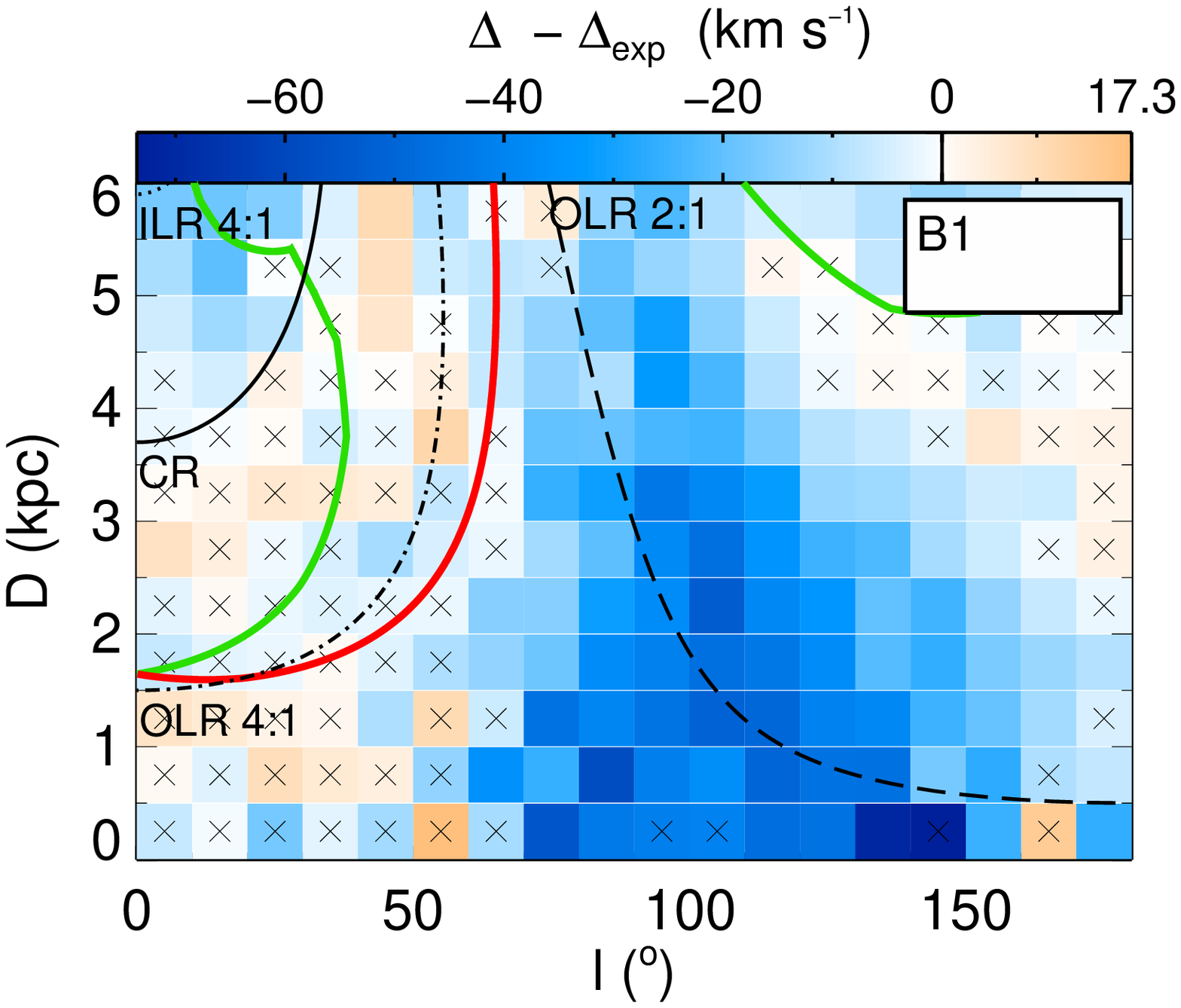}
\includegraphics[width=0.24\textwidth]{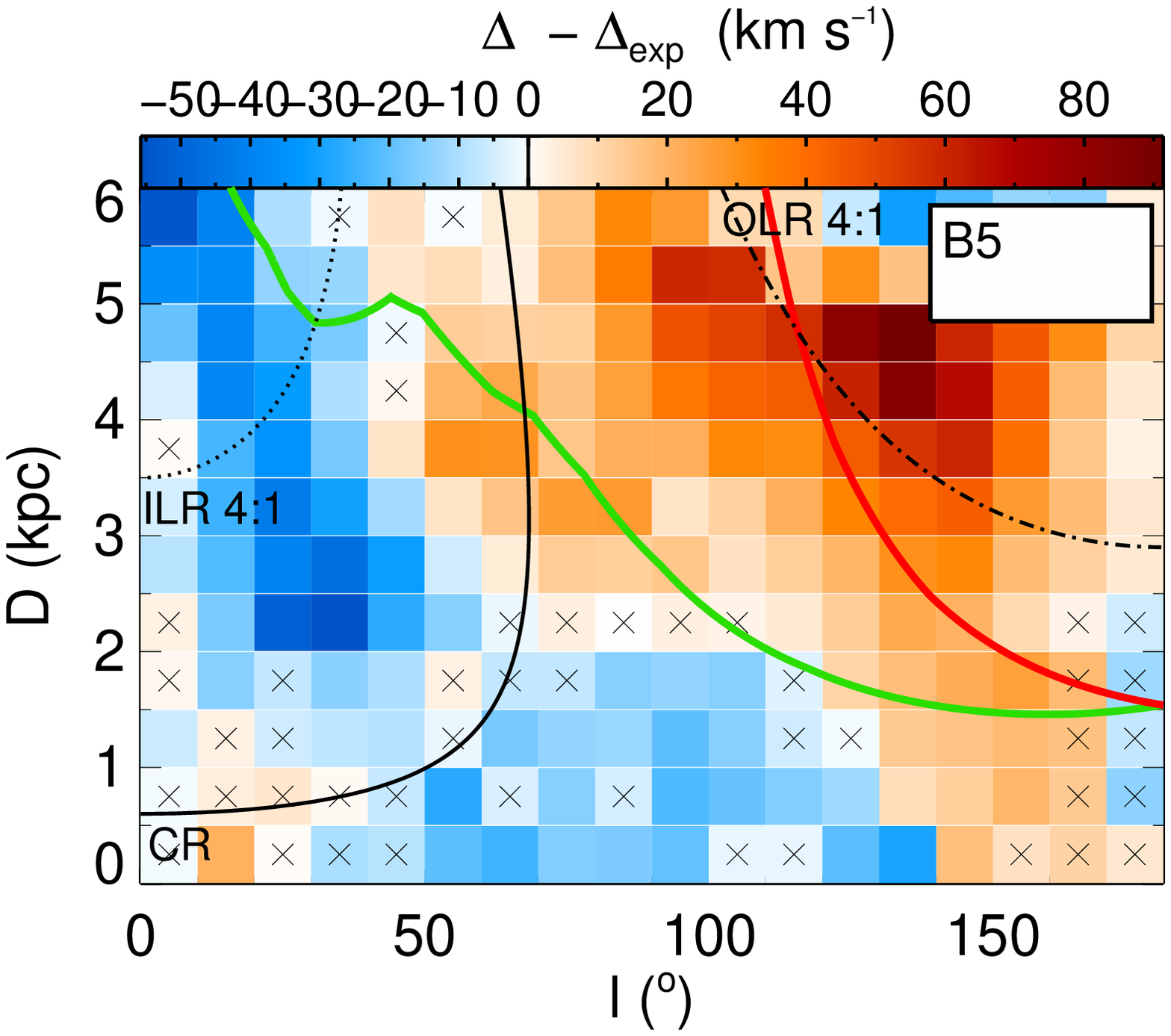}
\includegraphics[width=0.24\textwidth]{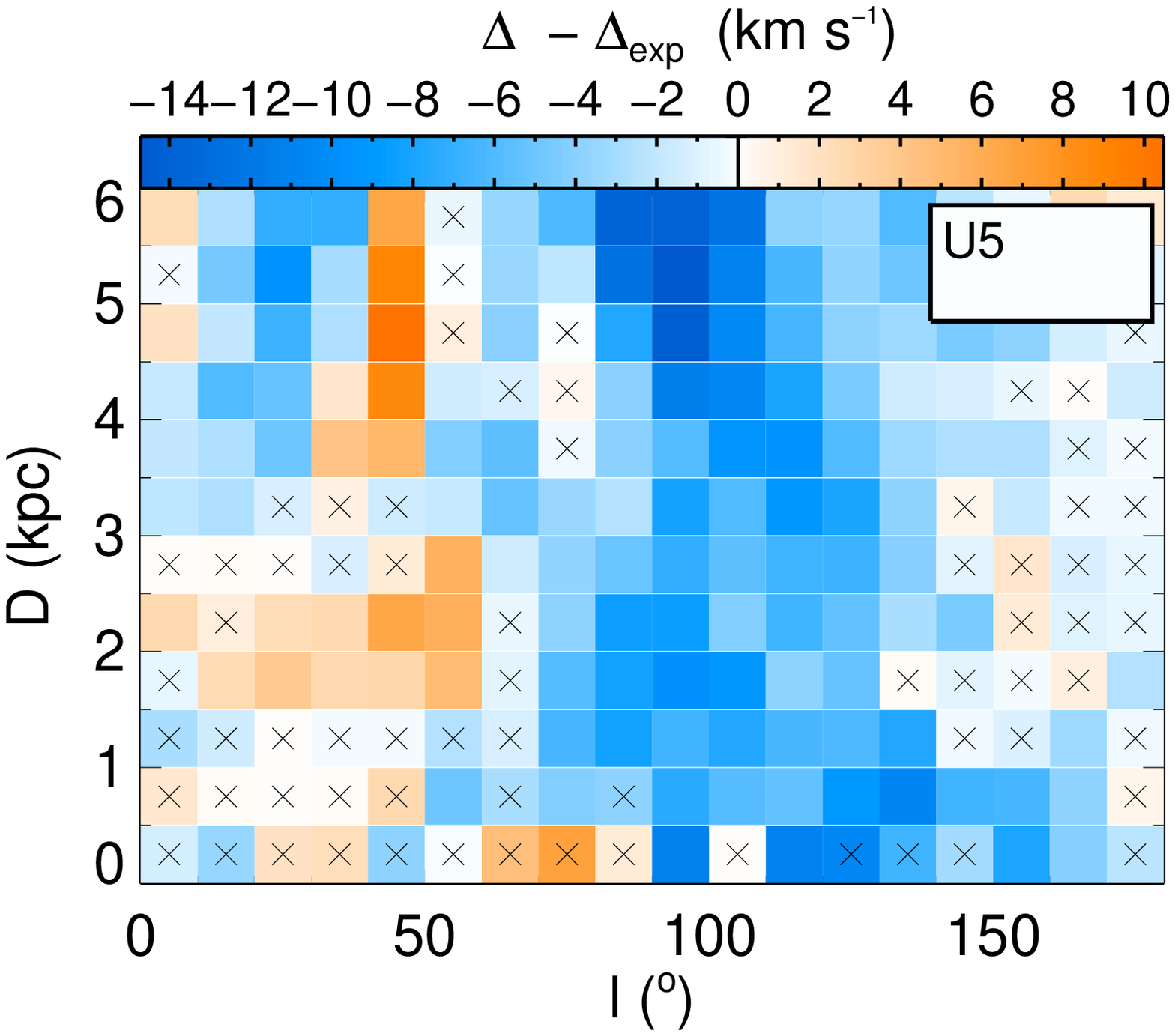}
\includegraphics[width=0.24\textwidth]{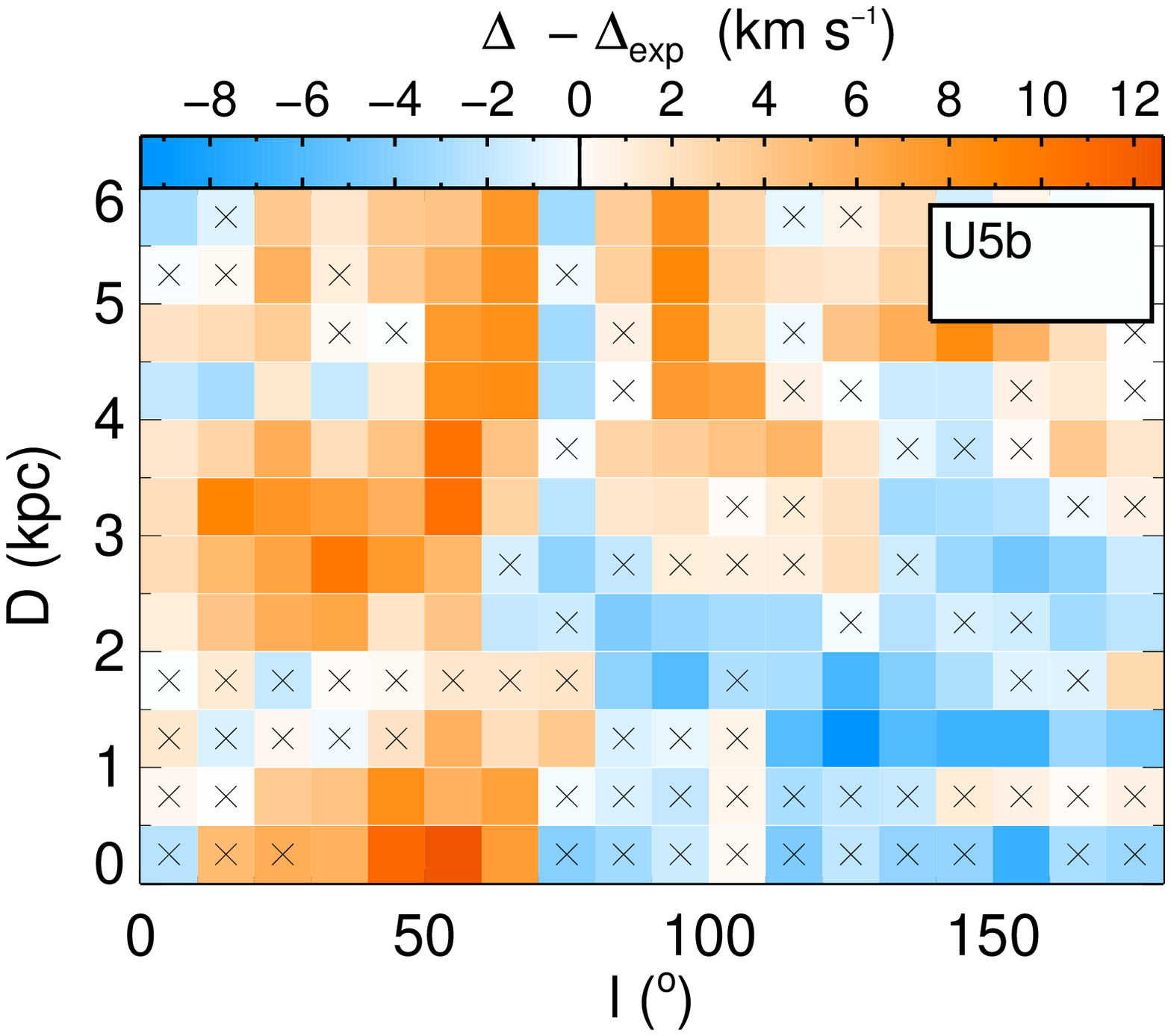}

  \caption{Values of $\D-\De$ as a function of $l$ and $D$, where $\D\equiv\vt\,(l>0)-\vt\,(l<0)$ is the difference between the median transverse velocity in symmetric bins ($l$,$d$) and ($-l$,$d$), and $\De=2\Us\sin l$ is the expected value for an axisymmetric model. The first two rows are for simulations with analytic models, and the last row is for N-body models. The colour scale is the same for all panels, except for models B1 and B5 that have higher values. We plot a black cross where the $\D-\De$ is  statistically consistent with 0 with a $75\%$ confidence, i.e. where the values of $\D$ are compatible with an axisymmetric model. The locations of the resonances follow the same line code as \fig{vgalTP}. The loci of the spiral arms are indicated in green and red for positive and negative $l$, respectively.  We use the loci estimated in \citet{RocaFabrega2013,RocaFabrega2014} for  N-body models B. }
         \label{vtsym2d}
   \end{figure*}

     In \fig{vtsym2d} we plot the discrepancy between the difference in $\vt$ at symmetric longitudes ($\D$) and the expected value in an axisymmetric case ($\De$) at each ($l$,$d$) bin and its symmetric counterpart at ($-l$,$d$) for all of our models, that is $\D-\De$. 
     In practise, we basically subtract the left part of the bottom middle panel of \fig{obs} from the right part, and then subtract the expected value. We estimate the statistical error on the median $\vt (l>0)$ and $\vt (l<0)$ with bootstrapping. We plot a black cross in bins where the $\D-\De$ is still compatible with 0 with a 75\% confidence, i.e. where $\D$ is compatible with the expected value for an axisymmetric model $\De$.  We overplot the locations of the resonances (thin black lines) and the locus of the arms in green and red for the part at positive and negative $l$, respectively. \Fig{vtsym2daxi} is equivalent to \fig{vtsym2d} but for an axisymmetric model. In this case, we expect $\D-\De$ to be 0 for all bins. This is indeed the case, except for the Poisson fluctuations. In $82\%$ of the bins of this plot, $|\D-\De|$ is smaller than $0.5\kms$.%\red{also a limit in number of particles per bin, limit to 50?, not done finally}
   
\subsection{Comparison of different models}\label{sym1}
   
\begin{figure}
   \centering
\includegraphics[width=0.24\textwidth]{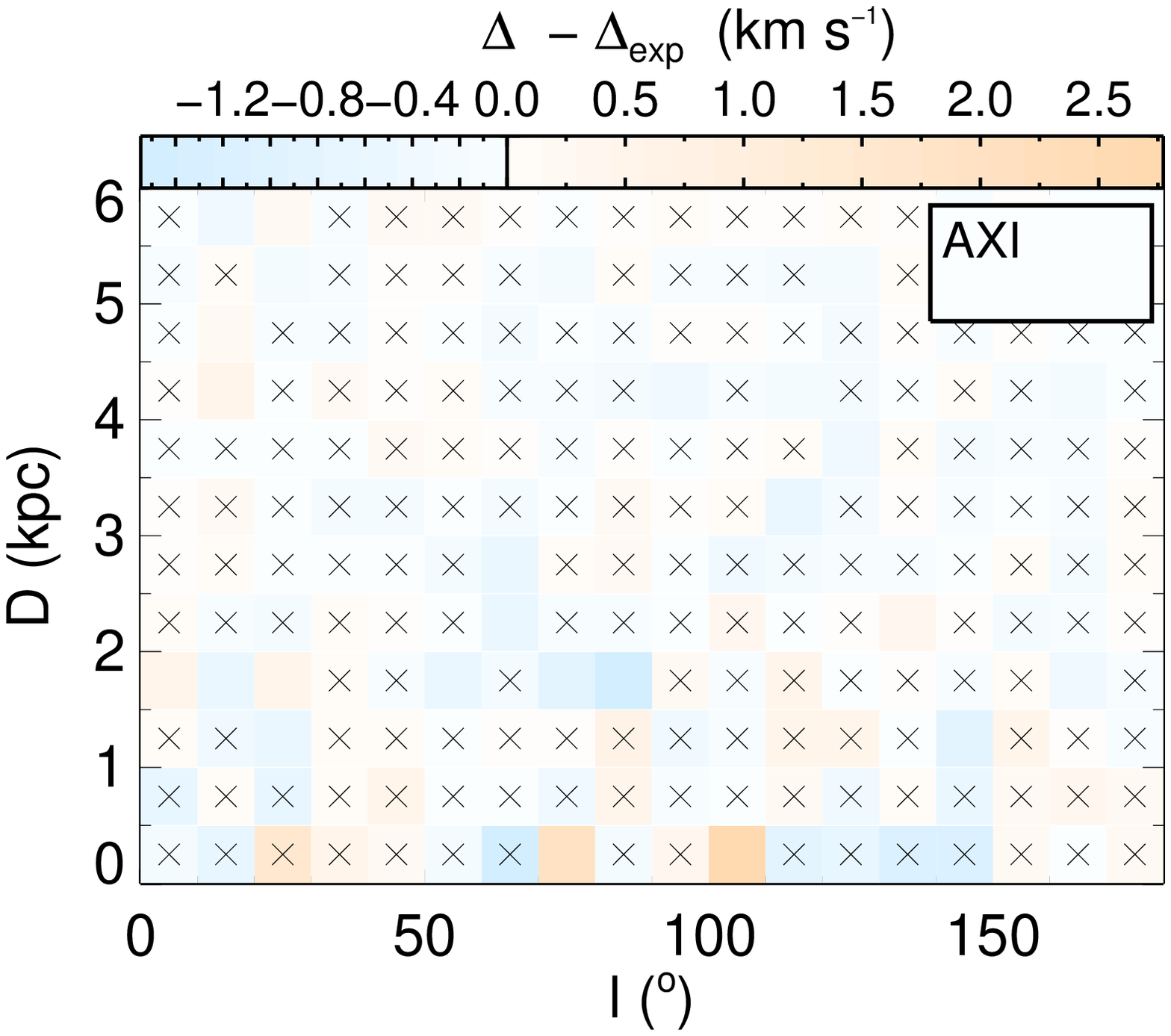}
  \caption{Same as \fig{vtsym2d} but for an axisymmetric model.}
         \label{vtsym2daxi}
   \end{figure}

   Models in \fig{vtsym2d} (with spiral arms) show higher values of $\D-\De$ compared to \fig{vtsym2daxi} (axisymmetric) and present defined patterns. In models TWA0 and TWA1 (first two panels in top row), we expect an excess of transverse velocity for positive $l$  compared to negative $l$ (red) at close distances for all longitudes and at large distances for the range $l\sim[70,120]\deg$. Towards the Galactic centre, we see three changes of sign: at $2.5\kpc$, slightly after crossing the spiral arms, and around the ILR 2:1. 
   %Note, however, that at inner Galactic regions, the effects of the bar that are not included here could be important. 
   In the anti-centre direction, according to these two models, we expect a change of sign around $3\kpc$. Models TWA0 and TWA1 have the same configuration of spiral arms, except for their pattern speed which differs by $5\kmskpc$. Even with this small difference, there are deviations between these two models: model TWA0 has larger perturbations and the last change of sign at $l=0$ related to the ILR 2:1 occurs at 4 instead of $5\kpc$ as in TWA1. 
   
   The differences between these two models and TWA2 and TW3 (\fig{vtsym2d} third and fourth panels from left in top row) are even more conspicuous. For TWA2 and TWA3, the spiral arms spatial configuration is the same as for TWA0 and TWA1 but the location of the resonances is very different with CR very close to the Sun. In TWA2, for distances up to 4 $\kpc,$ this model presents positive and negative values of $\D-\De$ with the limit approximately following  the CR resonance. For $l\sim0\deg$, 
  % slightly after crossing the arm there is also a change of sign from negative to positive, as in TWA0 and TWA1, but 
   there are now only two changes of sign up to $6\kpc$. In TWA3, negative values of $\D-\De$ dominate for all $l$ at nearby distances. We now see different changes of sign associated with the crossing of spiral arms and resonances.

   In the second row of \fig{vtsym2d} we present $\D-\De$ for the other simulations with the analytic potential (Table~\ref{t_sim}). The model TWA10 is the same as TWA1 (second panel in top row) but for a double velocity dispersion at the Sun radius. This allows us to see whether a similar kinematic perturbation appears for a hotter (older) population. In TWA11 we reduced the amplitude of the spiral potential. In these two cases the patterns in the velocity do not change compared to TWA1, but present smaller values of $\D-\De$. As expected, we observe larger kinematic perturbations for model
TWA12 with a higher spiral force amplitude.   
   Model TWA13 uses locus 2. In this case, the changes of sign follow the same patterns as in TWA1 but at different distances corresponding to the new location of the arms and resonances. For instance, the change of sign at $l\sim180\deg$ occurs closer to the Sun, exactly after crossing the arm.    
   While there could be doubt about whether  the cause of the sign change at $l\sim180\deg$ in models TWA0 and TWA1 was due to the arm crossing or to the resonance crossing (these two features overlap), after inspecting model TWA13, we conclude that the change of sign  seems to be due to the former.

   In the last row of \fig{vtsym2d} we show the results for our N-body models. Model B1 presents negative $\D-\De$ for most of the bins with some hints of positive values for small $l$ though statistically not very significant due to the low number of particles in this simulation. Interestingly, the kinematic features in this model resemble model TWA3. These two models have a large pattern speed and even though they have a very different nature, the kinematic effects on the stars look alike. Model B5 shares kinematic similarities with models TWA2 and TWA3\footnote{Despite the pattern speed of B5 and TWA2 being similar, the location of the resonances of B5 is more similar to TWA3 because of the different rotation curve of these models.}, especially the combination of positive and negative values of $\D-\De$.
   Model U5  is a multi-arm model that shows messier kinematic patterns that do not resemble any of the previous models, and even has patterns of different intensity inside blue regions. Finally, model U5b, which is the same as U5 but with a different orientation relative to the Sun, is similar to U5 in that it presents complex and smaller scale kinematic features but with a different pattern and sign. We notice here that models U5 and U5b have values of $\D-\De$ that are in the same range of our analytic models. By contrast, models B present much higher values. The interpretation of the panels in the last row is more complex since in they include other aspects that influence the disk kinematics apart from the arms (e.g. a bar or a lopsided disk).
      
  Plots such as those in \fig{vtsym2d} can be created directly with the \gaia observables and do not require any assumption on the axisymmetric potential of the Galaxy (but see Section~\ref{sym3}). With this advantage, this approach tells us directly about the effects of  spiral arms on disk kinematics. Moreover, the comparison with models, such as those shown here, allows us to retrieve properties of the spiral arms of the MW, namely the location of the arms and the main resonances, and thus, the pattern speed, strength, and nature of the arms (grand design versus corotating multi-arm configuration).

% \begin{figure*}
%   \centering
%   \includegraphics[width=0.33\textwidth]{/user_data/tantoja/ana_sim_t/moments_plots/dif_U_VT_l_dist_sym_r10_v18_a85_i15_nrev4_ic2my_a15000_rel_12_tf}
%\includegraphics[width=0.33\textwidth]{/user_data/tantoja/ana_sim_t/moments_plots/dif_U_VT_l_dist_sym_r10_v18_a13_i15_nrev4_ic2my_a15000_rel_tf}
%\includegraphics[width=0.33\textwidth]{/user_data/tantoja/ana_sim_t/moments_plots/dif_U_VT_l_dist_sym_r10_v18_a65_i12_vall_nrev4_ic2my_a15000_rel_tf}
%
%\includegraphics[width=0.33\textwidth]{/user_data/tantoja/ana_sim_t/moments_plots/dif_U_VT_l_dist_sym_r10_v15_a85_i15_nrev4_ic2my_a15000_rel_tf}
%\includegraphics[width=0.33\textwidth]{/user_data/tantoja/ana_sim_t/moments_plots/dif_U_VT_l_dist_sym_r10_v20_a85_i15_nrev4_ic2my_a15000_rel_tf}
%\includegraphics[width=0.33\textwidth]{/user_data/tantoja/ana_sim_t/moments_plots/dif_U_VT_l_dist_sym_r10_v30_a85_i15_nrev4_ic2my_a15000_rel_tf}
%  \caption{}
%         \label{polar}
%   \end{figure*}

%\green{other things that could be done:} examples of symmetrical \los 1d \fig{vtsym1d}. do we include them? before or after \fig{vtsym2d}

In Table \ref{t:results} we summarise the impact of effects of the spiral arms in symmetric longitudes in the explored region of the disk in the different models. The table shows the distance and longitude of the maximum absolute value of $\D-\De$ (columns 2, 3 and 4), that is the region of the disk surrounding the Sun that deviates more from its symmetric counterpart compared to the axisymmetric expectation. This region happens to be at inner longitudes and at a similar distance for the analytic models, but it is more variable for the N-body models. Columns 5 and 7 show the longitudes with maximum and minimum average $|\D-\De|$, respectively, and columns 6 and 8 indicate the corresponding average $|\D-\De|$. The Galactic longitude that is globally more affected by the arms is different for all models but the longitudes of $25\deg$ $75\deg$ seem recurrent in the analytic models. The direction of the anti-centre is clearly the least affected in the analytic models and in most of the N-body models.

Table \ref{t:results} also shows the fraction of bins of \fig{vtsym2d} (i.e. up to a distance of $6\kpc$) that have a value of $|\D-\De|$ larger than $2\kms$ (column 9) and $5\kms$ (column 10). The first fraction is large in most of the models except for TWA2 and TWA11 for which it is around $15\%$. It is even larger than $60\%$ for some models. But notice that, in the case of TWA11, we decreased the amplitude of the spiral arm force below the estimated ranges for the MW. We also note that model B5, which has most of the bins with a very large value of $|\D-\De|$, is not comparable to the MW. The fraction of bins with perturbations larger than $5\kms$ is small for most of the models. In  column 11 we show the median value of $|\D-\De|$.
 
From Table \ref{t:results} we also see that $\D-\De$ is of the order of $2\kms$ for an important fraction of the sphere around the Sun and, therefore, a kinematic precision smaller than $1\kms$ is needed in the measured median $\vt$ at each longitude and its symmetric counterpart.
 
 \begin{table*}
 \caption{Kinematic differences when comparing positive and negative Galactic longitudes in the spiral models. Columns show: 1) model; 2), 3), and 4) Galactic longitude with the bin of maximum $|\D-\De|$ in the explored range of distances ($6\kpc$), distance of this bin, and corresponding value of $\D-\De$; 5) and 6) longitude of maximum mean $|\D-\De|$ and corresponding maximum mean $|\D-\De|$ for this longitude; 7) and 8) longitudes of minimum mean $|\D-\De|$ and corresponding minimum mean; 9) fraction of bins with $|\D-\De|\ge2\kms$; 10) fraction of bins with $|\D-\De|\ge5\kms$; 11) median value of $|\D-\De|$; 12) derived value of $\Us$ from Equation \ref{e_d2} obtained through a median (see text); 13) same as 12) but only for bins at $l=$155$\deg$, 165$\deg$, 175 $\deg$.   }             
 \label{t:results}      
 \centering          
     \tabcolsep 3.pt
% \begin{tabular}{lrrrrrrrrrr}     % 11 columns 
 \begin{tabular}{l|ccr|cr|cr|cc|c|rr}     % 11 columns 
 \hline\hline       
% \multicolumn{2}{l}{}&($\deg$)&($\deg$)&($\deg$)&&&&&&($\kmskpc$)\\ 
 %&&&$(\kpc)$&$(\kms)$&$(\kms)$&&&$(\kmskpc)$&$(\deg)$\\ 
\multicolumn{1}{c|}{1}& \multicolumn{1}{c}{2}& \multicolumn{1}{c}{3}& \multicolumn{1}{c|}{4}& \multicolumn{1}{c}{5}& \multicolumn{1}{c|}{6}& \multicolumn{1}{c}{7}& \multicolumn{1}{c|}{8}& \multicolumn{1}{c}{9}& \multicolumn{1}{c|}{10}& \multicolumn{1}{c|}{11}& \multicolumn{1}{c}{12}& \multicolumn{1}{c}{13} \\ \hline  
Model& $l_{\rm max}$&$D_{\rm max}$&\multicolumn{1}{c|}{$(\D-{\De})_{\rm max}$}&$l$&\multicolumn{1}{c|}{$\overline{|\D-\De|}_{\rm max}$}&$l$&\multicolumn{1}{c|}{$\overline{|\D-\De|}_{\rm min}$}&$\%_{\ge2\kms}$&$\%_{\ge5\kms}$&$|\D-\De|_{med}$&\multicolumn{1}{c}{$\Us$}&\multicolumn{1}{c}{$\Us$($l=[155,175]\deg$)}\\     
& ($\deg$)&($\kpc$)&\multicolumn{1}{c|}{($\kms$)}&($\deg$)&\multicolumn{1}{c|}{($\kms$)}&($\deg$)&\multicolumn{1}{c|}{($\kms$)}&&&($\kms$)&\multicolumn{1}{c}{($\kms$)}&\multicolumn{1}{c}{($\kms$)}\\       \hline  
%             TWA0& 15&4.8&12.1+-0.4& 85& 4.0&175& 0.4\\
%TWA0 ($D<4 \kpc$)& 25&3.2&-8.3+-0.4& 85& 4.4&175& 0.5\\
%             TWA1& 25&4.8& 8.7+-0.4& 25& 4.2&175& 0.5\\
%TWA1 ($D<4 \kpc$)& 65&0.2&-6.0+-2.0& 65& 4.2&175& 0.5\\

             TWA0& 15&3.8& 22.9$\pm  0.3$& 75&11.4&175& 0.3& 68& 53& 5.5&12.6$\pm 0.4$& 9.3$\pm 0.7$\\
             TWA1& 15&4.8& 12.2$\pm  0.3$& 75& 4.1&175& 0.2& 49&  7& 1.9&10.1$\pm 0.2$& 9.1$\pm 0.3$\\
             TWA2&  5&5.8&-10.0$\pm  0.3$& 35& 2.2&175& 0.3& 13&  0& 0.9& 9.1$\pm 0.1$& 9.5$\pm 0.2$\\
             TWA3& 25&4.8&  8.9$\pm  0.3$& 25& 4.2&175& 0.4& 50&  6& 2.1& 8.6$\pm 0.2$& 8.9$\pm 0.4$\\            
            TWA10& 15&4.8&  8.7$\pm  0.5$& 25& 3.4&165& 0.5& 29&  3& 1.4& 9.5$\pm 0.1$& 9.5$\pm 0.3$\\
            TWA11& 15&4.8&  5.6$\pm  0.3$& 75& 2.1&165& 0.3& 16&  0& 0.8& 9.3$\pm 0.1$& 9.1$\pm 0.2$\\
            TWA12& 15&4.8& 14.8$\pm  0.3$& 75& 7.2&175& 0.3& 61& 36& 3.2&11.0$\pm 0.3$& 8.9$\pm 0.5$\\
            TWA13& 15&4.8& 15.5$\pm  0.5$& 75& 7.9&175& 0.4& 57& 27& 2.6&10.1$\pm 0.4$& 8.8$\pm 0.4$\\
               B1&145&0.2&-73.5$\pm 37.0$& 95&34.5& 35& 3.4& 86& 70& 9.9& 1.1$\pm 0.9$&-2.8$\pm 3.1$\\
               B5&135&4.8& 91.4$\pm  1.5$&135&37.2&175& 7.5& 94& 86&15.7&14.2$\pm 1.7$&39.9$\pm 3.1$\\
               U5& 95&5.2&-14.7$\pm  0.9$& 95& 9.9&175& 1.1& 67& 36& 3.6& 6.6$\pm 0.2$& 6.7$\pm 0.6$\\
              U5b& 55&0.2& 12.6$\pm  3.8$& 55& 6.9&175& 1.6& 60& 22& 2.7& 9.7$\pm 0.3$& 8.8$\pm 0.8$\\
          
\hline                  
 \end{tabular}
 \end{table*}

        \begin{figure*}
   \centering
\includegraphics[width=0.24\textwidth]{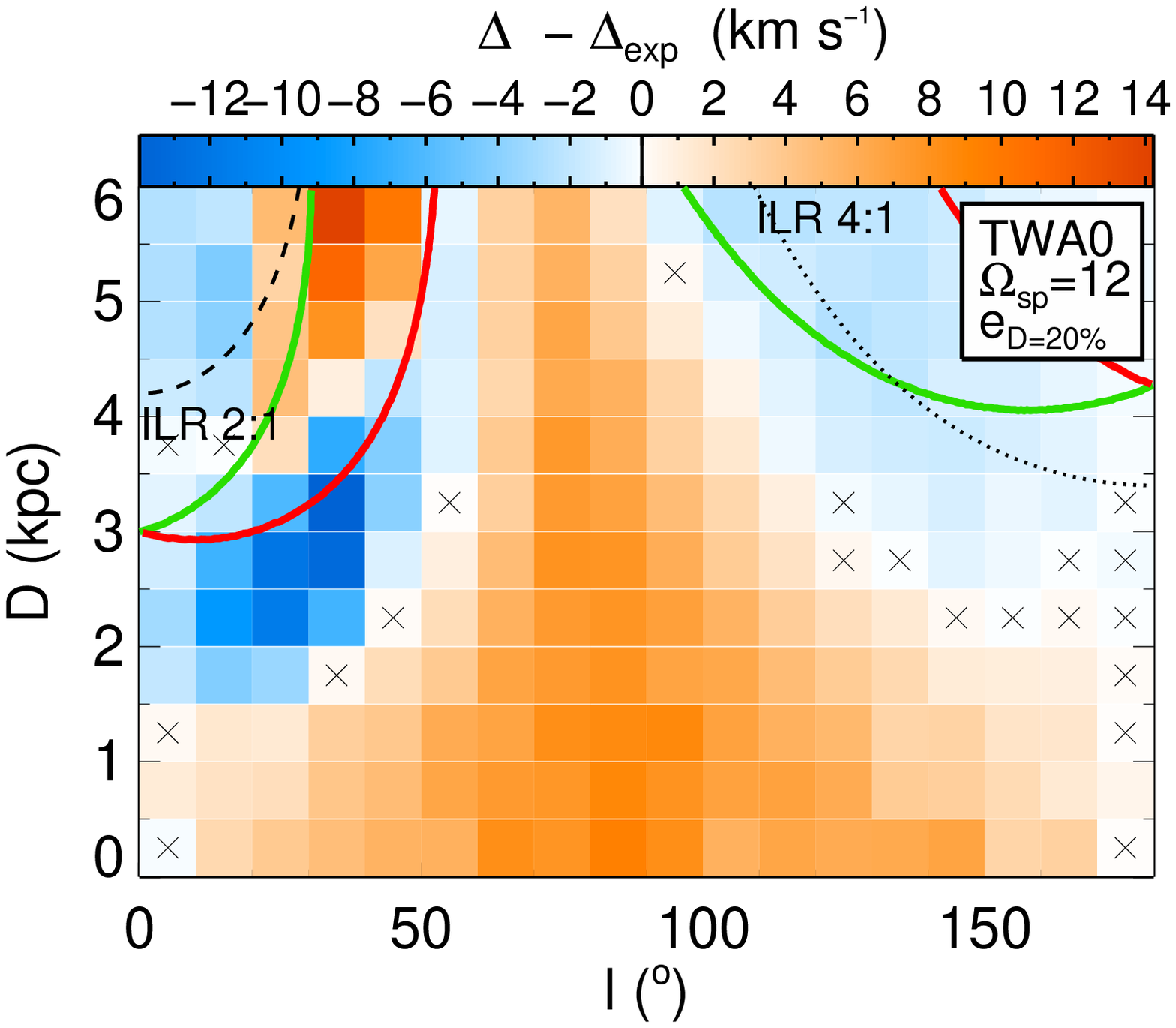}
\includegraphics[width=0.24\textwidth]{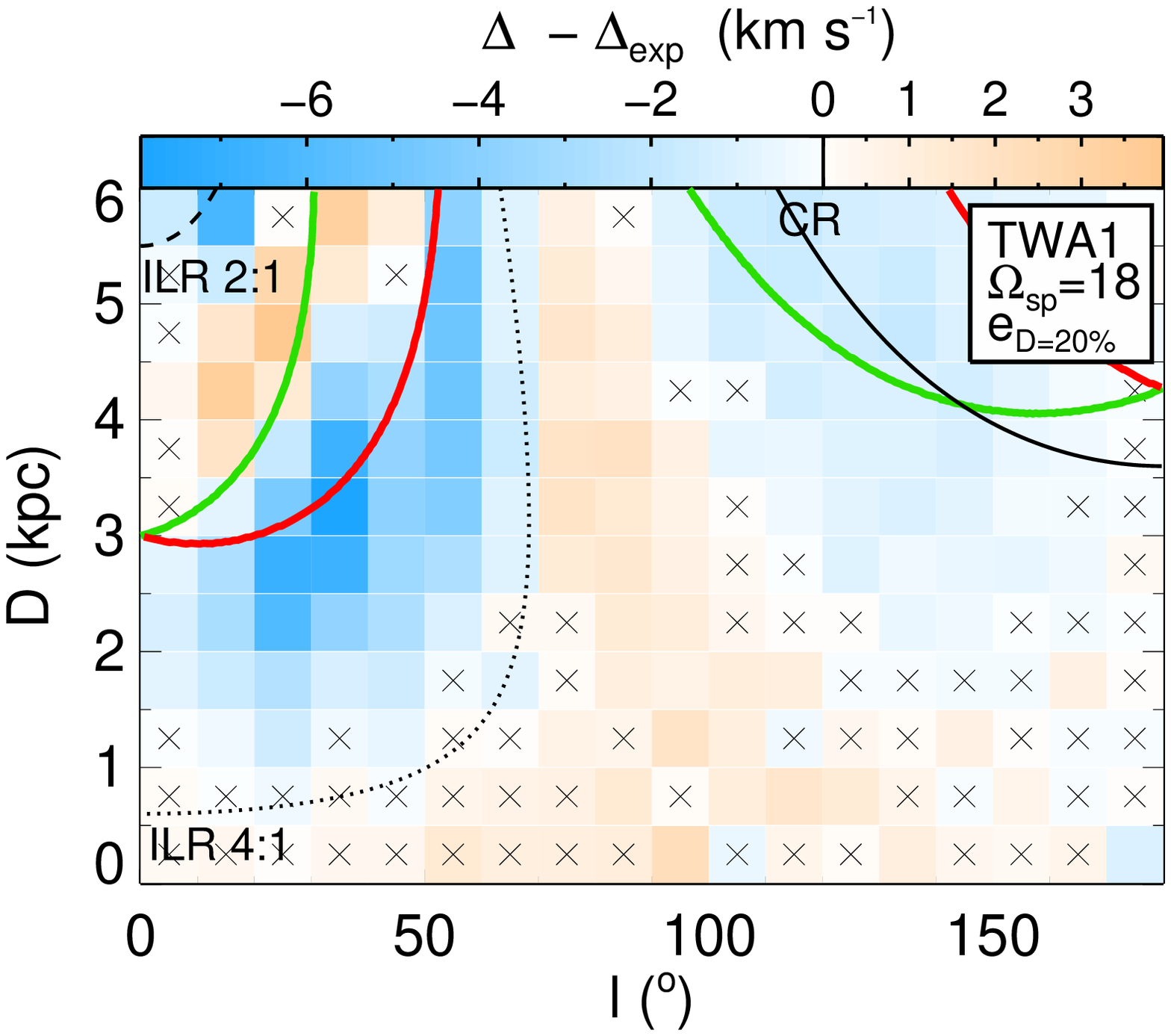}
\includegraphics[width=0.24\textwidth]{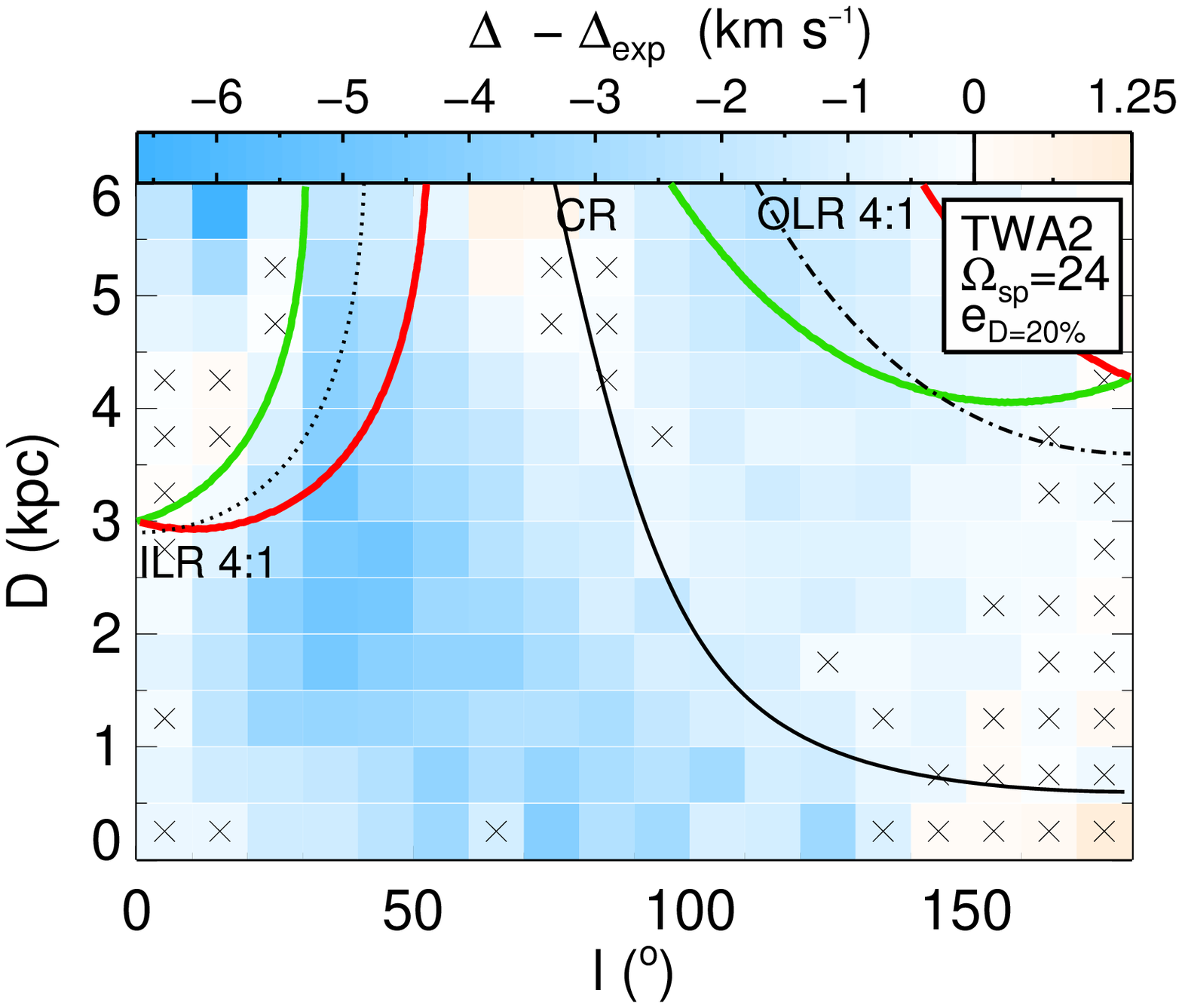}
\includegraphics[width=0.24\textwidth]{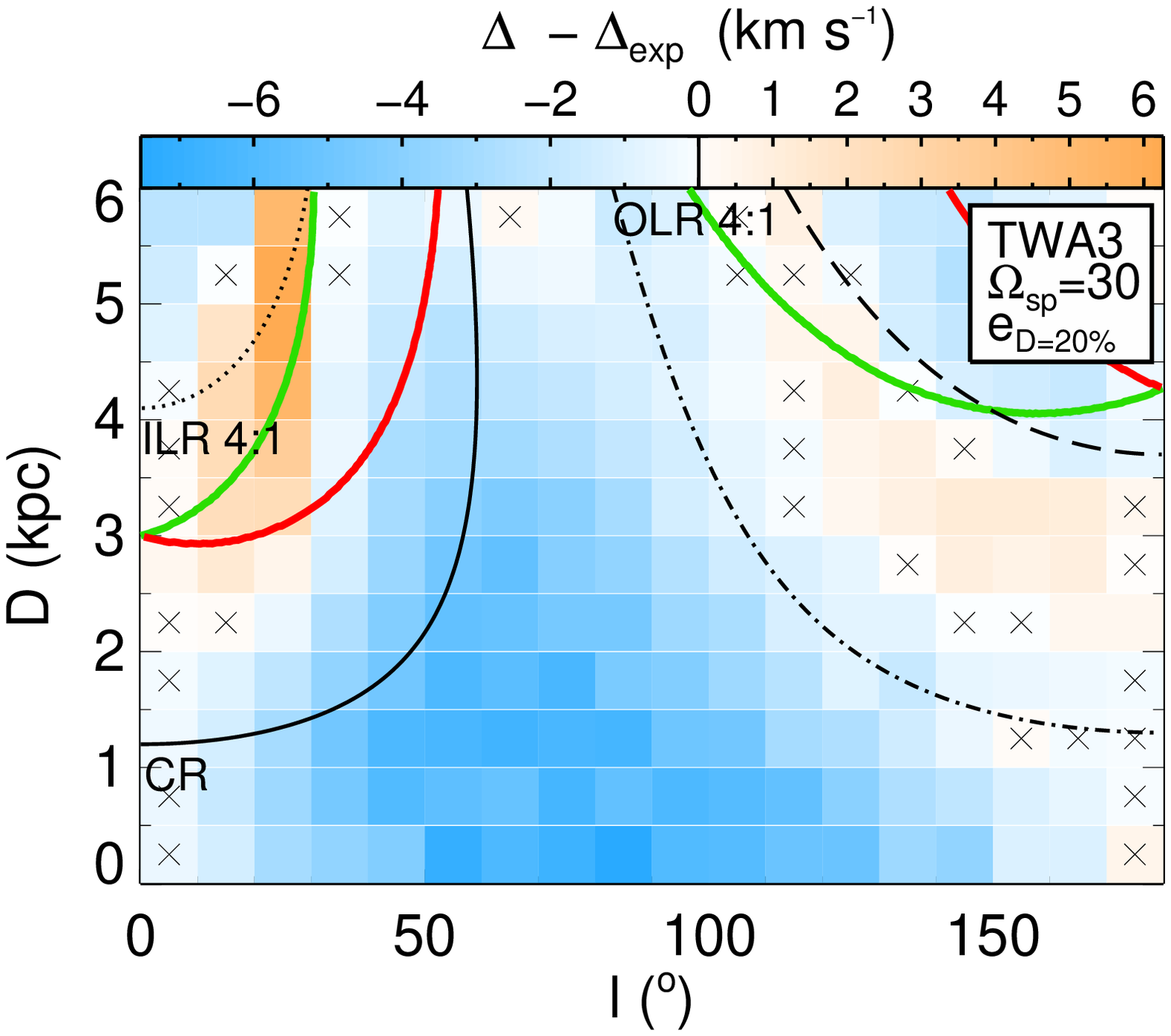}
  \caption{Same as \fig{vtsym2d} but introducing a random relative error in distance of $e_D=20\%$ in some models. The colour scale is the same as in Fig.~\ref{vtsym2d}.}
         \label{disterror}
   \end{figure*}      
      
\subsection{Distance error}\label{sym2}

In \fig{disterror} we show several models where we have introduced a random relative error in distance of $e_D=20\%$. They can be compared to the first row of \fig{vtsym2d} (without error). In general, the velocity patterns are preserved, but we do observe several introduced biases that we must bear in mind for the future analysis with real data. First, in the models TWA0 and TWA1 with errors, for large values of $l$, the negative values of $\D-\De$ are now located at closer distances. Thus, this could mislead us to conclude that the spiral arms are closer than they really  are in the anti-centre direction, since this is similar to what happens in model TWA13. Model TWA2 is now dominated by blue colours, which would lead us to derive a slightly higher pattern speed. Model TWA3 does not change significantly. We  explored the models with different distance errors and conclude that an error of $e_D=20\%$ should be our maximum tolerable error in order not to have important biases that dilute the kinematic features.

\subsection{The value of $\Us$}\label{sym3}

%In this methodology we assume that we know the correct value of $\Us$. 
The effects of the non-axisymmetries are not usually taken into account in the measurement of the solar peculiar velocity. We investigate  how much the measured value of $\Us$ is affected by the spiral arms in our models. It follows from Equation \ref{e_d2} that, if we assume that there is no effect of the spiral arms on the kinematics of stars, the comparison of the transverse velocities in symmetric longitudes give us the value of $\Us=(\D-\De)/(2\sin l) $. We computed $\Us$  from this formula to each bin in \fig{vtsym2d} and subsequently performed the median of all bins (column 12 of Table \ref{t:results}). We also provide the error on the median computed as the dispersion for 1000 bootstraps. The obtained value of $\Us$ is different from the true value used in the simulations ($\Us=9\kms$) by at most $2\kms$ for most of the analytic models. These differences are due to the bias introduced by the arms and not by the statistical error, which is much smaller. If we only use the bins at $l=155\deg$, $165\deg$, and $175\deg$, which are the least affected regions of the disk according to our models (column 7), the obtained $\Us$ (column 13) differs no more than $0.5\kms$ from the true value. On the contrary, the obtained $\Us$ for the N-body models B1 and B5 is significantly different from the true value. We have to keep in mind, however, that the  perturbations for these models are very high and other effects such as a lopsided disk could also contribute to these biased values of $\Us$. 

 Ideally, when fitting the data to the models presented here, one should also try to fit  for the value of $\Us$ or, at least, to include its uncertainty so it can be propagated in the fit. However,  in our method, adopting a wrong value of 
 $\Us$ would change the term $2\Us\sin l$ in Equation \ref{e_d2}, which would shift the colour scale of our plots. This would not change the colour patterns in \fig{vtsym2d} except if the assumed value of $\Us$ is significantly wrong and changes $2\Us\sin l$ by an amount comparable to the typical values of $\De-\D$ ($2\kms$ or higher). We  demonstrated above that, in most cases, it is possible to determine the value of $\Us$ with better accuracy. 
 %Since these are of the order of $2\kms$ or higher , that would require an error in $\Us$ of at most $\sim1\kms$ (the one required for $l=90$). In \fig{wrongU} we show the results for model TWA1 having assumed wrong values of $\Us$ of 8 and $10\kms$, where we see how the global colour patterns do not change significantly. 

%\begin{figure}
%   \centering
%\includegraphics[width=0.24\textwidth]{/user_data/tantoja/ana_sim_t/moments_plots/difmed_VT_l_ldistr10_v18_a85_i15_nrev4_ic2my2d_a15000_r_tf_h_exp_U8}
%\includegraphics[width=0.24\textwidth]{/user_data/tantoja/ana_sim_t/moments_plots/difmed_VT_l_ldistr10_v18_a85_i15_nrev4_ic2my2d_a15000_r_tf_h_exp_U10}
%%\includegraphics[width=0.24\textwidth]{/user_data/tantoja/ana_sim_t/moments_plots/difmed_VT_l_ldistr10_v30_a85_i15_nrev4_ic2my2d_a15000_r_tf_h_exp_U8}
%%\includegraphics[width=0.24\textwidth]{/user_data/tantoja/ana_sim_t/moments_plots/difmed_VT_l_ldistr10_v30_a85_i15_nrev4_ic2my2d_a15000_r_tf_h_exp_U10}
%%\includegraphics[width=0.24\textwidth]{/user_data/tantoja/ana_sim_t/moments_plots/difmed_VT_l_ldistNB_B1_tf_h_exp_U8}
%%%\includegraphics[width=0.24\textwidth]{/user_data/tantoja/ana_sim_t/moments_plots/difmed_VT_l_ldistNB_B1_tf_h_exp_U10}
%%\includegraphics[width=0.24\textwidth]{/user_data/tantoja/ana_sim_t/moments_plots/difmed_VT_l_ldistNB_U5_o2_tf_h_exp_U8}
%%\includegraphics[width=0.24\textwidth]{/user_data/tantoja/ana_sim_t/moments_plots/difmed_VT_l_ldistNB_U5_o2_tf_h_exp_U10}
%  \caption{Same as \fig{vtsym2d} but assuming a wrong value of $\Us$ of 8 and $10\kms$ for model TWA1.}
%         \label{wrongU}
%   \end{figure}

\subsection{Extension of the method}\label{others}

\begin{figure*} 
   \centering

\includegraphics[width=0.24\textwidth]{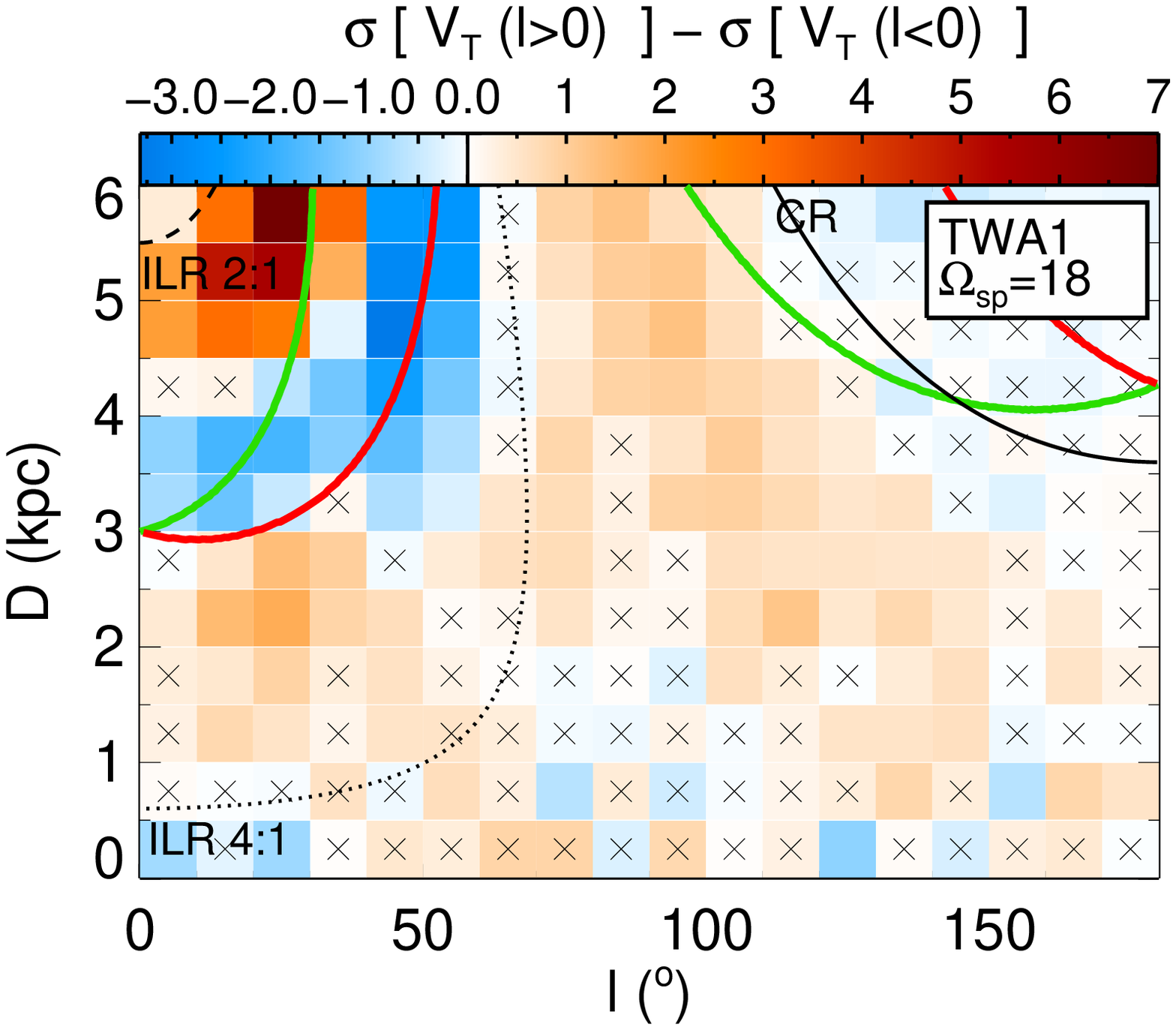}
\includegraphics[width=0.24\textwidth]{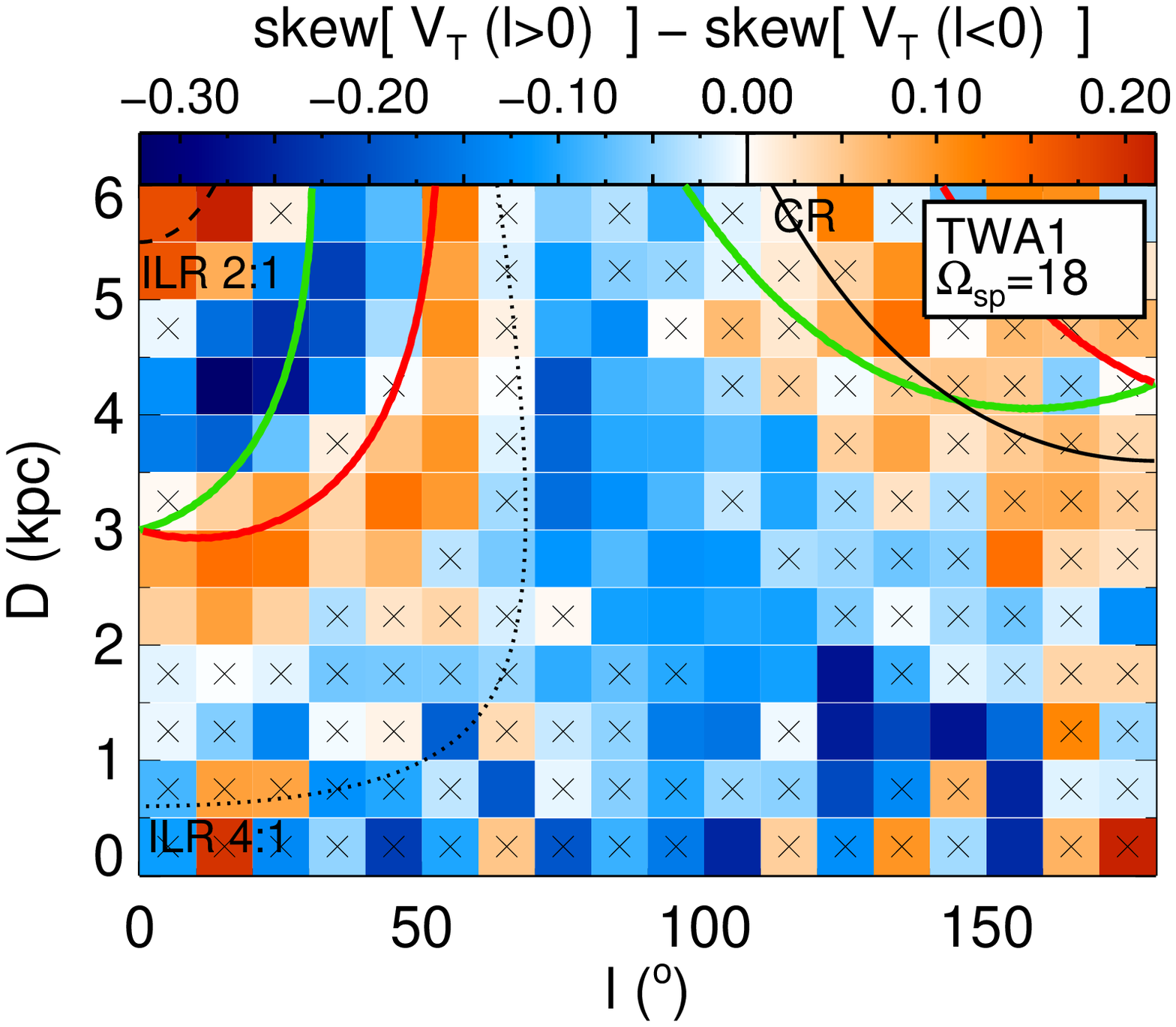}
\includegraphics[width=0.24\textwidth]{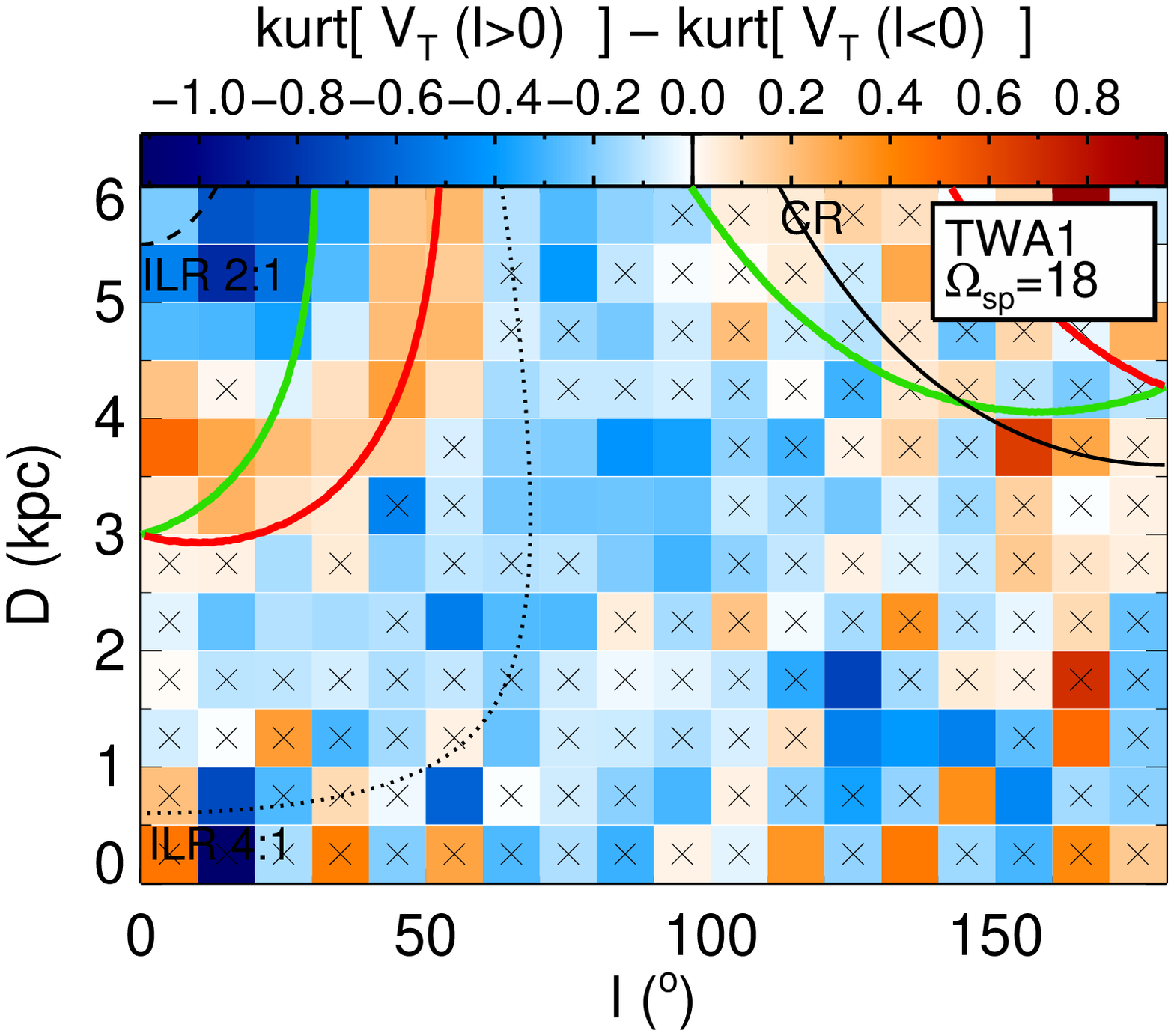}
\includegraphics[width=0.24\textwidth]{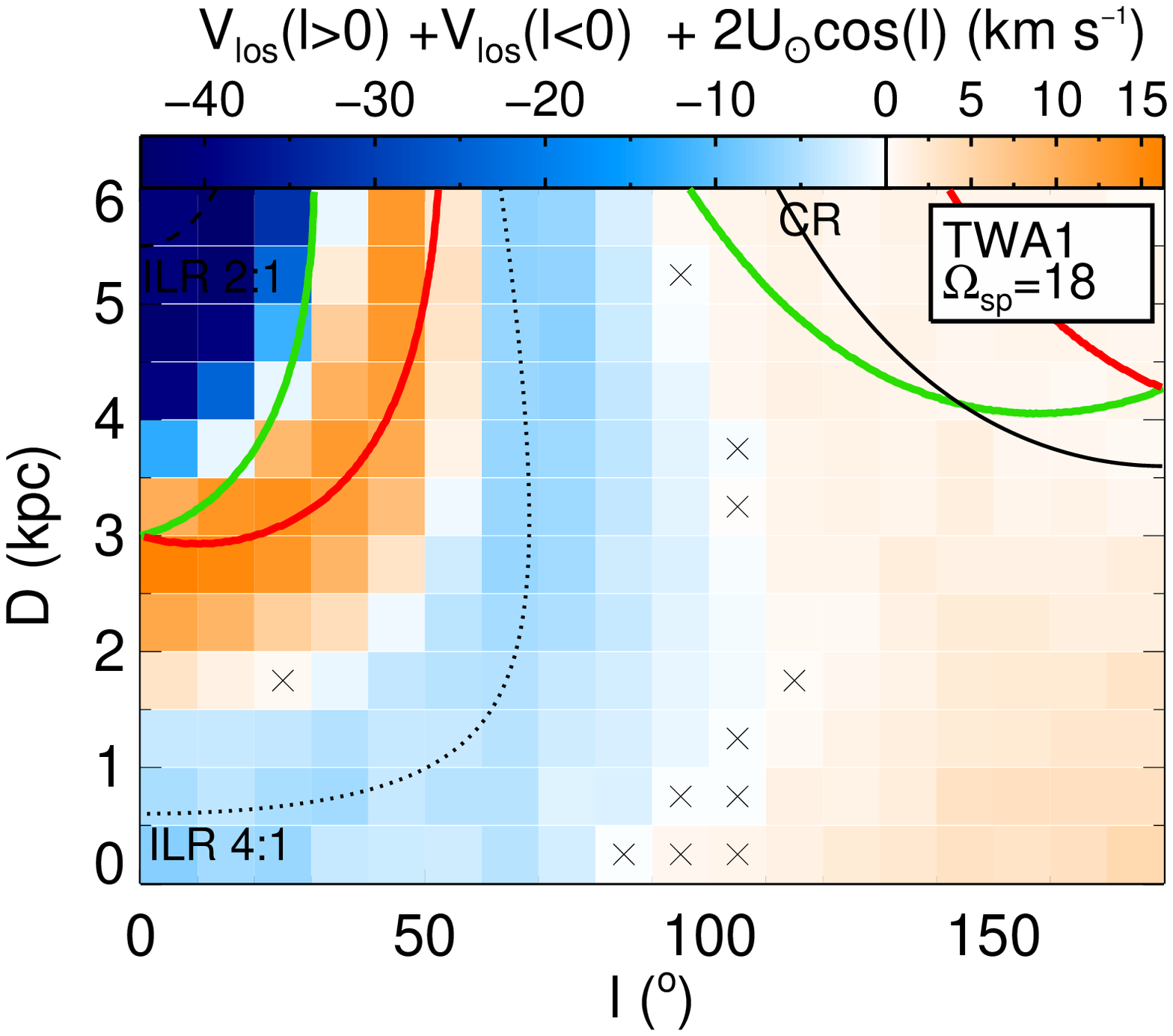}
  \caption{Difference between several kinematic quantities in symmetric bins ($l$,$d$) and ($-l$,$d$) for model TWA1. The first three plots show the transverse velocity dispersion $\sigma(\vt)$, skewness, and  kurtosis of the transverse velocity distribution. The last plot shows the difference between the \los velocity $\vlos$ compared to the expected value $-2\Us\cos l$. We plot a black cross in bins where these quantities are  statistically consistent with 0 with a $75\%$ confidence, i.e. where they are compatible with an axisymmetric model. The rest of the notation is as in \fig{vtsym2d}. 
}
         \label{variablessym}
   \end{figure*}

The approach of comparing the kinematics of symmetric longitudes in the Galaxy can be used with  quantities other than the median transverse velocity. We show in \fig{variablessym} the difference in symmetric longitudes between the dispersion, skewness,  and kurtosis of the $\vt$ distribution. The expected difference of these quantities for an axisymmetric model is 0, but we show  that they present patterns  related to the location of the arms and main resonances, similar to the median $\vt$. The differences are small but noticeable. In the last panel of \fig{variablessym} we show the differences between the median \los velocity in the symmetric longitudes. In this case, we compute $\vlos\,(l>0)+\vlos\,(l<0)$ and we subtract the expected value for an axisymmetric model, which is $-2\Us\cos l$. We see again similar patterns and, in particular, we observe high values up to $-40\kms$, which are even higher than the maximum values seen for the same model for transverse velocities (\fig{vtsym2d}).

\section{\gaia performance}\label{gaia}

 \begin{figure*}
   \centering
\includegraphics[width=0.95\textwidth]{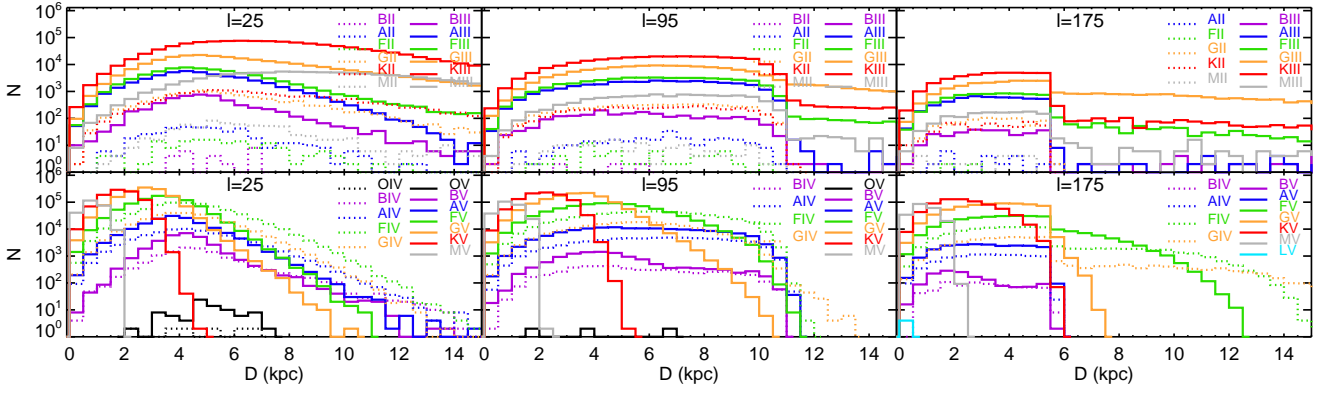}
  \caption{Number of stars in the GUMS model for three different directions: $l=25\deg$ (\emph{left}), $l=95\deg$ (\emph{middle}) and $l=175\deg$ (\emph{right}) in regions of angular size of  $\Delta l=\pm5\deg$ and $\Delta b=\pm5\deg$.  The top row corresponds to super-giant (dotted lines) and giant (solid) stars,  and the bottom row to sub-giant (dotted) and dwarf stars (solid). }
         \label{NGUMS}
   \end{figure*}

   \begin{figure*}       
      \centering\includegraphics[width=0.95\textwidth]{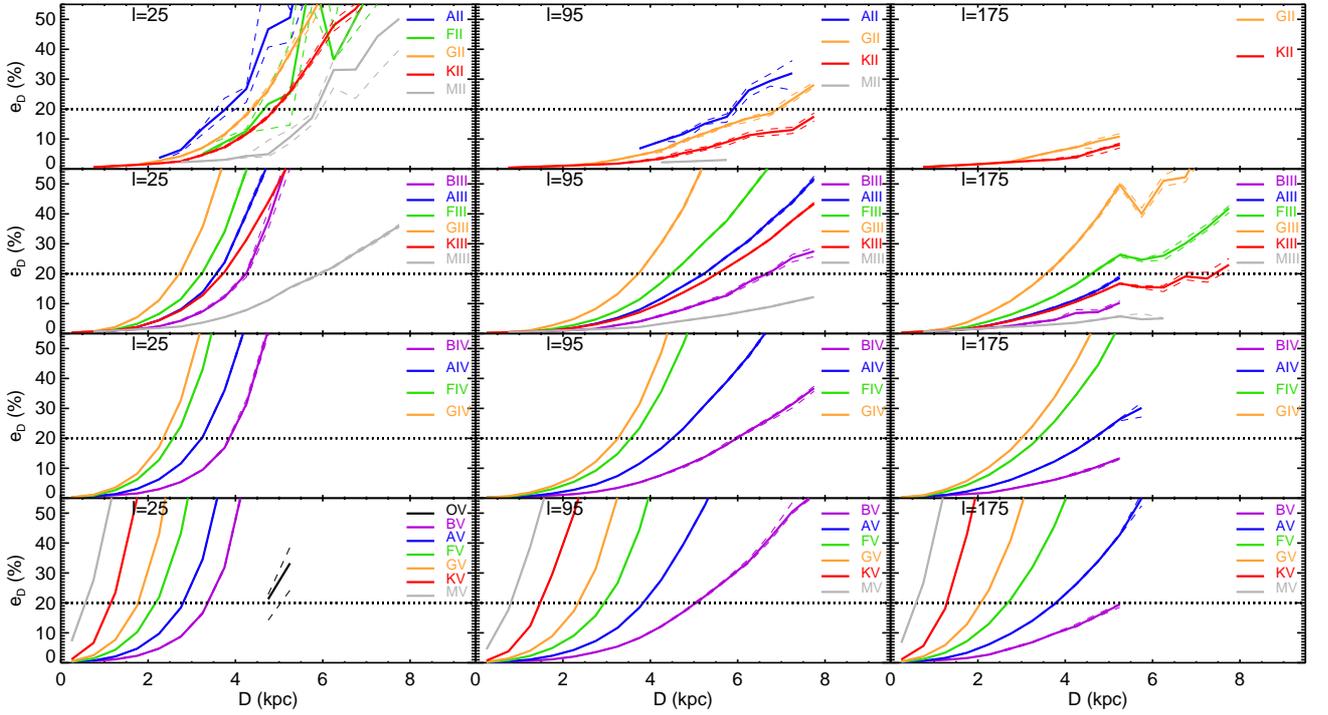}
    \caption{Median relative error in distance as  a function of distance for stars in the GUMS model for the different spectral types and luminosity classes for three different directions indicated in the panel labels. We only plot bins with at least ten stars. We use bootstrapping to estimate the error on the median ($75\%$ confidence limit), which we indicate with dashed lines.}
   \label{rpareGUMS}
   \end{figure*}
    
  \begin{figure*}
  \centering\includegraphics[width=0.95\textwidth]{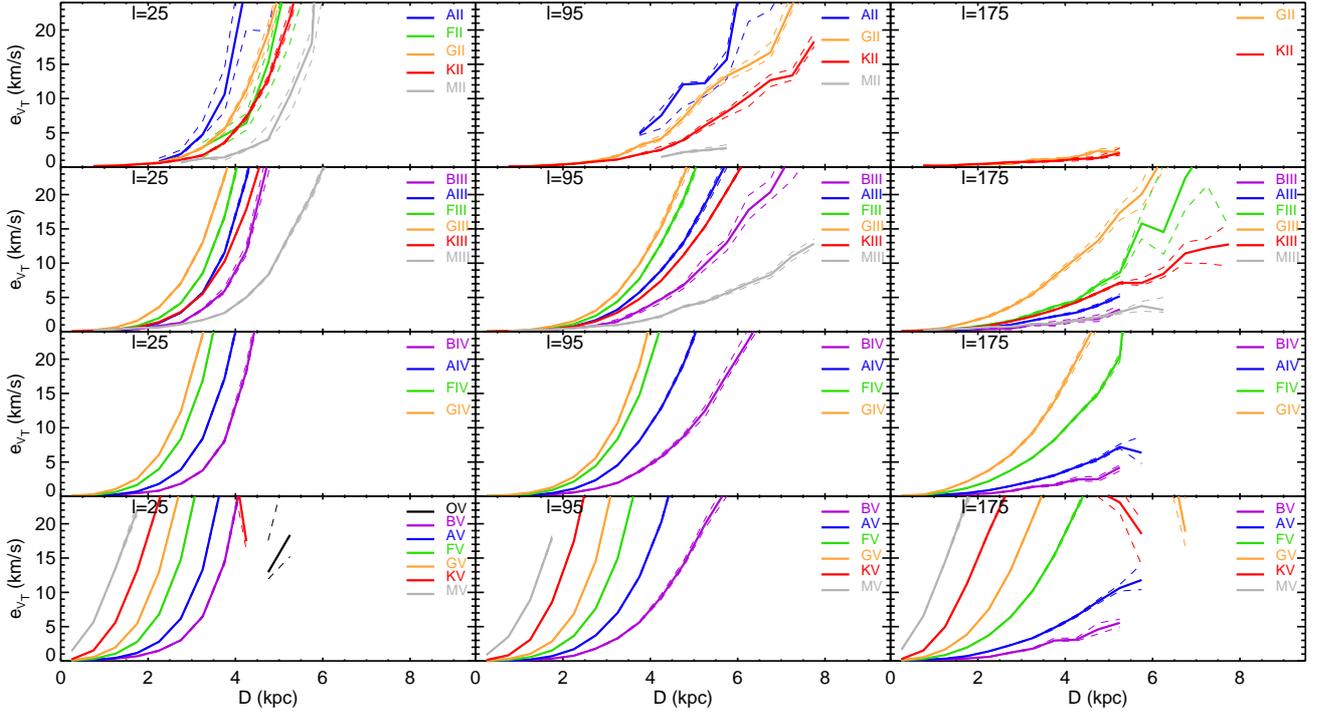}
    \caption{Median error in transverse velocity $\vt$ as  a function of distance for stars in the GUMS model for the different spectral types and luminosity classes for three different directions indicated in the panel labels. The dashed lines show the $75\%$ confidence limit of the median.}
   \label{eVTGUMS}
   \end{figure*}

 \begin{figure*}
  \centering\includegraphics[width=0.95\textwidth]{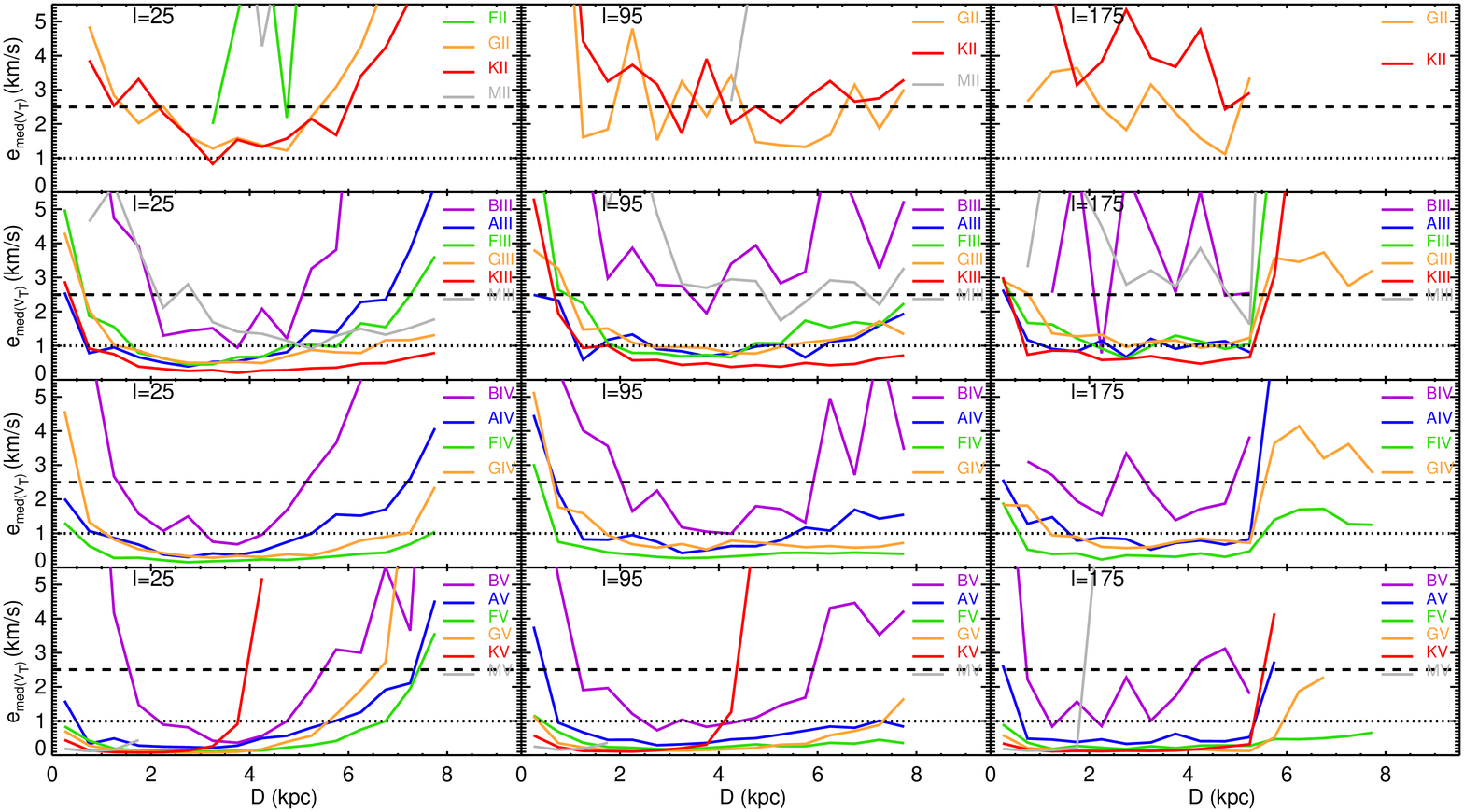}
  
  \caption{Error in the median transverse velocity $\vt$ as a function of distance for stars in the GUMS model for the different spectral types and luminosity classes, and for three different directions indicated in the panel labels. The two horizontal lines show the limits for errors of 1 and $2.5\kms$. }
   \label{emedVTGUMS}
   \end{figure*}

 In Section~\ref{sym} we established the conditions for which the signatures of spiral arms can be detected with our new approach of studying symmetric Galactic longitudes. These are: i) a maximum distance error of $e_D=20\%$ (Section~\ref{sym2}), and ii) a maximum kinematic error in the determination of $\D-\De$ of 2$\kms$ (Section~\ref{sym1}), which in turn requires an error in the median $\vt$ in a single longitude of $e_{{\rm med}(\vt)}=1\kms$. In this section we estimate whether some \gaia tracers meet these conditions.% and thus, if the kinematic signatures that we see in the models can be detected with \gaia data.  

To do this, we use the GUMS model that is a simulated catalogue of the sources expected to be observed by {\em Gaia}. The Galactic sources (stars) in GUMS are generated based on the Besan\c{c}on Galactic Model, which includes Galactic thin and thick disks, bulge and halo, based on appropriate density laws, kinematics, star formation histories, enrichment laws, initial mass function, and total luminosities for each of the populations, described in \citet{Robin2012}. GUMS gives us the simulated true values for \gaia observables.
These are the five astrometric parameters ($l$, $b$, $\varpi$, $\mu_l$, $\mu_b$), the \los velocity $\vlos$,  and the \gaia photometry (including the $G$ \gaia magnitude, and the two broadband magnitudes $G_{\rm{BP}}$ and $G_{\rm{RP}}$). The final \gaia catalogue will also provide atmospheric parameters (metallicity, surface gravity, and effective temperature) and extinction. 

The GUMS model includes multiple systems \citep{Arenou2011}. To determine which of these systems will be resolved by {\em Gaia}, we use a prescription employed in the Data Processing and Analysis Consortium\footnote{\url{http://www.cosmos.esa.int/web/gaia/dpac}}. The minimum angular separation on the sky that \gaia can resolve depends on the apparent magnitudes of the stars in the system, and, in the best case, it is $\sim38$ mas. We only consider stars that are resolved for which we have a reliable model for \gaia performance. %For the unresolved cases, a single detection is considered by computing the total integrated magnitude, averaging positions and taking the atmospheric parameters (such as surface gravity) of the primary star in the system.Lacking a model of the \gaia performances for unresolved systems, we use the same prescriptions as for single or resolved stars. 

To simulate {\em Gaia}-like errors for the GUMS sources, we use the code presented in \citet{RomeroGomez2015}\footnote{Available at \url{https://github.com/mromerog/gaia-errors}.}, updated to the post-launch performance,
as described in \cite{deBruijne2015}. %Up to date information is available from the \gaia web pages\footnote{\url{http://www.cosmos.esa.int/web/gaia/science-performance}}. 
The uncertainties on the astrometry, photometry, and spectroscopy are mainly a  function of the stellar magnitudes and colours.
The geometrical factors and the effect of the scanning law are also taken into account\footnote{\url{http://www.cosmos.esa.int/web/gaia/table-6}}. 
%The errors on the atmospheric parameters are computed following \citet{BailerJones13}.
%For the surface gravity we take a constant error of 0.25 dex, based on table~4 of \citet{BailerJones13}. 
Only stars with magnitude $G<20$ (the \gaia magnitude limit) are considered. We include the Galactic extinction given by GUMS, which is based on \citet{Drimmel2003}.

We extracted from GUMS the stars in three different longitudes , $l= 25\deg, 95\deg$, and $175\deg$, in regions of $\Delta l=\pm5\deg$ centred in these longitudes and a range in latitude of $b=[0,5]\deg$. 
 We  binned the data of these directions in distance with $\Delta D=0.5\kpc$ as in our simulations. \Fig{NGUMS} shows histograms of the number of stars as a function of true distance (i.e. without errors) in each of the longitudes (columns) for different spectral types (colours) and luminosity classes (rows). We  doubled the number of stars to account for the sky region below the Galactic plane $b=[-5,0]\deg$. With this we assume that the distribution and properties of stars is the same for $b=[0,5]\deg$ and $b=[-5,0]\deg$. Although this is not strictly true since the extinction can vary with latitude,  we only aim to estimate of the number of stars observed by \gaia in these directions.% and the uncertainty of the GUMS model for $b=[0,5]\deg$ is probably larger than the one made with our assumption. 

%in GUMS and also for the symmetric \los's $l= -25\deg, -95\deg$ and $-175\deg$, which is true except for the differences in extinction. 

As expected, giant and super-giant stars (top row) can be detected up to farther distances. The increase in the number of stars as a function of distance for small distances, subsequent plateau and decrease is due to a combination of aspects. Firstly, our bins, which have a fixed bin $\Delta D$, have a volume that increases with distance. For $l=25\deg$ the density of stars also increases with distance because of the density laws of the model.  There is a decrease in the number of stars for the same reason for the other directions. The sharp cut-off at around $11\kpc$ for $l=95\deg$ and at $5.5\kpc$ for $l=175\deg$ is due to the size of the disk of the GUMS models with radius of $14\kpc$. The stars that are seen beyond those distances are from the halo component. There is a limiting distance for which a star of a certain spectral type can be seen given the magnitude limit of {\em Gaia} for dwarf and sub-giant stars (bottom row). In most cases, this limit is not sharp but progressive. This is because of the spread in absolute magnitudes of each spectral type and the different extinction across our fields.

   In \fig{rpareGUMS} we show the median relative error in parallax for the different spectral types and luminosity classes as a function of the true distance for the three longitudes. The differences between the errors in different longitudes are due to extinction, which increases the astrometric errors in the inner fields. We indicate the limit of a relative error $e_D=20\%$ with a horizontal line (condition i). To meet this condition, for longitudes of less extinction ($l=95\deg$ and $l=175\deg$) we could use giant stars (second row) up to 4 and $6\kpc,$ depending on  the spectral type. The MIII, BIII, and KIII stars are the best tracers in this case. Most of the supergiant stars (top row) also fulfil this condition up to $6\kpc$. For $l=25\deg$ the high extinction puts the limits for giant stars at 3-$4\kpc$. For dwarfs stars (bottom row), the limit of $e_D=20\%$ is between $\sim1$ and $4\kpc$. The OV, BV, AV, and FV stars are the best tracers. Some sub-giant stars also fulfil the distance error condition up to $\sim2$-$4\kpc$.

  We  show in  \fig{eVTGUMS} the median error in transverse velocity. These errors influence the precision for which the median $\vt$ can be determined. The errors for giant stars are smaller than $5\kms$ up to $\sim3$-$5 \kpc$, depending on the direction. For dwarf stars this limit is achieved at $\sim1$-$4\kpc$.

      The precision on the determination of the median transverse velocity of a certain population $e_{{\rm med}(\vt)}$ (not to be confused with the median error of the population ${e}_{\vt}$) depends on the number of observed stars and the dispersion in $\vt$. The latter, in turn, depends on the intrinsic dispersion of the population and the error of the measurements ${e}_{\vt}$.  To estimate the error $e_{{\rm med(}\vt)}$, we compute the median $\vt$ of the GUMS velocities after the addition of {\em Gaia}-like errors at each bin in distance, and compute its error with the bootstrapping technique. We take  the limits of the $75\%$ confidence interval as the error on the median. Since these limits are not necessarily symmetric, we use the maximum absolute difference between the median and the lower or upper limit. We also have to account for the fact that we only extracted the region of $b=[0,5]\deg$ from GUMS. To double the number of stars and thus also include the negative latitude, we assume that the error on the median scales similarly as the error on the mean by a factor $1/\sqrt{2}$. Also, strictly speaking, this would be the error of the  median of the $\vt$ in the distribution of the GUMS model. This is not necessarily similar to that shown in our spiral arms models, but we assume that the statistical and measurement errors would yield similar numbers.
      
      In \fig{emedVTGUMS} we show $e_{{\rm med}(\vt)}$. We indicate the error of $1\kms$ (dotted, condition ii) and, additionally, the error of $2.5\kms$ (dashed),     
      which correspond to the limit where a signal of $\D-\De=5\kms$ could be detected with confidence. 
       For super-giant stars (top row), the number of observed stars is too small to fulfil our kinematic condition ($e_{{\rm med}(\vt)}<1\kms$).  For giant stars (second row), the error in the median $\vt$ is $\lesssim1\kms$ up to $6\kpc$ for most of the spectral types except for BIII and MIII stars. For dwarf stars (bottom row), the error is well below $1\kms$ up to $4\kpc$ and even beyond for the directions with less extinction, except for B stars. The errors for sub-giants stars happen to be in between giant and dwarf stars and for most of the spectral types (except BIV) the error is $e_{{\rm med}(\vt)}< 1\kms$. 
      
      To conclude, we find that the condition of the error in distance (i) necessary for the application of our method to \gaia data is more demanding than the kinematic precision condition (ii). Both conditions are met for most of the spectral types of giant stars up to distances of $\sim6\kpc$ (especially KIII stars) and for dwarf stars (especially AV and FV stars) up to closer distances $\sim3-4\kpc$ and with better precision (down to $0.5\kms$), but slightly smaller distances for the longitudes close to $l=0\deg$.

\section{Discussion and conclusions}\label{conclusions}

 We have shown that comparing the stellar kinematics in symmetric longitudes (i.e. $+l$ and $-l$) is a useful strategy to put constraints on the properties of spiral arms. We have also seen that the Gaia catalogue will allow us to measure the disk kinematics with enough precision for this approach to be successful.   
 
 To do this, we modelled the effects of the arms on the stellar kinematics  via controlled orbital integrations in analytic potentials and self-consistent simulations. We studied the trends and magnitudes of the difference between the median transverse velocity as a function of distance in symmetric longitudes. Whereas this difference is expected to be constant and predictable ($2\Us\sin l$) in an axisymmetric disk, we find that in our models it oscillates in a pattern related to the location of the arms and their resonances. The typical discrepancies between the model values and the axisymmetric predictions are of $\sim2\kms$. The detection of this pattern will allow us to quantify the importance of the effect of spiral arms on the orbits of stars in different regions of the disk. Also, it will enable us to put constraints on some properties of the spiral structure, namely the location of the arms, their main resonances and thus, their pattern speed, as well as on their dynamical nature (e.g. grand design versus transient and floculent multiple arms) directly related to the different origin scenarios of spiral structure.
 
 Furthermore, we showed with the GUMS model that the number of stars and the distance and kinematic precision of certain stellar tracers is excellent for detecting the kinematic signatures that we see in the models. With giant stars, we will be able to measure the median transverse velocity with precision smaller than $1\kms$ up to a distance of $\sim4$-$6\kpc$ and with dwarfs up to $\sim$2-$4\kpc$ and even better kinematic precision ($<0.5\kms$). Although KIII, AIV, AV, and FV stars seem to be the best tracers, an optimum approach would be to examine  several spectral types, combined and individually, since each tracer offers different qualities in terms of reached distance and precision. By adding different tracers, we can definitively improve the precision of the measured median transverse velocities, but we must be cautious since different spectral types are composed of populations with different intrinsic dispersions that might have different responses to the potential of spiral arms. 
 
 Even more promising, for certain spectral types, we could obtain more precise distances from photometric determinations, e.g. for red clump stars \citep{Bovy2014}, which are also discussed in \citet{RomeroGomez2015} as good candidate tracers for studies of the Galactic bar. The M0III stars are the stellar tracer chosen in \citet{Hunt2014} that allows them to recover the structure and kinematics of the disk with M2M modelling.
  Although these intrinsically bright stars have the smallest \gaia errors in parallax and proper motions, we find here that their smaller number compared to, for example KIII stars, is not enough to determine the median transverse velocity with the precision sufficient for our approach.
 
 Although there has currently been some effort made in producing good axisymmetric models to describe the MW kinematics, such as in \citet{Bienayme2015} and \citet{Sanders2015}, our approach does not require the assumption of an axisymmetric model as in previous studies. Our strategy does require, however, knowledge of the value of $\Us$.  There is currently some debate about this value \citep{Schonrich2010,Sharma2014,Huang2015}, which oscillates between $7$ and $14\kms$. Nevertheless, some studies such as \citet{Schonrich2012} presented new methods for measuring the peculiar solar motion that, together with the exceptional \gaia catalogue, are expected to deliver the Sun's peculiar velocity accurately. However, the kinematic effects of spiral arms are not considered in these methods. Here, with our models we quantified that the determination of $\Us$ by a simple median of the velocities 
 %would be biased by few $\kms$. However, we have seen that a median of the velocities only 
 in the regions less affected by the arms would yield a value of $\Us$ that differs from the true value by less than $0.5\kms$. We  determined that, in all of our models, this is in the direction of the anti-centre ($l=180\deg$). We also checked that assuming a $\Us$ that is wrong by $\sim1\kms$ would not significantly bias the results obtained with our approach.

A caveat of our study is that we assumed the same extinction in symmetric longitudes. A different extinction would yield different accuracy in parallax and proper motion. This would not create a strong bias, but could compromise the capabilities of detecting the predicted signatures if the extinction in one longitude is much higher than that estimated here. However, a good extinction map will be also a product of the \gaia data and will allow us to evaluate the practical implications of this issue.

Most of the simulations in our study have spiral arms, but no bar. While it is clear that our Galaxy has a bar, this approach allow us to explore the isolated effects of the spiral structure. We are aware that the comparison of the kinematics of symmetric longitudes in real observations will be harder to interpret because of this and other additional effects. A kinematic asymmetry between the first and third quadrant, presumably associated with the effects of the bar, has already been reported in \citet{Humphreys2011}. Other effects that we neglect  are the passage of satellite galaxies nearby the disk \citep{Quillen2009} or structures such as the Gould Belt that could distort our maps. We also tested our approach with self-consistent N-body models with the purpose of studying more complete models (for instance including a bar), and we have already seen the complexity that these cases can involve.

We have also seen that the approach of comparing the kinematics of symmetric longitudes could also be useful with other moments of the transverse velocity distribution function such as the dispersion, skewness, and kurtosis, and also with the \los velocities. Although it remains to be seen whether these signatures will be detected with statistical significance by {\em Gaia}, we will definitively explore this avenue with the data. We also suggest that our proposed approach can be used with surveys of radial velocities such as APOGEE, 4MOST, and WEAVE. The handicap is that these surveys do not cover a wide range or longitudes and their symmetric counterparts. The use of different surveys to compare symmetric longitudes can give rise to additional biases due to differences in the selection function of the surveys or a distance that is difficult to determine, which will need to be evaluated.

Our proposed method can be applied to the second \gaia data release\footnote{\url{http://www.cosmos.esa.int/web/gaia/release}}
 scheduled for summer 2017, which will contain the five astrometric parameters for most of the single stars in the final catalogue as well as integrated BP/RP photometry and astrophysical parameters for stars with appropriate standard errors. This release will not enable us to separate the different spectral types completely, but that will be possible with the third release (2017/2018).
 Earlier searches can be conducted 
 %using earlier \gaia releases like the HTPM (a Hundred Thousand Proper Motions from the combined solution Hipparcos-Gaia) although this would require the use of Hipparcos parallaxes. More interestingly is the possible release of 
 using the {\em Tycho}-{\em Gaia} astrometric solution \citep[TGAS;][]{Michalik2015}, which will contain proper motions and parallaxes for 2.5 million stars (summer 2016).

\begin{acknowledgements}
 We thank the anonymous referee for his/her comments. TA is supported by an ESA Research Fellowship
in Space Science.

\end{acknowledgements}

\bibliographystyle{aa} % style aa.bst
%\bibliography{/data/users/antoja/bib/mybib} % your references Yourfile.bibQbÿ¿¿¿¿¿bÿ¿¿¿¿¿b
%\bibliography{/user_data/tantoja/publications/mybib/mybib}
%\bibliography{/home/tantoja/Desktop/MyCollection}
\bibliography{mybib}

\end{document}